\documentclass[a4paper,10pt]{article}

\usepackage[a4paper,margin=1in]{geometry}

\usepackage{graphicx}      
\usepackage{amsmath}       
\usepackage{amssymb}       
\usepackage{bm}            
\usepackage{physics}       
\usepackage{mathtools}     
\usepackage{hyperref}      
\usepackage[numbers,sort&compress]{natbib}
\usepackage{booktabs}      
\usepackage{caption}       
\usepackage{subcaption}    
\usepackage{xcolor}        
\usepackage{float}         
\usepackage{enumitem}
\usepackage{titletoc} 

\title{Physics constraints and response validation in discrete-time reduced-order modeling: from idealized turbulent systems to climate dynamics}
\author{
Fabrizio Falasca\textsuperscript{1,*} and Laure Zanna \textsuperscript{1}\\[1ex]
\textsuperscript{1}Courant Institute School of Mathematics, Computing and Data Science \\ New York University, New York, NY, USA \\
\textsuperscript{*}Corresponding author: \texttt{fabrifalasca@gmail.com}
}%

\begin{document}
\maketitle

\begin{abstract}
A central challenge across science and engineering is to build data-driven reduced-order models of turbulent dynamical systems that reproduce stationary statistics, predict responses to external perturbations, and remain practical for real-world applications. To this end, we introduce an abstract discrete-time formulation of turbulent dynamical systems with exact energy-conserving nonlinearities. Parameterizing this structure with neural networks yields stable, physics-constrained reduced-order models. We then use the fluctuation-dissipation theorem (FDT) to validate the emulators' forced responses from
unperturbed data alone, testing models beyond stationary statistics. The FDT also identifies candidate direct causal links, which are used as regularization terms only when they pass validation. We first test the framework on two idealized models of geophysical turbulence: the proposed emulators reproduce stationary statistics and accurately predict responses to weak and strong forcings, despite being trained solely on unperturbed data. We then move beyond idealized systems to model tropical climate dynamics from reanalysis data, where data scarcity, partial observability, and sensitivity to choices of stochastic parameterizations become central challenges. The resulting physics-constrained model reproduces key statistics of the El Ni\~no-Southern Oscillation (ENSO) and qualitatively captures its cumulative responses to perturbations; augmenting it with a non-Markovian stochastic closure substantially improves quantitative agreement with the FDT benchmark. This response-validated model is then used to characterize long-term causal drivers of ENSO variability. The proposed methodology establishes a modular framework for stable reduced-order models capable of probing causal mechanisms in realistic, partially observed turbulent systems.
\end{abstract}


\startcontents[main]

\section*{Contents}
\begingroup
\setcounter{tocdepth}{3}
\printcontents[main]{}{1}{}
\endgroup

\section{Introduction} \label{sec:intro}

Turbulent dynamical systems are ubiquitous across science and engineering, spanning fluid dynamics, climate science, materials science, and neuroscience \cite{MajdaIntroTurbulence}. Modeling and predicting their complex multiscale dynamics often requires replacing high-dimensional descriptions with reduced-order models, in which relevant coarse-grained variables evolve according to effective stochastic dynamics \cite{nonEqStatMech,Izvekov,Prinz,Harrison,Rangan,ROBERTS200812,Roberts2006,Hasselmann,MTV1,NormalForms,GhilLucarini,LucariniChekroun,NanChenBook,KITSIOS,KITSIOSOcean}. Useful reduced-order models should capture relevant dynamical features of such systems in statistical equilibrium and approximate how key statistics, such as means and variances, respond to external perturbations \cite{majdaSIAM}. Capturing forced responses is essential for advancing scientific understanding, as it allows one to probe causal mechanisms and explore counterfactual \textit{``what if?''} scenarios when experiments on the natural system are impractical or impossible. This is particularly relevant for the Earth's climate system, where predicting forced responses is essential for both practical projection (i.e., ``climate change-like'' questions) and fundamental understanding (i.e., causal questions) \cite{FabriCoarseGraining}. In practice, this program poses two intertwined challenges: (i) identifying an appropriate coarse-grained representation $\mathbf{x}_t$ of a partially observed, high-dimensional physical system; and (ii) constructing a model for the evolution of $\mathbf{x}_t$ that remains stable over long times and faithfully predicts responses to perturbations \cite{Penland89,Horenko,Pavliotis,majdaConstraints,KRAVTSOV,KONDRASHOV201533,brunton,Andre1,CGChecnMajda,nanChen,GiorginiScore1,GiorginiScore2,Azencott,Ferretti}. The first challenge is a major bottleneck in data-driven modeling of real-world complex systems, as it depends strongly on the processes under consideration and the scales of interest \cite{FabriCoarseGraining}. Crucially, the selected coarse-grained state inherently dictates the causal pathways and responses that any resulting model can capture. Even after a physically meaningful coarse-grained state is selected, memory effects may persist in the deterministic residuals; these can optionally be modeled through non-Markovian stochastic closures \cite{KRAVTSOV,KONDRASHOV201533}. We discuss this first challenge throughout the paper and provide a concrete instance of it in our concluding climate application. The main methodological proposal of this paper tackles the second challenge: given a set of coarse-grained variables $\mathbf{x}_t$, our goal is to learn a stable model that faithfully reproduces their statistical properties and predicts responses to external forcing. Recent advances in machine learning contribute to this inverse modeling task by providing flexible neural network architectures for data-driven emulation, with promising applications in fluid dynamics \cite{GENEVA2020109056,MILANO20021,Vlachas}, weather \cite{Pathak2022,fourcastnet3}, and climate science \cite{Watt-Meyer2023,Watt-Meyer2024,Lucie,Camulator,Samudra,WillG}. However, neural network-based models are often not constrained by the physical and causal structure of the underlying system. This can lead to several failure modes, such as violations of energy conservation \cite{majdaConstraints,KONDRASHOV201533,Nikolaj} and consequent blow-up of solutions \cite{Chris}, as well as incorrect responses to perturbations \cite{DAemulators,Senne,Bosong}. Indeed, a model may achieve low training error and reproduce unperturbed stationary statistics while still becoming unstable, or producing physically incorrect responses, when probed by external perturbations.\\ 

Here, we address these challenges by introducing a flexible strategy for constructing physics-constrained neural models, while using statistical response theory to (i) rigorously evaluate the resulting models beyond stationary statistics and (ii) infer causal constraints to guide model training. We focus on the less-explored context of discrete-time modeling, which avoids practical inconsistencies that arise when continuous-time conservation principles are applied to temporally coarse-grained data. To this end, we propose an abstract, discrete-time formulation of turbulent dynamical systems that embeds exact energy-conserving nonlinearities. Parameterizing this discrete mapping with neural networks yields stable, physics-constrained reduced-order emulators. Building on this physical foundation, we leverage the Fluctuation-Dissipation Theorem (FDT) \cite{FALCIONI,MajdaBook} as a tool for response validation. The time-dependent response operator derived from the FDT establishes a response-theory benchmark to validate emulator responses to small external perturbations using only unperturbed stationary data, circumventing the need for numerical ground truth. The same operator, when evaluated at the shortest resolved time step, identifies candidate direct causal couplings across the system's degrees of freedom \cite{Baldovin}. When these inferred links pass validation, they can be incorporated into model training through a regularization term that penalizes spurious, non-causal interactions. In this work, we refer to this last step as ``causal regularization'' or ``causal constraints'' interchangeably. The proposed physics and causal constraints operate differently. The physics constraints, specifically formulated for discrete settings, provide the foundation of our approach: defined and imposed \textit{a priori} from theoretical arguments, they promote robust inference in data-scarce regimes and stable long-time simulations. In contrast, the causal constraints are inferred from data, making their accuracy dependent on data availability and system complexity. We therefore outline systematic validation criteria that must be met before enforcing these causal constraints. To evaluate this framework, we first test it on two idealized models of geophysical turbulence: a stochastic version of the Charney-DeVore model \cite{charneyDeVore} and a symmetry-broken variant of the Lorenz-96 system \cite{Lorenz96} in a strongly turbulent regime. The resulting neural emulators are stable, capture stationary statistics, and reproduce responses to both weak and strong external forcings (including time-dependent shifts of the attractor) despite being trained exclusively on unperturbed data. We then move beyond idealized models and target the setting of primary interest: reduced-order modeling of real-world turbulent systems whose governing equations are unknown. As a concrete instance, we focus on large-scale tropical climate dynamics using reanalysis data, with the goal of quantifying the cumulative causal sensitivity of the El Ni\~no--Southern Oscillation (ENSO) to perturbations across physical fields and ocean basins. To this end, we construct a minimal multivariate physics-constrained stochastic model. This application exposes several practical challenges of realistic reduced-order modeling, often absent in idealized models: the available record is short, the system is only partially observed, and unresolved atmospheric and oceanic processes can result in temporally correlated residuals. We validate the model's perturbation-response skill against empirical cumulative responses estimated through the FDT over a 12-month horizon, where the Markovian physics-constrained model captures the qualitative structure of the responses. The model is then augmented with a non-Markovian stochastic closure, substantially improving quantitative agreement with the FDT benchmark. The response-validated non-Markovian model is then used to predict 10-year cumulative sensitivity maps, quantifying ENSO's sensitivity to different physical variables and tropical basins. This final application demonstrates the flexibility and practical value of the proposed framework for modeling and understanding the dynamics of complex, real-world turbulent systems.

\section{Physics constraints for discrete-time turbulent dynamical systems} \label{sec:physics_constrained}
\subsection{Previous work} 
We start from the abstract representation of turbulent dynamical systems adopted by Majda and collaborators \cite{MajdaBook,MajdaIntroTurbulence} as
\begin{equation}
\begin{aligned}
\dot{\mathbf{x}} = \mathbf{F} + \mathbf{A}\mathbf{x} + \mathbf{B}(\mathbf{x},\mathbf{x}) + \mathbf{\Sigma}\bm{\xi}(t),
\end{aligned}
\label{eq:majdaQuadratic}
\end{equation}
which arises when a large class of fluid flows is projected onto orthogonal basis functions. In Eq.~\eqref{eq:majdaQuadratic}, $\mathbf{F}$ and $\mathbf{\Sigma}\bm{\xi}(t)$ respectively represent deterministic and stochastic forcings, with $\bm{\xi}(t)$ denoting a standard Gaussian white-noise process. The linear operator $\mathbf{A}$ is typically decomposed as $\mathbf{A} = \mathbf{L} + \mathbf{D}$, with $\mathbf{L}$ skew-symmetric ($\mathbf{L}^{\mathrm T} = -\mathbf{L}$), representing dispersion processes, and $\mathbf{D}$ symmetric and negative definite ($\mathbf{D}^{\mathrm T} = \mathbf{D} < 0$), representing dissipative processes (e.g., surface drag, viscosity).\\
The quadratic interactions $\mathbf{B}(\mathbf{x},\mathbf{x})$ arise from projecting the conservative advective nonlinearities of the underlying fluid equations onto the retained basis functions. Crucially, the quadratic operator conserves energy by itself, satisfying $\mathbf{x} \cdot \mathbf{B}(\mathbf{x},\mathbf{x}) = 0$ \cite{Kwasniok,majdaConstraints}, with the system energy defined as $E = (1/2)\mathbf{x}^\mathrm{T}\mathbf{x}$. Turbulent dynamical systems as represented in Eq.~\eqref{eq:majdaQuadratic} are often characterized by high-dimensional phase spaces and many unstable directions, reflected, for example, in a large number of positive Lyapunov exponents on the attractor \cite{MajdaIntroTurbulence}. The energy-conserving quadratic nonlinear interactions $\mathbf{B}(\mathbf{x},\mathbf{x})$ play a central structural role: they mitigate these linear instabilities by redistributing energy from unstable modes to stable ones, where dissipation dominates, thereby supporting a statistical steady state \cite{SapsisMajdaPhysicaD,SapsisMajdaPNAS,MajdaEnergy,SapsisPoF}. Building on \cite{VANGASTELEN2024113003,SANDERSE2020109736,PortHamiltonianNN}, we extend the framework in Eq.~\eqref{eq:majdaQuadratic} to general conservative nonlinearities as:
\begin{equation}
\begin{aligned}
\dot{\mathbf{x}} = \mathbf{F} + \mathbf{A}\mathbf{x} + \mathbf{S}(\mathbf{x})\mathbf{x} + \mathbf{\Sigma}\bm{\xi}(t).
\end{aligned}
\label{eq:majda_generalized}
\end{equation}
The extension in \eqref{eq:majda_generalized} is obtained by substituting the quadratic term $\mathbf{B}(\mathbf{x},\mathbf{x})$ with general nonlinearities $\mathbf{S}(\mathbf{x})\mathbf{x}$, where $\mathbf{S}(\mathbf{x})$ is constrained to be skew-symmetric, $\mathbf{S}^\mathrm{T}(\mathbf{x})=-\mathbf{S}(\mathbf{x})$. This formulation generalizes the quadratic term in Eq.~\eqref{eq:majdaQuadratic} to a broader class of energy-conserving nonlinearities of the form $\mathbf{S}(\mathbf{x})\mathbf{x}$, since $\mathbf{x}^{\mathrm T}\mathbf{S}(\mathbf{x})\mathbf{x}=0$.

\subsection{An abstract discrete-time formulation for turbulent dynamical systems}
Observational data are fundamentally discrete and temporally coarse-grained, motivating discrete-time mapping models of the form $\mathbf{x}_{t+1}=\mathbf{f}(\mathbf{x}_t)$. However, physics constraints are typically formulated in the continuous-time limit. To bridge the gap between finite-time data-driven modeling and continuous-time physics constraints, we propose a discrete, abstract formulation of turbulent dynamical systems that preserves exact energy-conserving nonlinearities at arbitrary sampling intervals. Given observations of an $n$-dimensional stochastic nonlinear system $\mathbf{x}_t \in \mathbb{R}^{n}$, we model the discrete-time forward dynamics as:
\begin{equation}
\begin{aligned}
\mathbf{x}_{t+1} = \mathbf{F} + \mathbf{M}\mathbf{Q}(\mathbf{x}_t)\mathbf{x}_t + \mathbf{\Sigma}\bm{\xi}_t,
\end{aligned}
\label{eq:discrete_case}
\end{equation}
where $\mathbf{F}$ and $\mathbf{\Sigma}\bm{\xi}_t$ represent the effective deterministic and stochastic forcings integrated over the finite time step. The nonlinear operator $\mathbf{Q}(\mathbf{x}_t)$ is constrained to be strictly orthogonal ($\mathbf{Q}^\mathrm{T}\mathbf{Q} = \mathbf{I}$). In our implementation, $\mathbf{Q}(\mathbf{x}_t)$ is parametrized as the exponential of a skew-symmetric matrix (see Section \ref{sec:fitting}); hence $\det \mathbf{Q}(\mathbf{x}_t)=1$, so that $\mathbf{Q}(\mathbf{x}_t)\in SO(n)$, i.e. the special orthogonal group.\\ 

At each time step, the state-dependent term $\mathbf{M}\mathbf{Q}(\mathbf{x}_t)\mathbf{x}_t$ factors into two sequential operations:  an energy-preserving rotation, $\mathbf{v}_t = \mathbf{Q}(\mathbf{x}_t)\mathbf{x}_t$, followed by a linear transformation $\mathbf{M}\mathbf{v}_t$. The isolated nonlinear term redistributes energy across modes while contributing no net growth or decay to the energy budget, strictly preserving the $L_2$-norm of the state:
$$ \|\mathbf{Q}(\mathbf{x}_t)\mathbf{x}_t \|^2 = \mathbf{x}_t^\mathrm{T}\mathbf{Q}^\mathrm{T}(\mathbf{x}_t)\mathbf{Q}(\mathbf{x}_t)\mathbf{x}_t = \|\mathbf{x}_{t}\|^2. $$
Consequently, energy growth in the deterministic forward map is controlled by the dominant singular value of $\mathbf{M}$. Optional constraints can be imposed on $\mathbf{M}$, as further discussed at the end of this section; however, we often find in practice that leaving the linear operator unconstrained yields more flexible yet still stable data-driven models, provided the nonlinear energy-conservation constraint is strictly enforced.\\

In the context of data-driven modeling, the discrete-time formulation in Eq.~\eqref{eq:discrete_case} offers distinct advantages over its continuous-time counterpart in Eq.~\eqref{eq:majda_generalized}. In the limit of small \(\Delta t\), Eq.~\eqref{eq:discrete_case} recovers Eq.~\eqref{eq:majda_generalized} as it can be shown to be consistent with a numerical splitting approximation \cite{LieTrotter}. However, unlike the continuous formulation, Eq.~\eqref{eq:discrete_case} can learn effective dynamics from coarse-grained observations at arbitrary time steps without violating the exact energy-conservation constraint of the nonlinear operator. In practical applications, the operators $\mathbf{M}$ and $\mathbf{Q}$ should be interpreted as effective coarse-grained quantities to be learned that need not coincide with operators derived from a small-$\Delta t$ discretization of the underlying continuous-time dynamics. In Section~2 of the SM we provide a derivation of Eq.~\eqref{eq:discrete_case} via a numerical splitting procedure, we recast a stochastic triad model in the proposed formulation, and we demonstrate that the framework enables stable learning from severely sub-sampled data where continuous-time data-driven models fail. To provide geometric intuition for the proposed discrete-time framework, we also recast the deterministic Lorenz-63 model in the form of Eq.~\eqref{eq:discrete_case} in Appendix~\ref{app:app_A}.

\subsubsection{Neural model fitting} \label{sec:fitting}
Given a stationary trajectory $\mathbf{x}_t$, we parametrize the deterministic drift $\mathbf{f}(\mathbf{x}_t) = \mathbf{F} + \mathbf{M}\mathbf{Q}(\mathbf{x}_t)\mathbf{x}_t$ by training a multilayer perceptron (MLP) to minimize the mean squared error $\text{MSE}(\mathbf{x}_{t+1},\mathbf{f}(\mathbf{x}_t))$. States are standardized to zero mean and unit variance before training. Importantly, since the $L_2$ norm is not preserved under shifting and scaling, the energy-conserving constraint is imposed in the original physical coordinates: the MLP takes the standardized state as input, but the resulting rotation acts on the unstandardized physical state; see SM Sec.~1.1 for details. We enforce the orthogonality of $\mathbf{Q}(\mathbf{x}_t)$ by predicting the matrix exponential $\mathbf{Q}(\mathbf{x}_t) = \exp(\mathbf{S}(\mathbf{x}_t))$ of a skew-symmetric matrix $\mathbf{S}(\mathbf{x}_t)$. For high-dimensional systems we instead use the Cayley transform $\mathbf{Q}(\mathbf{x}_t) = (\mathbf{I}-\mathbf{S}(\mathbf{x}_t))^{-1}(\mathbf{I}+\mathbf{S}(\mathbf{x}_t))$ which provides a memory-efficient orthogonal parametrization. Because both the matrix exponential and the Cayley transform map a skew-symmetric generator into the special orthogonal group $SO(n)$, $\mathbf{Q}(\mathbf{x}_t)$ has determinant equal to one, ensuring the network learns pure rotations. Crucially, the linear terms are initialized with a ``first guess'' ordinary least squares (OLS) solution, while the MLP is initialized to yield $\mathbf{Q}(\mathbf{x}_t) = \mathbf{I}$. This ensures the joint optimization starts from a Linear Inverse Model (LIM) \cite{Penland89} baseline before learning the nonlinear corrections, and ultimately enables model fitting even in severely data-scarce regimes. Further implementation details, including careful data rescaling, are discussed in SM Sec.~1.\\

\textit{Choice of stochastic parametrization.} Once the physics-constrained deterministic drift $\mathbf{f}$ is fitted, we complete the model with a stochastic closure. First, we compute the residuals of the deterministic drift as $\mathbf{r}_t = \mathbf{x}_{t+1} - \mathbf{f}(\mathbf{x}_t)$. For the idealized numerical examples (the Charney-DeVore and Lorenz-96 models), we simply define the diffusion matrix $\mathbf{\Sigma}$ as a diagonal matrix containing the standard deviations of these residuals. For the real-world climate application, where cross-correlations among variables are significant, we employ two distinct closures. First, for a purely Markovian closure, we compute the empirical residual covariance matrix $\mathbf{C} = \langle \mathbf{r} \mathbf{r}^\mathrm{T} \rangle$. We then derive a lower triangular diffusion matrix $\mathbf{\Sigma}$ as the Cholesky factor of $\mathbf{C}$, such that $\mathbf{C} = \mathbf{\Sigma}\mathbf{\Sigma}^\mathrm{T}$. Second, to account for unresolved memory effects, we implement a non-Markovian closure by integrating our framework with the multilevel stochastic modeling strategy of \cite{KONDRASHOV201533}.

\subsubsection{Optional constraints on the linear operator $\mathbf{M}$} Finally, we note that it is possible to add constraints to the linear operator $\mathbf{M}$ in analogy with the continuous-time decomposition $\mathbf{A}=\mathbf{L}+\mathbf{D}$ introduced above. Assuming a unit discrete time step $\Delta t = 1$, this parameterization takes the form $\mathbf{M}=\exp(\mathbf{L}+\mathbf{D})$, where $\mathbf{L}^{\top}=-\mathbf{L}$ and $\mathbf{D}^{\top}=\mathbf{D} < 0$ (see above for the continuous case). The resulting mapping constitutes a discrete-time analogue of the Majda-type turbulent dynamical system in Eq.~\eqref{eq:majdaQuadratic}: energy is injected via deterministic and stochastic forcing, conservatively redistributed across degrees of freedom by the nonlinear term, and dissipated through the linear operator. This fully constrained formulation provides strong structural stability guarantees \cite{MajdaIntroTurbulence}. In practice, however, enforcing this additional structure on $\mathbf{M}$ can limit model flexibility and degrade predictive accuracy. We therefore treat these additional constraints on $\mathbf{M}$ as an optional extension and focus primarily on the nonlinear energy-preserving constraint $\mathbf{Q}(\mathbf{x}_t)$ as the central stability prior of our architecture.\\

Finally, while energy-conserving nonlinearities originate from idealized ``dry'' fluid equations, enforcing this property acts as a powerful inductive bias even in complex real-world systems where the strict physical correspondence is relaxed. In these settings, the proposed constraint serves as a structural proxy that prevents the nonlinear component from injecting energy, thereby promoting stable long-time simulations. We demonstrate this in a real-world application in Section~\ref{sec:application}.

\section{The Fluctuation–Dissipation Theorem: Model Validation and Causal Inference} \label{sec:FDT-Causal}
Given an $n$-dimensional stochastic nonlinear system $\mathbf{x}_t = (x^{(1)}_t, x^{(2)}_t, \dots, x^{(n)}_t)$, the FDT \cite{FALCIONI} states that the ensemble-averaged linear response of an observable $\mathcal{A}(x^{(k)}_t)$ to a small impulse perturbation $\delta x^{(j)}_0$ applied to $x^{(j)}_0$ is given by:
\begin{equation}
R^{k,j}_t =  \lim_{\delta x^{(j)}_0\to0} \frac{\delta \langle \mathcal{A}(x^{(k)}_t) \rangle}{\delta x^{(j)}_0} = - \langle \mathcal{A}(x^{(k)}_t) s_j(\mathbf{x}_0) \rangle, \quad s(\mathbf{x}) = \nabla \log \rho(\mathbf{x}).
\label{eq:response_general}
\end{equation}
Here, $\rho(\mathbf{x})$ is the invariant density of the unperturbed system, $s(\mathbf{x})$ is the associated \textit{score function}, and $s_j$ denotes its $j$-th component.  The angle brackets $\langle \cdot \rangle$ denote ensemble averages over the stationary density. Eq.~\eqref{eq:response_general} bridges spontaneous fluctuations and forced responses, enabling rigorous data-driven causal inference in the \textit{interventional} sense \cite{Baldovin}: $R^{k,j}_t$ quantifies the effect of an imposed perturbation rather than a statistical association. In particular, its time integral measures the cumulative effect of an intervention on $x^{(j)}$. We will often refer to this quantity as a \textit{sensitivity} but stress that it remains a causal object as explained above.\\
We use the FDT primarily as a response benchmark for model evaluation; the same response operator additionally provides candidate direct causal links, which we use only when they pass the validation criteria of Section~\ref{sec:validation}:
\begin{enumerate}
    \item \textit{Model evaluation.} The primary strength of the FDT is that it yields a response-theory benchmark for validating emulator responses to forcing using only unperturbed stationary data, thereby bypassing the need for explicit perturbation experiments or access to an underlying numerical solver. This evaluation is valid strictly within the linear response regime, obtained by perturbing the system with small external forcings. For this validation, we consider observables $\mathcal{A}(x^{(k)}_t) = x^{(k)}_t$ and $\mathcal{A}(x^{(k)}_t) = (x^{(k)}_t - \mu^{(k)})^2$, where $\mu^{(k)}$ is the stationary mean. These observables capture how perturbations alter the time-dependent mean and variance of the state distribution, respectively. These comparisons are often performed by direct inspection of the full response operators rather than through a single score; see the discussion at the end of Section~\ref{sec:validation}.
    In real-world applications, finite sample sizes introduce a statistical bottleneck: the statistical error of the FDT estimator grows rapidly with lead time \cite{nonEqStatMech}. This error growth is especially pronounced for higher-order observables like variance. Consequently, for real-world scenarios, the FDT analysis is inherently limited and validation in such regimes is most reliable when focused on the ensemble mean response evaluated over short timescales. We provide a practical example in the real-world application presented in Section~\ref{sec:application}.
    \item \textit{Direct causal links.} Responses at the shortest resolved time scale, $R^{k,j}_1$, identify candidate direct causal links $x^{(j)} \rightarrow x^{(k)}$, which we use to construct causal constraints. In this context, it is sufficient to focus on the identity observable $\mathcal{A}(x^{(k)}_t) = x^{(k)}_t$, which quantifies the ensemble mean response to perturbations.
\end{enumerate}

\subsection{FDT from data: qG-FDT and score-based FDT} \label{sec:FDT-estimation}
The main challenge in evaluating Eq.~\eqref{eq:response_general} lies in estimating the score function $s(\mathbf{x})$. Depending on data availability, we employ two complementary estimation strategies:
\begin{itemize}
    \item \textit{Quasi-Gaussian FDT (qG-FDT) for data-scarce regimes.} When the available data are limited, the quasi-Gaussian approximation (qG-FDT) provides a robust and computationally efficient baseline \cite{Majda2010}. It approximates the invariant density by a multivariate Gaussian $\rho^G(\mathbf{x})$ \cite{Leith,MajdaBook,GRITSUN}, yielding:
    \begin{equation}
    s^{G}(\mathbf{x}) = \nabla \ln \rho^{G}(\mathbf{x}) = - \mathbf{C}^{-1} (\mathbf{x} - \bm{\mu}),
    \label{eq:quasi-Gaussian} 
    \end{equation}
    where $\bm{\mu} \in \mathbb{R}^{n}$ and $\mathbf{C} \in \mathbb{R}^{n, n}$ are the sample mean and covariance matrix. For the ensemble-mean observable $\mathcal{A}(\mathbf{x}_t) = \mathbf{x}_t$, Eq.~\eqref{eq:response_general} reduces to $\mathbf{R}_t = \mathbf{C}_t\mathbf{C}^{-1}$, with $\mathbf{C}_t$ the lagged covariance at lag $t$. This coincides with the propagator of Linear Inverse Models (LIMs) \cite{Penland89}, and we therefore highlight two differences to avoid confusion. First,  while $s^G(\mathbf{x})$ assumes a Gaussian score, the expectation in Eq.~\eqref{eq:response_general} is evaluated over the empirical invariant distribution. Consequently, unlike LIMs, which have Gaussian dynamics and cannot capture variance responses, the qG-FDT retains skill in capturing responses in ensemble variance \cite{Majda2010,GritsunMajda}. Second, a LIM fits the operator $\mathbf{C}_\tau\mathbf{C}^{-1}$ at a short lag (e.g. $\tau = 1$) and constructs a Markovian model which can be perturbed to obtain a response operator; in contrast, the qG-FDT evaluates $R^{k,j}_t$ independently at each lag $t$. As a result, the qG-FDT retains high skill in predicting responses even when an emulator requires a non-Markovian closure, as we show in Section~\ref{sec:application}. This final property is not specific to qG-FDT and it is shared by any FDT estimator, including the score-matching estimator described next.
    \item \textit{Score-matching FDT for data-rich regimes.} In data-rich settings, the score function can instead be learned directly from stationary data using advances in score-based modeling \cite{ResponseScore,LudovicoFabriAndre}. These previous applications used denoising score matching \cite{song_sde}, which learns the score of a noise-perturbed density. Here, we instead employ the original Hyv\"arinen score-matching loss \cite{Aapo}, which learns the score directly on the system's attractor. We parameterize $s(\mathbf{x})$ with a multilayer perceptron $s_{\theta}(\mathbf{x})$ and minimize the loss:  
    $$ \mathbb{E}_{\rho(\mathbf{x})}\left[\mathrm{tr}(\nabla_{\mathbf{x}} s_{\theta}(\mathbf{x})) + \frac{1}{2}\lVert s_{\theta}(\mathbf{x}) \rVert_{2}^{2}\right]. $$
    The divergence term $\mathrm{tr}(\nabla_{\mathbf{x}} s_{\theta}(\mathbf{x}))$ is computed via automatic differentiation for low-dimensional systems, or via a Hutchinson trace estimator \cite{Hutch} for high-dimensional ones. The expectation $\mathbb{E}_{\rho(\mathbf{x})}$ is approximated by a time average over the stationary data. The resulting estimate is then refined with the correction procedure detailed in SM Sec.~3.
\end{itemize}

\subsection{Candidate direct causal links and causal regularization} \label{sec:causal_protocol}
Consider a discrete Markovian stochastic system $x^{(i)}_{t+1}=f_i(\mathbf{x}_t)+\sigma_i\xi^{(i)}_t$, with $i=1,\ldots,n$. At the resolved sampling interval, the direct link $x^{(j)}\rightarrow x^{(k)}$ is absent if $\partial f_k/\partial x^{(j)}\equiv0$.  Given the identity observable $\mathcal{A}(x^{(k)}_t) = x^{(k)}_t$, the corresponding one-step response satisfies $R^{k,j}_1 = \langle \partial f_k(\mathbf{x})/\partial x^{(j)}\rangle$. Therefore, an absent direct link implies $R^{k,j}_1=0$. The converse can in principle fail through exact cancellations in the ensemble averages, but, as discussed later in this section, such cases are rarely restrictive in realistic systems. We therefore use near-zero values of $R^{k,j}_1$ as a practical proxy for absent direct links. We use $\mathbf{R}_1$ to construct a binary adjacency matrix $\mathbf{A} \in \mathbb{R}^{n, n}$, where $A^{k,j}=1$ denotes a retained candidate link $x^{(j)} \rightarrow x^{(k)}$ and $A^{k,j}=0$ denotes the absence of such a candidate. To construct this matrix, we use manual thresholding as the primary verification step and also provide an objective, automated rule.
\begin{enumerate}[label=\roman*., topsep=0pt, itemsep=0pt]
\item \textit{Self-interactions.} We retain all self-interactions, setting $A^{j,j}=1$. This is a conservative modeling choice for dissipative dynamical systems that preserves local autoregressive dependence independently of the statistical threshold. This assumption can be relaxed depending on the system or physical priors.
\item \textit{Cross-interactions.} For the off-diagonal entries $R^{k,j}_1$ ($j\neq k$), we apply a logarithmic transformation $\ln |R^{k,j}_1|$, separating potential causal links (larger responses) from the bulk of spurious, near-zero interactions. The absolute response $|R^{k,j}_1|$ may alternatively be used. Then we proceed in two complementary ways:
\begin{itemize}
    \item \textit{Manual thresholding.} A visual inspection of the flattened 1D distribution of $\ln |R^{k,j}_1|$ often reveals a clear separation between significant links and noise. In this case, thresholding can be performed manually. If no such separation exists, we recommend withholding causal regularization entirely: the absence of a gap implies that the system is either densely coupled or that the data length is insufficient for reliable causal discovery. This manual verification mirrors the procedure proposed in \cite{nanChen} for a different causality method.
    \item \textit{Automatic thresholding.} A $k$-means algorithm \cite{scikit-learn} ($k=2$) can be used to partition these transformed response values $\ln |R^{k,j}_1|$ into two clusters. Entries assigned to the cluster with the larger centroid are retained ($A^{k,j}=1$); all others are set to $0$.
\end{itemize}
\end{enumerate}
Threshold selection is problem dependent: performance depends on data length, sampling interval, dimensionality, coupling strength, and possible block structure in the Jacobian matrix, for which a single global threshold is inadequate (see, e.g., \cite{ChecnMajdaPNAS}). Consequently, manual thresholding remains an effective practical baseline, while developing new statistical procedures for threshold selection is an important avenue for future work. The matrix $\mathbf{A}$ is then used to regularize neural emulators. Specifically, the MSE minimization is regularized with a causal penalty:
\begin{equation}
\begin{aligned}
\mathcal{L} = \mathcal{L}_{\text{MSE}} + \mathcal{L}_{Causal} 
= \text{MSE}(\mathbf{x}_{t+1}, \mathbf{f}(\mathbf{x}_t)) + \lambda \sum_{(k,j) \in \mathcal{S}} (\frac{\partial f_k}{\partial x^{(j)}})^2~,
\end{aligned}
\label{eq:causal_loss}
\end{equation}
where $\mathcal{S}$ is the set of pairs $(k,j)$ for which $A^{k,j}=0$, and $\lambda$ determines the strength of the regularization.
\paragraph{Imperfect causal estimates.} False negatives in the adjacency matrix $\mathbf{A}$ (true links penalized as noncausal) can degrade the emulator when training via Eq.~\eqref{eq:causal_loss}. Physics-constrained architectures mitigate the impact of imperfect causal estimates through rigid structural priors. Additionally, if false negatives are suspected (e.g., via validation with stationary and perturbed statistics (see below)) a more robust penalty can be employed:
\begin{equation}
\begin{aligned}
 \mathcal{L}_{Causal}^{Robust} = \lambda \sum_{(k,j) \in \mathcal{S}} 
\min\!\Big((\frac{\partial f_k}{\partial x^{(j)}})^2,\gamma\Big),
\end{aligned}
\label{eq:causal_loss_robust}
\end{equation}
which bounds the penalty once $(\partial f_k/\partial x^{(j)})^2 \ge \gamma$, allowing the MSE term to dominate when interactions are mistakenly excluded. A heuristic estimate for $\gamma$ based on the FDT operator $R^{k,j}_1$ is given in Sec. 1.2.1 in the SM. In practice, we recommend starting with the standard quadratic regularization in Eq.~\eqref{eq:causal_loss} and implementing the robust loss in Eq.~\eqref{eq:causal_loss_robust} only after (a) detecting potential degradation in the MSE loss relative to the physics-only baseline, and (b) validating the model via stationary and perturbed statistics as outlined in Section~\ref{sec:validation}.
\paragraph{Possible limitations for idealized systems.} Because $R^{k,j}_1$ represents an ensemble-averaged Jacobian, nonzero local couplings may cancel in the average, yielding $\langle \partial f_k(\mathbf{x})/\partial x^{(j)} \rangle=0$ even when $\partial f_k/\partial x^{(j)}\neq0$. Systems with exact symmetries provide an important example: in such cases the system may require a symmetry-reduction step \cite{Predrag,MIRANDA1993105}. However, exact cancellations are nongeneric and are therefore not restrictive in realistic, partially observed turbulent flows, such as weather and climate, which lack exact symmetries. More generally, this possibility motivates the empirical validation procedure introduced in Section~\ref{sec:validation}.

\subsection{Validation of causal regularization} \label{sec:validation}

Unlike physics constraints, which are defined \textit{a priori} by theoretical arguments, causal constraints are inferred from finite data; therefore, their reliability depends on data availability, estimator accuracy, and the system's complexity. Importantly, the causal links (encoded as a binary adjacency matrix $A^{k,j}$) are inherently more sensitive to finite-data biases than the continuous response operator $R^{k,j}_t$, as binary thresholding amplifies small estimation errors. Consequently, the use of such constraints must be validated empirically in two successive steps:
\begin{enumerate}
    \item We verify that the emulator accurately reproduces stationary statistics, such as invariant densities and autocorrelation functions;
    \item We compare the emulator's mean and variance response operators against the FDT-based estimates obtained directly from the unperturbed data through Eq.~\eqref{eq:response_general}. As noted at the start of Section~\ref{sec:FDT-Causal}, for observational datasets with limited samples, this comparison is most reliable when restricted to short-time ensemble-mean responses.
\end{enumerate}
We retain causal regularization provided that, relative to the physics-only baseline, it does not degrade stationary statistics or linear-response predictions. An informative comparison of densities, autocorrelation functions, and response operators should account for structural discrepancies, e.g., a wrong decay time or spurious oscillations in an autocorrelation function, or a response of the right amplitude but the wrong sign. We therefore compare these functions directly, as is standard in the reduced-order modeling literature \cite{FranzkeMajda,STROUNINE2010145}, and use aggregate scores only when the number of comparisons is too large for individual inspection (see, e.g., Appendix~\ref{app:app_C} and SM Sec.~7 for aggregate MSE analyses).

\section{Physics constraints and causal regularization in practice} \label{sec:numerical_examples}

We evaluate the proposed framework on two canonical geophysical turbulence models where exact numerical ground truths allow direct validation of forced dynamics. First, we consider a six-dimensional stochastic Charney-DeVore system \cite{charneyDeVore,CROMMELIN}. In this low-dimensional test case, we recover the exact causal graph and the physics- and causality-constrained model accurately captures forced responses in both ensemble mean and variance, outperforming the model constrained by physics alone. Second, we test the framework on a high-dimensional, symmetry-broken Lorenz-96 system \cite{Lorenz96} in a strongly turbulent regime. Here, the dynamics evolves on a high-dimensional state space with many unstable directions \cite{MajdaQiControl}. This contrasts with many rigorous causal-discovery benchmarks for dynamical systems, which often focus on low-dimensional chaos \cite{Docquier,Nathaniel}. The physics-constrained model gives accurate responses to very large forcings, while the model constrained by both physics and causality further leads to significant corrections in the ensemble variance response, of central importance for uncertainty quantification \cite{SapsisMajdaPNAS}.\\
Together, these benchmarks rigorously test the framework under controlled conditions: Markovian dynamics, long trajectories, and access to direct numerical ground truth for forced responses. Because such ground truth is available, the FDT is not strictly needed here to validate the emulator responses; rather, these experiments allow us to verify and motivate the FDT validation strategy itself. We close this section by formalizing the additional methodological challenges posed by real-world turbulent systems, where these favorable conditions generally do not hold. This sets the stage for our main application in Section~\ref{sec:application}, where we construct and validate a reduced-order model of tropical climate dynamics.

\subsection{Stochastic Charney-DeVore model} We consider a stochastic version of the Charney--DeVore model \citep{charneyDeVore}, ``CdV'' hereafter, in the formulation of De Swart \cite{de1988low}. The model is a six-dimensional Galerkin truncation of the barotropic vorticity equation on a $\beta$-plane channel with topography and has been widely used to study large-scale atmospheric circulation \cite{CROMMELIN,crommelin2004mechanism,dorrington2023interaction,grafke2019numerical}. It is given by
\begin{equation}
\begin{aligned}
\dot{x}_1 &= \tilde{\gamma}_1 x_3 - C(x_1 - x_1^*) + \sigma \xi_1(t), \\
\dot{x}_2 &= -(\alpha_1 x_1 - \beta_1)x_3 - C x_2 - \delta_1 x_4 x_6 + \sigma \xi_2(t), \\
\dot{x}_3 &= (\alpha_1 x_1 - \beta_1)x_2 - \gamma_1 x_1 - C x_3 + \delta_1 x_4 x_5 + \sigma \xi_3(t), \\
\dot{x}_4 &= \tilde{\gamma}_2 x_6 - C(x_4 - x_4^*) + \varepsilon (x_2 x_6 - x_3 x_5) + \sigma \xi_4(t), \\
\dot{x}_5 &= -(\alpha_2 x_1 - \beta_2)x_6 - C x_5 - \delta_2 x_4 x_3 + \sigma \xi_5(t), \\
\dot{x}_6 &= (\alpha_2 x_1 - \beta_2)x_5 - \gamma_2 x_4 - C x_6 + \delta_2 x_4 x_2 + \sigma \xi_6(t).
\end{aligned}
\label{eq:CdV}
\end{equation}
We integrate Eq.~\eqref{eq:CdV} with a Euler-Maruyama scheme for $T=10^7$ time steps at $dt=0.01$, corresponding to $10^5$ model time units (MTU). The resulting trajectory is treated as a discrete time series $\mathbf{x}_t=(x^{(1)}_t,\ldots,x^{(6)}_t)$ from which we learn the finite-time flow map in Eq.~\eqref{eq:discrete_case}. Model parameters and their physical interpretation are reported in SM Sec.~5. The chosen stochastic regime has rapidly decaying autocorrelations and regular, unimodal stationary densities, properties relevant to coarse-grained climate dynamics and reduced-order modeling of turbulent systems \cite{FranzkeMajda,Judith,Sura,MajdaStructuralStability,MajdaIntroTurbulence}. In the main text, we focus on the causal adjacency matrix and on responses to weak and strong forcing of variable $x^{(6)}$, which are representative of the overall response behavior observed across forcing experiments; additional diagnostics across all variables are provided in the SM.
\begin{itemize}
    \item \textit{Causal Adjacency matrix from data.} Given the unperturbed CdV trajectory, we estimate the one-step mean response operator
    \[
    R^{k,j}_1 = - \langle x^{(k)}_1 s_j(\mathbf{x}_0) \rangle,
    \]
    using the score-based FDT estimate outlined in Section \ref{sec:FDT-estimation}. We then apply the transformation $\log |R^{k,j}_1|$ and show the result in Fig.~\ref{fig:Causal_Framework}(a). Large responses separate clearly from the near-zero bulk, so the threshold can be placed manually; we additionally verify this partition using $k$-means clustering with $k=2$. The resulting binary adjacency matrix is shown in Fig.~\ref{fig:Causal_Framework}(b). 
    \begin{figure}
    \centering
    \includegraphics[width=0.6\linewidth]{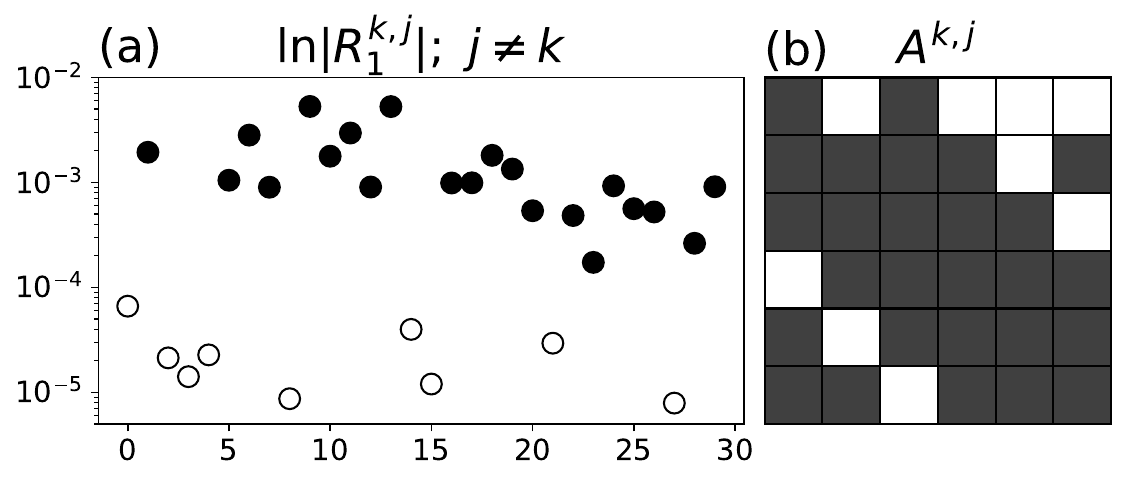}
    \caption{Causal Adjacency matrix of the CdV model (Eq.~\eqref{eq:CdV}). Panel (a): The 30 off-diagonal responses $\ln|R^{k,j}_1|$, flattened and plotted against a dummy index. Causal links are identified via k-means clustering ($k=2$), shown as filled and open circles. Panel (b): Resulting causal adjacency matrix $A^{k,j}$: $A^{k,j}=1$ (dark) for responses in the cluster with the larger centroid, and $A^{k,j}=0$ (white) otherwise. Example: the first row $A^{1,j}$ implies $x^{(1)}_{t+1} = f_1(x^{(1)}_t,x^{(3)}_t)$ as expected from Eq.~\eqref{eq:CdV}.}
    \label{fig:Causal_Framework}
    \end{figure}
    \item \textit{Neural emulators.} We train three discrete neural emulators from the long CdV trajectory: a physics-constrained emulator, denoted as ``Physics''; a physics- and causality-constrained emulator, denoted as ``Physics $\&$ Causal''; and an unconstrained baseline emulator, denoted as ``Vanilla''. All three emulators achieved comparably small training MSEs.
    \item \textit{Stationary statistics.} We simulate each emulator for $T=10^7$ time steps and compare its stationary statistics with those of the numerical model. All data-driven models remain numerically stable in unperturbed simulations and reproduce the invariant densities and autocorrelation functions; see SM Sec.~6.
    \item \textit{Perturbed statistics: linear regime.} A small impulse perturbation $\delta x^{(j)}_0$ is applied at $t=0$, and the response of observables $\mathcal{A}(x^{(k)}_t)=x^{(k)}_t$ (mean) and $\mathcal{A}(x^{(k)}_t)=(x^{(k)}_t-\mu^{(k)}_t)^2$ (variance) is measured, $\bm{\mu}_t$ representing the time-dependent mean. Appendix~\ref{app:app_A_1} describes the numerical estimation of these response operators. As a representative example, we perturb $x^{(6)}_0$ and report the statistical response of $x^{(1)}_t$ in mean and variance (Fig.~\ref{fig:responses_CdV}(a,c)). We report all responses in the SM Sec.~7. First, the score-matching based FDT (Eq.~\eqref{eq:response_general}) predicts responses with high accuracy in both mean and variance, relying solely on stationary dynamics. This underscores that in applications, the FDT provides a useful response-theory benchmark for validating responses to perturbations. All FDT predictions are reported in Figures 4 and 5 in the SM. The constrained emulators reproduce the response operator well; however, the causality-constrained model outperforms the physics-only model. This improvement is robust across perturbation-response pairs, as shown in Figure~9 in SM Sec.~7, where we report time-dependent MSEs aggregated over all entries of $R^{k,j}_t$. Notably, the unconstrained ``vanilla'' emulator becomes unstable under impulse perturbations, despite achieving a training MSE comparable to that of the physics-constrained model and remaining stable in unperturbed simulations. In other words, stability and accurate stationary statistics do not by themselves imply correct responses to perturbations. Because the unconstrained baseline is already unstable under small impulse perturbations, we exclude it from the following large-forcing experiment and focus on the constrained emulators.
    \item \textit{Perturbed statistics: nonlinear regime.} We next test nonlinear response predictions by applying a large step forcing on the right hand side of both the numerical model and the constrained emulators. The forcing acts on $x^{(6)}$, $\mathbf{F}=(0,0,0,0,0,\sigma_6),~ t\geq 0,$ where $\sigma_6$ is the stationary standard deviation of $x^{(6)}$. We simulate an ensemble of $10^5$ realizations and measure the responses of ensemble mean and variance of variable $x^{(1)}$. Results are shown in Fig.~\ref{fig:responses_CdV}(b,d). The numerical model exhibits a pronounced transient before relaxing to a new equilibrium after $\sim40$ model time units, reflecting a shift of the attractor in state space. Both emulators remain stable under this strong forcing and predict the response well, with the physics and causality constrained emulator generally outperforming the physics-only model. This improvement is robust across forcing experiments as reported in Section 8 of the SM. This analysis shows that causal constraints inferred from the shortest-time linear response can improve nonlinear response predictions over long time horizons, well beyond the linear-response regime.
    \begin{figure}
    \centering
    \includegraphics[width=0.8\linewidth]{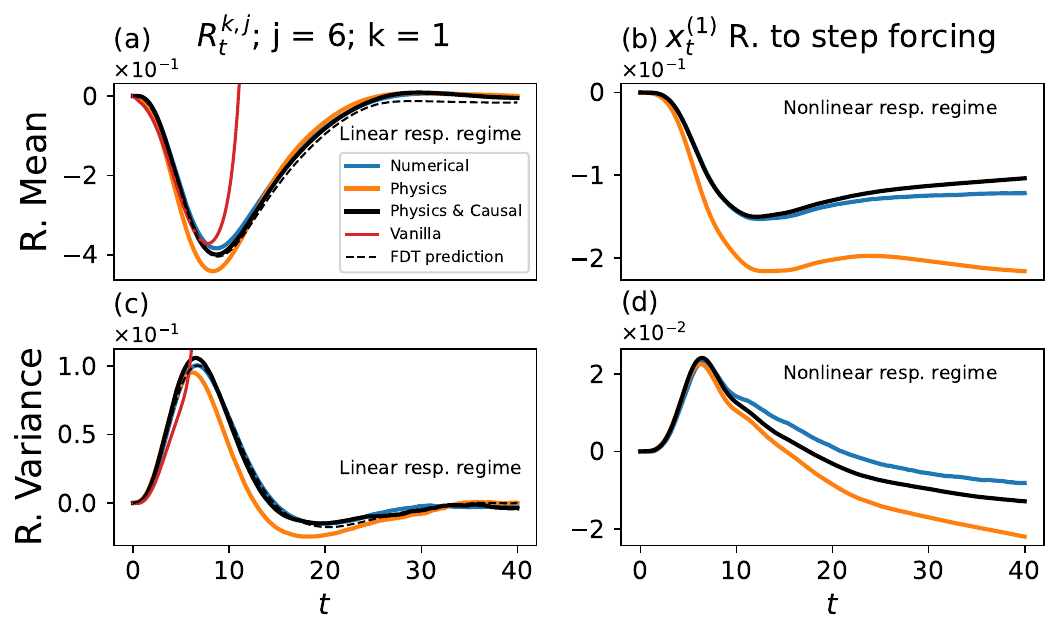}
    \caption{Charney-DeVore model: Perturbed statistics. ``Physics'': physics-constrained model; ``Physics \& Causal'': physics- and causality-constrained model; ``vanilla'': unconstrained model; ``FDT prediction'': response \textit{predicted} via Eq.~\eqref{eq:response_general} using unperturbed stationary statistics. Panels (a) and (c): linear response regime, showing the time-dependent response of the ensemble mean (a) and variance (c) of $x^{(1)}_t$ to an impulse perturbation $\delta x^{(6)}_0$ applied at $t = 0$. Panels (b) and (d): Nonlinear response regime, showing the response of the ensemble mean (b) and variance (d) of $x^{(1)}_t$ to a step-function forcing $\mathbf{F} = (0,0,0,0,0,\sigma_6)$ applied to the right-hand side of each model for $t \ge 0$; $\sigma_6$ is the standard deviation of $x^{(6)}$. The unconstrained vanilla model is unstable under small impulse perturbations (panels (a,c)) and it was then excluded from the large-forcing experiment}
    \label{fig:responses_CdV}
    \end{figure}
\end{itemize}

\subsection{Symmetry-broken Lorenz-96 system} \label{sec:AL96}

\begin{figure}
\centering
\includegraphics[width=0.6\linewidth]{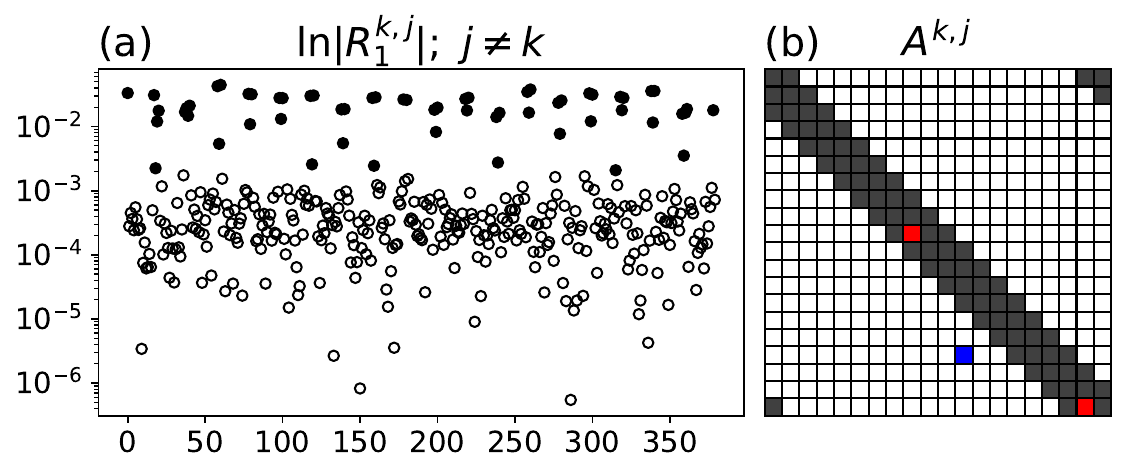}
\caption{Same as Figure \ref{fig:Causal_Framework}, but for the L96 system defined in Eq. \eqref{eq:L96}. Red squares indicate false negatives: true physical links erroneously labeled as noncausal. The blue square indicates a false positive: a spurious link erroneously labeled as causal. Example: the first row $A^{1,j}$ indicates a functional dependency $x^{(1)}_{t+1} = f_1(x^{(19)}_t, x^{(20)}_t, x^{(1)}_t, x^{(2)}_t)$ as expected from Eq. \eqref{eq:L96}.}
\label{fig:Causal_Framework_L96}
\end{figure}

We define a symmetry-broken version of the Lorenz-96 (L96) \cite{Lorenz96} system as
\begin{equation}
\dot{x}_j = (x_{j+1} - x_{j-2})x_{j-1} - x_{j} + F_j + \sigma \xi_j(t), 
\label{eq:L96}
\end{equation}
with $j=1,\dots,J$ and periodic boundary conditions $x_j = x_{j+J}$. We set $J=20$. The time-independent forcing is defined as $F_j = F + \eta_j$, where $\eta_j \sim \mathcal{N}(0,\sigma_F^2)$ with $\sigma_F=10$. To maintain positive forcing $F_j$, we impose a minimum value of $4$. The resulting site-dependent forcing, ranging from $4$ to $\sim31$, breaks the translational invariance of the original Lorenz-96 model and induces spatially heterogeneous turbulence. The stochastic term $\sigma \xi_j(t)$, with $\sigma=5$, represents unresolved fast processes. The system in Eq.~\eqref{eq:L96} provides a standard prototype of a turbulent dynamical system, with strong mixing, rapid decay of autocorrelations, many positive Lyapunov exponents, and regular distributions even in the deterministic limit \cite{majdaSIAM,MajdaIntroTurbulence}. For reproducibility, the specific values of $F_j$ are reported in SM Sec.~9. We simulate a trajectory of length $T=10^6$ time steps with $dt=0.01$, corresponding to $10^4$ model time units. The resulting 20-dimensional trajectory is treated as a discrete time series $\mathbf{x}_t=(x^{(1)}_t,\ldots,x^{(20)}_t)$ from which we learn the finite-time flow map in Eq.~\eqref{eq:discrete_case}. We infer the causal adjacency matrix using the protocol introduced in Section \ref{sec:causal_protocol}. As shown in Fig.~\ref{fig:Causal_Framework_L96}, the inferred graph largely recovers the true causal structure, including the periodic boundary conditions, despite the strongly turbulent nonlinear dynamics. The inferred graph contains two false negatives and one false positive (red and blue entries in Fig.~\ref{fig:Causal_Framework_L96}). Because the inferred causal structure is imperfect, we impose causal constraints through the robust penalty in Eq.~\eqref{eq:causal_loss_robust}.

\begin{itemize}
    \item \textit{Emulators and stationary statistics.} As for the Charney-DeVore model, we train three neural emulators: a physics-constrained emulator (``Physics''), a physics- and causality-constrained emulator (``Physics \& Causal''), and an unconstrained baseline (``Vanilla''). All three reach comparably small training MSEs, although the Vanilla model requires approximately 10 times as many epochs to do so, indicating that the physics constraints also ease the optimization problem. All emulators remain stable in unperturbed simulations and reproduce the invariant densities and
    autocorrelation functions (SM Sec.~9).
    \item \textit{Perturbed statistics: linear regime.} We consider the linear response regime, focusing on reproducing the response operator by computing the system's time-dependent mean and variance responses to an impulse perturbation applied in the middle of the domain at $t = 0$. The numerical estimation of this response operator is described in Appendix~\ref{app:app_A_1}. Results are shown in Figure~\ref{fig:Response_impulse}. First, consistent with the Charney--DeVore model in the previous section, the FDT (Eq.~\eqref{eq:response_general}) predicts responses with high accuracy in both mean and variance. The system's linear response is accurately captured by all tested emulators: (a) physics-constrained, (b) physics and causality constrained, and (c) the unconstrained vanilla emulator. However, while the unconstrained baseline accurately reproduces stationary statistics and weak linear responses, the absence of physical priors provides no guarantee of stability or generalization. We examine its behavior under strong forcing next.
    \begin{figure}
    \centering
    \includegraphics[width=0.9\linewidth]{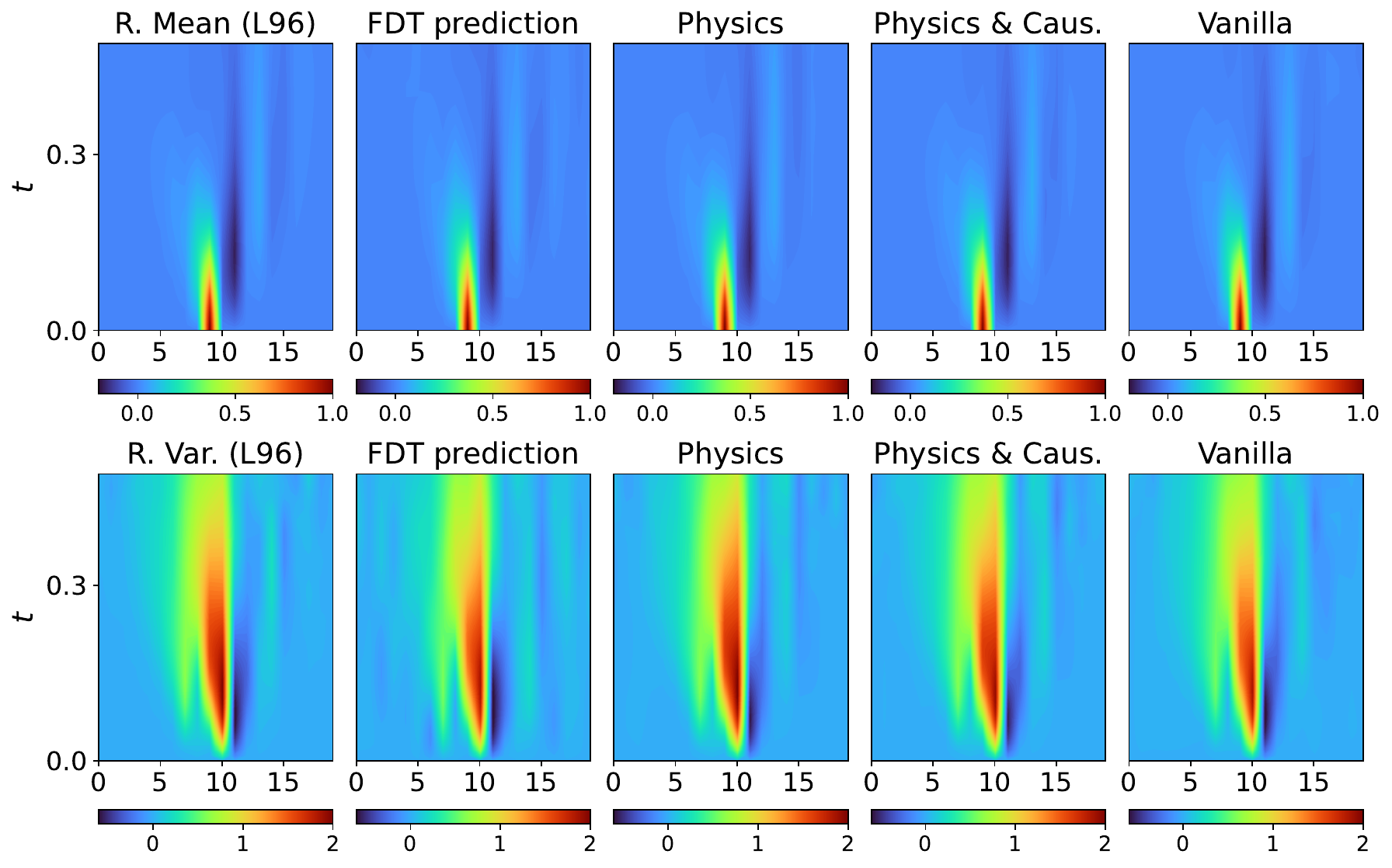}
    \caption{Lorenz-96 system: Linear response regime. Response operator for the L96 system: a small impulse perturbation is imposed at the center of the domain at time $t = 0$. Top row: ensemble-mean response. Bottom row: ensemble-variance response. Left: ground truth. Second column: response \textit{predicted} via the FDT in Eq.~\eqref{eq:response_general} using only stationary statistics of the unperturbed system.  Third column: response of the physics-constrained model. Fourth column: response of the model constrained by both Physics and Causality. Fifth column: response of the unconstrained (vanilla) model.}
    \label{fig:Response_impulse}
    \end{figure}
    \item \textit{Perturbed statistics: nonlinear regime.} We probe the nonlinear response regime by applying a Gaussian-shaped forcing to the right-hand side of Eq.~\eqref{eq:L96}, held constant for all $t \ge 0$. The forcing is centered at lattice site $j=10$,
    \begin{equation}
    \begin{aligned}
    G_j = w \exp\!\left(-\frac{d(j,\mu)^2}{2\ell^2}\right), ~ \text{with} ~ d(j,\mu) = \min(|j-\mu|,\, J-|j-\mu|),
    \end{aligned}
    \label{eq:gaussian_forcing}
    \end{equation}
    with $\mu=10$ and $\ell=3$. The distance function $d(j,\mu)$ is needed to account for the periodic boundary conditions. To set the forcing amplitude, we compute the total standard deviation $\sigma_{\mathrm{Tot}}$, which is of the same order as the single-site standard deviations $\sigma_j$, and choose $w = 10 \sigma_{\mathrm{Tot}}$. This defines a severe perturbation, roughly an order of magnitude larger than the natural variability of any degree of freedom $x_j$, and therefore outside the range of states encountered during training. As shown in Fig.~\ref{fig:Response_Gaussian_F}, this forcing produces a substantial shift in both the mean and variance of the attractor. Differences across emulators are small, and are therefore analyzed in terms of differences from the numerical ground truth. The physics-constrained emulator already captures these changes accurately. Adding causal constraints yields only modest improvements for the mean response, but it produces a clear and systematic improvement for the variance response, which is central for uncertainty quantification in turbulent systems \cite{SapsisMajdaPNAS}. Importantly, the causality constrained emulator retains high performance despite the small errors in the inferred graph. Finally, while the vanilla emulator remained stable under the forcing in Eq.~\eqref{eq:gaussian_forcing}, it exhibited much larger response errors, as reported in Fig.~\ref{fig:Response_Gaussian_F}. This further underscores that stability in unperturbed simulations and accurate stationary statistics are necessary but not sufficient conditions for accurately capturing responses to perturbations.
    \begin{figure}
    \centering
    \includegraphics[width=0.8\linewidth]{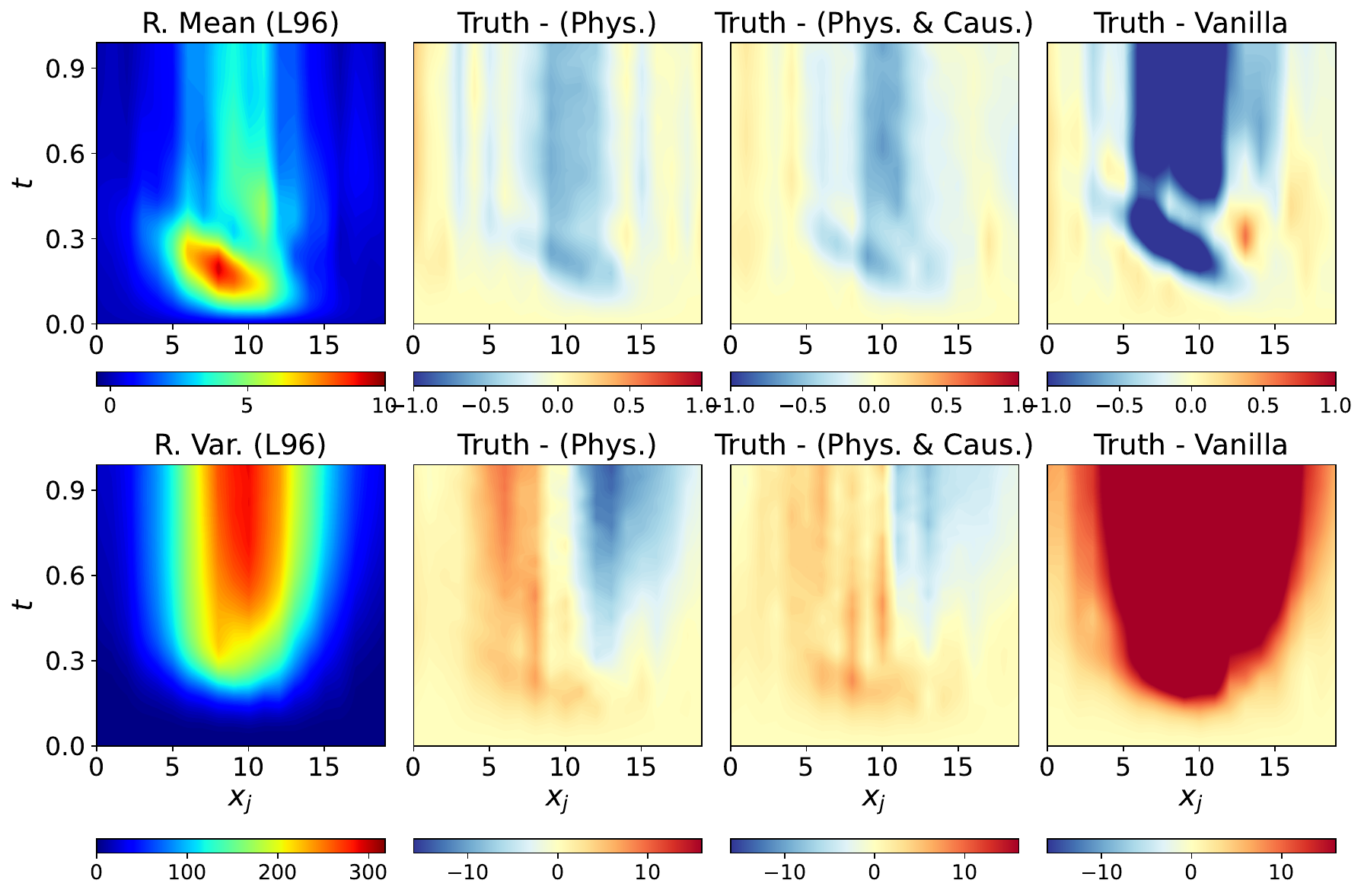}
    \caption{Lorenz-96 system: Nonlinear response regime. Response of the L96 system to a large Gaussian-shaped forcing [Eq.~\eqref{eq:gaussian_forcing}] applied and held constant for $t \ge 0$. Top row: ensemble-mean response. Bottom row: ensemble-variance response. Left: ground truth. Second column: difference between the ground truth and the physics-constrained emulator. Third column: difference between the ground truth and the physics- and causality constrained emulator. Fourth column: difference between the ground truth and the unconstrained vanilla model.}
    \label{fig:Response_Gaussian_F}
    \end{figure}
\end{itemize}
We emphasize that, in this experiment, no spatial locality was imposed in the learning procedure. The emulators were implemented as MLPs \footnote{In the standard Lorenz-96 setup, a convolutional neural network (CNN) naturally encodes the
local interactions and periodic boundary conditions through shared local filters. Our site-dependent forcing $F_j$ breaks the translational symmetry of the system, so a CNN trained only on observations $\mathbf{x}_t$ would impose inappropriate weight sharing unless supplemented with positional or forcing information. This motivates the use of an MLP.}, and the causal discovery step was not supplied with any \textit{a priori} information about the local coupling structure of Eq.~\eqref{eq:L96}. We made this design choice to rigorously evaluate the data-driven causal inference algorithm in a generalized context where the underlying interaction topology is presumed entirely unknown. However, we recommend in practical applications to use as much physical knowledge as possible before the causal inference step: any knowledge of locality, or of specific couplings among variables, should inform the causal algorithm \textit{a priori}, as any statistical inference solution is bound to have spurious results because of finite data. We expand upon this philosophy of physics-guided causal discovery in the following section on limitations and future directions.

\subsection{Challenges and practical considerations} \label{sec:discussions}

The controlled experiments above tested the framework under known Markovian dynamics, long trajectories, and access to forced-response numerical ground truth. For real-world turbulent systems, these conditions generally do not hold, raising methodological and conceptual challenges. We formalize such challenges below, motivating the modeling choices in the tropical climate application of Section~\ref{sec:application}.
\begin{itemize}
    \item \textit{Methodological challenges: Finite data and the importance of physical priors.} Model fitting inevitably depends on the length of the available time series. Physics constraints, specified \textit{a priori} from theory, alleviate this limitation by restricting the space of admissible dynamics. In contrast, candidate causal graphs inferred through the FDT, or any other statistical method, remain sensitive to finite-data errors. This sensitivity increases with system dimensionality and complexity. We therefore recommend leveraging all available physical and domain knowledge, such as known couplings, to identify and correct inconsistencies in statistically inferred causal graphs. In gridded emulators, the problem simplifies because classical physics is local: this spatial locality directly provides a sparse adjacency mask that can be imposed without inferring the graph from data.  In reduced-order or coarse-grained contexts, physical knowledge may provide only partial structural information. Nevertheless, such priors could still be used to validate and refine inferred graphs \textit{a posteriori}; for instance, one can remove minor artifacts that disrupt an otherwise structured adjacency matrix, as in Figure~\ref{fig:Causal_Framework_L96}. This underscores a core principle of our approach: physical priors should precede and guide statistical inference.
    \item \textit{Methodological challenges: High-dimensionality and score-based modeling.} The FDT-based causal discovery step requires estimating the system's score function. For the low- and moderate-dimensional systems considered here, this can be achieved via the qG-FDT approximation or via the Hyvärinen score-matching estimator introduced in Sec.~\ref{sec:FDT-estimation}. For complex spatiotemporal turbulent systems, our framework is optimally deployed after projecting the dynamics onto a reduced latent space (e.g., \cite{PierreLatent,latentPierre}). However, if the system's attractor remains fundamentally high-dimensional, then the score estimation becomes a computational bottleneck, and different strategies for inferring the score could be preferred. For instance, architectures that explicitly exploit spatial structure, such as U-Nets, can be better suited if the analysis needs to be carried out on gridded data \cite{LudovicoFabriAndre}.
    \item \textit{Conceptual challenges: Choices of coarse-grained representations.} The fundamental conceptual challenge in data-driven modeling of multiscale systems is the selection of coarse-grained variables, $\mathbf{x}_t = (x^{(1)}_t, x^{(2)}_t, \dots, x^{(n)}_t)$, whose effective deterministic dynamics are approximately Markovian at the chosen resolution. The difficulty is not specific to neural emulation: any attempt to construct a reduced model of a natural phenomenon, whether mechanistic or data-driven, first requires identifying the variables in which the phenomenon admits a useful closed description \cite{BaldovinEntropy}. The issue is central in statistical mechanics, as emphasized in classical studies \cite{Onsager} and recent work \cite{Lucente2025}, and is particularly acute in multiscale turbulent systems such as Earth's climate, where the dynamics span over 15 orders of magnitude in space and time with no clear space- and time- scale separation \cite{GhilLucarini}. Defining these coarse-grained variables requires specifying the physical fields of interest alongside a restricted range of spatial and temporal scales, including specific averaging and preprocessing protocols \cite{FabriCoarseGraining}. \footnote{We focus on data-driven models of multiscale turbulent systems for which only a projection of the state is observed or the relevant scales cannot all be resolved. This includes multiscale natural systems whose effective governing equations are unknown (e.g., interactions among large-scale climate modes) and complex numerical models that remain high-dimensional, parameterized, and potentially biased despite their known equations. A complementary line of theoretical work in applied mathematics studies prototype stochastic systems and, when the governing equations are known, develops systematic and, under suitable assumptions, rigorous methods for separating macroscopic and microscopic dynamics \cite{Roberts2006,ROBERTS200812}.} Such choices affect any causal inference method, including FDT-based approaches, because they require a Markovian dynamics to capture how perturbations travel across the systems (i.e. the effect of interventions), see e.g. \cite{Baldovin}. Consequently, they also affect the construction of data-driven emulators: their ability to correctly respond to forcings depends critically on whether the chosen variables close the effective dynamics at the selected spatiotemporal resolution.\footnote{We note that projections onto latent spaces, whether through linear methods or nonlinear autoencoders \cite{PierreLatent}, do not circumvent this issue. Such projections undoubtedly provide ways to find optimal bases for efficient dynamical modeling for a chosen dataset, but the learned latent space remains tied to choices of the initial input variables and to their spatial and temporal resolution. For example, a representation learned from hourly snapshots need not coincide with one learned from weekly averages, because the two datasets encode different effective dynamics and, therefore, different slow manifolds.} Importantly, the choice of an appropriate coarse-grained representation is non-unique and it must be guided by the questions of interest, the physical processes under study, and the relevant spatial and temporal scales. In climate dynamics, this goal-oriented perspective builds on a long successful tradition of reduced-order models targeting specific aspects of low-frequency variability (e.g., \cite{Penland89,Penland95,PENLAND1996534,LudoWettlaufer,KondraSeasonality,ChenENSO,ChenMJOgrl,ChenMJOmwr,Keyes} among many others) as first motivated by the theoretical work of Hasselmann \cite{Hasselmann,Hasselmann2}. We refer to \cite{FabriCoarseGraining} for a more detailed discussion of this conceptual limitation. Finally, we note that non-Markovian stochastic parameterizations (e.g. \cite{KONDRASHOV201533}) can compensate for memory in the deterministic residuals, but they cannot substitute for an inadequate choice of coarse-grained, slow variables; the model's skill still depends critically on that choice, as illustrated in Section~\ref{sec:application}.
\end{itemize}


\section{A real-world application: large-scale tropical dynamics} \label{sec:application}

We apply the proposed framework to large-scale tropical dynamics using a relatively short ocean-reanalysis record. This is the realistic setting the framework is ultimately aimed at, and it exposes both its practical value and its limitations. The main focus is on quantifying causal drivers of the El Ni\~no-Southern Oscillation (ENSO) over long time scales and across different tropical basins. The first major challenge is therefore to choose a coarse-grained representation tailored to this goal.  The results are then analyzed and interpreted in light of the extensive literature on ENSO feedbacks.

\subsection{Goals and coarse-grained representation} \label{sec:Goals}

The El Ni\~no--Southern Oscillation (ENSO) is the dominant mode of climate variability at interannual timescales \cite{ENSOcomplexity}. ENSO is a recurrent spatiotemporal pattern in the tropical Pacific: its warm phase (``El Ni\~no'') is characterized by anomalously warm sea surface temperatures in the eastern-to-central equatorial Pacific accompanied by cooler anomalies in the west. The cold phase (``La Ni\~na'') exhibits the opposite pattern. ENSO drives climate variability beyond the tropical Pacific through teleconnections, affecting rainfall, tropical cyclone activity, and the likelihood of droughts and floods around the world \cite{dimRedClimate,Harries,PRX,AndreouChenENSO}. While theoretical understanding has traditionally focused on local ocean–atmosphere interactions within the Pacific (see \cite{ChenENSO} and references therein), it is increasingly recognized that variability in other tropical basins significantly impacts ENSO \cite{Fonseca,dimRedClimate,Pantropical,Harries,Capotondi}. Disentangling and quantifying these non-local, time-dependent interactions is crucial for improving seasonal forecasts and understanding long-term changes in tropical variability \cite{TBIMIP}. Here, we focus on the ENSO component of this interacting tropical network and raise the following central question:
\begin{quote}
\textit{To which physical variables and tropical basins is ENSO variability most sensitive at long timescales?}
\end{quote}
To address this, we construct a stochastic reduced-order model of tropical variability.\footnote{We note that this approach simplifies the problem by approximating tropical dynamics, to first order, as a closed system, neglecting extratropical drivers at high latitudes. Future work will focus on extending this to larger domains.} Building on Zhao and Capotondi (2024) \cite{Capotondi}, we construct a minimal representation of tropical climate dynamics over ($0^\circ$--$360^\circ$ E, $30^\circ\text{S}$--$30^\circ\text{N}$) based on two fields:
\begin{itemize}
    \item \textit{Sea surface temperature (SST). Units [K].} SST anomalies reflect the combined effects of ocean advection, upwelling, mixing, and air-sea heat exchange. SST variations alter atmospheric heating and pressure gradients, triggering large-scale teleconnections \cite{Pantropical}. Consequently, SST has been traditionally used to build reduced-order models of tropical variability \cite{Penland89,Penland95,PENLAND1996534}.
    \item \textit{Sea surface height (SSH). Units [m].} SSH anomalies integrate changes in water-column mass and density \cite{FalascaEDS}. In the tropics, SSH provides a useful proxy for thermocline depth and subsurface ocean dynamics \cite{Capotondi}. Because the subsurface ocean evolves more slowly than the surface, SSH carries oceanic ``memory'' that serves as a precursor to large-scale SST anomalies.
\end{itemize}
To further filter out fast weather noise we use monthly averaged data. We obtain monthly fields from the Ocean Reanalysis System 5 (ORAS5) \cite{ORAS5} from 1960 to 2025 at $1^\circ$ resolution. This grid yields approximately $16,000$ spatial cells per field over $T = 792$ months. We focus on monthly anomalies and remove the seasonal cycle. Following \cite{FabCausal,FDTPatternEffect}, we high-pass filter the data with a cut-off frequency of $10^{-1}\text{ yr}^{-1}$. This removes the anthropogenic trend and (possible) low-frequency signals that are poorly sampled in a dataset with only $\sim 60$ years. This step yields a stationary dataset focused on interannual variability alone. We use $90\%$, approximately $712$ monthly samples, of this record for model fitting. Each field is then standardized by its total spatiotemporal standard deviation:
\begin{equation}
\hat{\mathbf{X}}^\text{SST} = \mathbf{X}^\text{SST}/\sigma_{\text{SST}} ~ \text{and} ~ \hat{\mathbf{X}}^\text{SSH} = \mathbf{X}^\text{SSH}/\sigma_{\text{SSH}}
\end{equation}
where $\sigma_{\text{SST}}$ ($\sigma_{\text{SSH}}$) represents the total standard deviation of the SST (SSH) field. Finally, each standardized field is projected onto its leading $m = 10$ Empirical Orthogonal Functions (EOFs), yielding the reduced components $\hat{\mathbf{x}}^\text{SST}$ and $\hat{\mathbf{x}}^\text{SSH}$, resolving $\sim 60\%$ and $\sim 51\%$ of the variance of the respective fields. The resulting reduced-order state vector at time $t$ is:
\begin{equation}
\mathbf{x}_t = [\hat{\mathbf{x}}^{\text{SST}}_t, \hat{\mathbf{x}}^{\text{SSH}}_t]^\mathrm{T},
\label{eq:state_vector}
\end{equation}
where $\mathbf{x} \in \mathbb{R}^{n,T}$ with $n = 20$ and $T=792$. In the following section, we evaluate the validity of treating this low-dimensional state vector as an approximately Markovian system.

\subsection{Stationary statistics of ENSO} \label{sec:stationary_statistics}

We fit the physics-constrained model introduced in Eq. \eqref{eq:discrete_case} to the state vector $\mathbf{x}_t$. Given the severely limited sample size ($T = 792$ months), the linear inverse modeling initialization proposed in Section~\ref{sec:fitting} is essential for robust training. The short record prevents robust identification of direct causal links from short-time responses. Following the conservative protocol of Section~\ref{sec:causal_protocol}, we therefore withhold causal regularization and retain the simpler physics-constrained model for the remainder of the analysis. As outlined previously, our primary focus is the representation of ENSO dynamics: we use the leading EOF of the SST field, encoded in $x^{(1)}_t$, as a proxy for ENSO variability. We refer to this quantity hereafter as the ``ENSO mode''.\\

\begin{figure}
\centering
\includegraphics[width=0.9\linewidth]{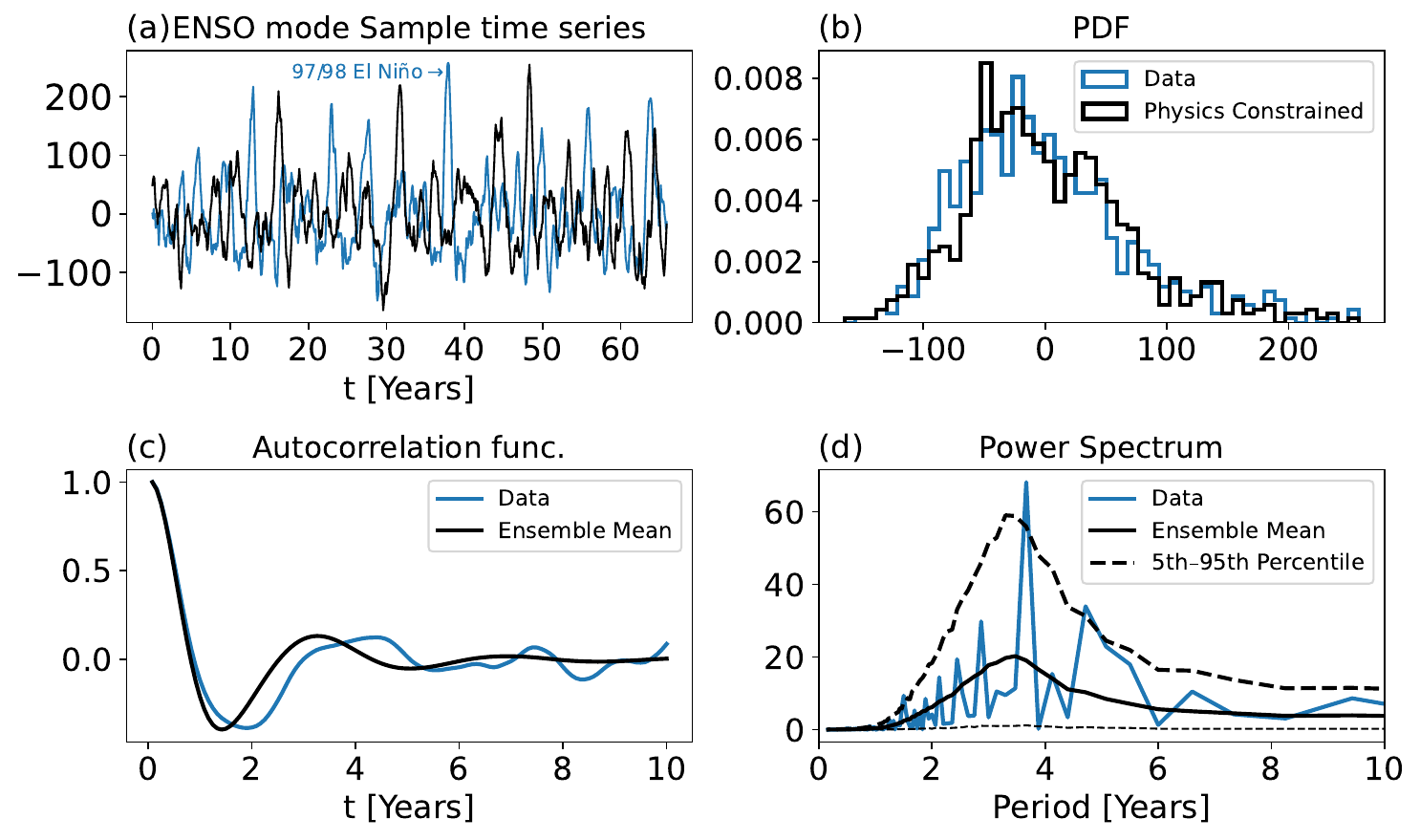}
\caption{Stationary statistics of the ENSO mode. Black lines denote results from the reduced-order neural model; blue lines represent the reanalysis data. Panel (a): A representative simulated trajectory plotted alongside the observed ENSO mode (unitless). The historic 1997/1998 extreme event in the observational record is highlighted. Panel (b): Probability density functions. Panel (c): Autocorrelation functions. The model curve shows the ensemble mean computed over 1,000 independent simulations. Panel (d): Fourier spectra. The solid black line indicates the ensemble mean, while the upper bound represents the 95th percentile across the 1,000-member ensemble, demonstrating that the observed spectral peaks fall within the model's statistical spread.}
\label{fig:ENSO_stationary}
\end{figure}

Figure~\ref{fig:ENSO_stationary} evaluates the emulator's ability to reproduce key stationary statistics of the ENSO mode. Figure~\ref{fig:ENSO_stationary}(a) compares a representative model simulation against the observed time series. Notably, the emulator generates extreme El Ni\~no anomalies, comparable in magnitude to the historic 1997/1998 event, even though such large-amplitude events appear only once in the training record. Consequently, the model's probability density function (PDF) closely aligns with the observed non-Gaussian distribution in Figure~\ref{fig:ENSO_stationary}(b). To rigorously assess the temporal dynamics, we evaluate the autocorrelation function and Fourier spectra across a 1,000-member model ensemble in Figure~\ref{fig:ENSO_stationary}(c,d). The ensemble mean autocorrelation approximately tracks the observed decay at short and intermediate lags. More importantly, Figure~\ref{fig:ENSO_stationary}(c) highlights a key advantage of utilizing ensemble statistics from neural emulators: at long lead times, the ensemble mean correctly decays to zero, reflecting the physical loss of memory in the system. In contrast, the empirical autocorrelation exhibits spurious noise at long lags due to the limited observational sample size. Overcoming these finite-sample artifacts is a primary motivation for employing reduced-order models, and it becomes very relevant in the study of long-term responses to perturbations. Finally, Figure~\ref{fig:ENSO_stationary}(d) confirms that the emulator largely captures the ENSO spectra. The observed spectrum is highly consistent with the emulator's statistical spread, bounded by the ensemble's 95th percentile. Crucially, the modeled spectral power correctly peaks in the 2- to 5-year band, matching the periodicity of the observed ENSO cycle.\\

\subsection{Perturbed statistics: Efficient sensitivity analysis from reduced-order models} \label{sec:sensitivity}

We now leverage the trained reduced-order model to investigate the mechanisms controlling tropical variability. Specifically, we address the question posed in Section~\ref{sec:Goals}: \textit{to which physical variables and tropical basins is ENSO variability most sensitive at long timescales?} This causal question is naturally framed within response theory by quantifying the cumulative (in time) response of the ENSO mode to impulse perturbations applied to other fields across different geographic locations. Traditionally, this analysis would rely on linear Green's function experiments with comprehensive climate models \cite{barsugli2002global,blochjohnsonetal2024,hydroSens}. While powerful, this approach has known limitations: (a) high computational costs restrict experiments to single trajectories rather than robust statistical ensembles, (b) overcoming signal-to-noise ratios requires large perturbations that violate the linear assumptions of the Green's function framework \cite{williams2023circus}, and (c) perturbations are typically imposed over \textit{ad hoc} spatial ellipsoids, making the resulting sensitivities highly dependent on this choice \cite{Leif}.\footnote{A further distinction concerns the nature of the response itself. In atmosphere-only Green's function experiments, perturbations are applied to a prescribed boundary condition (usually SST), which therefore cannot adjust dynamically; so the measured response excludes coupled ocean-atmosphere feedbacks. Here all state variables are prognostic, and responses predicted by the FDT are therefore those of the coupled system, in which a perturbation to one field propagates through all state variables and can feed back on the field perturbed. Overlooking this distinction can make the FDT appear to fail when its coupled-system predictions are compared with responses from atmosphere-only experiments~\cite{FDTPatternEffect}.}\\
Reduced-order models can alleviate these limitations by enabling large-ensemble response experiments with weak perturbations. However, the emulator does not evolve the full gridded fields: it evolves a finite-dimensional latent state, here given by the leading EOF coefficients. Therefore, a localized physical-space perturbation is not itself a resolved model coordinate. Once projected onto the truncated EOF basis, such a perturbation is spread across the retained modes and depends on the chosen truncation. Following \cite{FabriCoarseGraining}, we instead perturb the system directly along the coordinates in which the reduced-order dynamics is defined: the retained large-scale modes. We compute responses in this latent space and project the results back into physical space via the EOFs to obtain sensitivity maps. Crucially, this backward projection serves solely as a tool for visual interpretability: the actual physics and computation of the responses reflect purely the interactions across the large-scale EOF modes. This strategy avoids choosing arbitrary physical perturbation patches and it preserves consistency with the reduced-order dynamics. In \cite{FabriCoarseGraining}, we first demonstrated this approach in the context of the ``pattern effect'', i.e., the sensitivity of radiative fluxes to perturbations in sea surface temperatures.\\

Formally, the strategy is executed in four steps:
\begin{enumerate}
    \item \textit{Choice of observable.} We select an observable of interest, $\mathcal{A}(\mathbf{x}_t)$. Here, we simply choose $\mathcal{A}(\mathbf{x}_t) = x^{(1)}_t$, which corresponds to the first SST mode, i.e. the ENSO mode considered in the previous section.
    \item \textit{Response operator.} We compute the time-dependent linear response operator $R^{(1,j)}_t$ for all EOF modes $j = 1,\dots,n$, with $n = 20$. $R^{(1,j)}_t$ is inferred via the FDT using Eq. \eqref{eq:response_general} and via the data-driven models using the method in Appendix \ref{app:app_A_1}. The operator quantifies the ensemble-mean response of the ENSO mode to a small impulse perturbation in any other mode $x^{(j)}$.
    \item \textit{Cumulative responses/sensitivities.} We integrate the response over time to define the cumulative sensitivity,
    \begin{equation}
     D_j = \sum_{t=0}^{\tau_\infty} R^{(1,j)}_t,
    \label{eq:sensitivity}
    \end{equation}
    which measures the cumulative response of the ENSO mode $x^{(k=1)}_{t}$ to a small impulse perturbation applied to $x^{(j)}_{t=0}$ at time $t = 0$. Given that the state vector is $\mathbf{x} \in \mathbb{R}^{n,T}$, the cumulative sensitivity will be $\mathbf{D} \in \mathbb{R}^{n}$. As discussed in Section~\ref{sec:FDT-Causal}, $D_j$ is therefore a causal quantity in the interventional sense: it measures the accumulated response of the ENSO mode to an \textit{external} perturbation imposed on $x^{(j)}$.
    \item \textit{Spatial projection.} Because the latent modes are grouped by physical variable (e.g., $j=1,\dots,10$ for SST, $j=11,\dots,20$ for SSH), the subset of sensitivities $D_j$ corresponding to a specific variable can be projected back onto its respective EOFs. This reconstructs continuous spatial sensitivity maps for each physical field.
\end{enumerate}

While we focus here on the ensemble mean of a specific mode, the framework is general. For example, one could instead define $\mathcal{A}(\mathbf{x}_t)$ to measure higher-order moments (e.g., variance). Because the latent modes are standardized, the resulting sensitivity maps are nondimensional. Their relative magnitudes can be compared across fields as responses per standardized EOF-coordinate perturbation. Alternatively, sensitivity maps in physical units can be obtained by evaluating a dimensional observable, or rescaling the spatial projection accordingly, as done in \cite{FabriCoarseGraining}. This step is not necessary for the present analysis. Note that Eq.~\eqref{eq:sensitivity} is a time-integrated quantity: transient fluctuations in $R^{(1,j)}_t$ that alternate in sign can cancel, so $D_j$ may be small even when short-time sensitivities are large. 

\subsubsection{Response validation against the FDT} \label{sec:FDT-validation}

\begin{figure}
\centering
\includegraphics[width=1\linewidth]{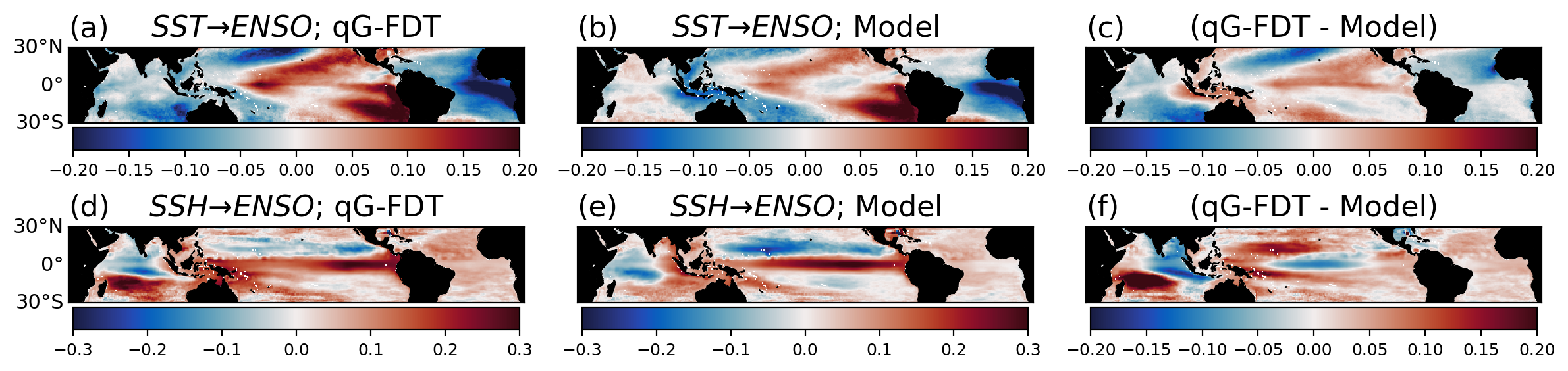}
\caption{Validation of the Markovian model's cumulative ENSO sensitivity predictions against the FDT benchmark over $\tau_\infty = 12$ months. Sensitivities are computed with Eq. \eqref{eq:sensitivity} and projected over the gridded map for interpretability. First row: ENSO sensitivity to perturbations in the SST field. Second row: ENSO sensitivity to perturbations in the SSH field. Left column: sensitivities estimated by the FDT. Center column: sensitivities predicted by the model. Right column: bias of the model; i.e., FDT - model prediction. Values are dimensionless, spatially projected from the latent EOF basis. Interpretation: as an example, negative values of SST in the tropical Atlantic (top-left panel) indicate that a positive (i.e. warming) SST perturbation in that region drives a long-term negative change in the mean of the ENSO mode; in physical space, such a response would correspond to cooler SST anomalies in the eastern-to-central equatorial Pacific.}
\label{fig:validation_ENSO_FDT}
\end{figure}

As a first step, we validate the responses of the reduced-order model against the independent FDT benchmark. The short record imposes three restrictions. First, to ensure robust evaluation, we compute both the quasi-Gaussian approximation of the FDT (qG-FDT) and the score matching (SM) estimator (Sec.~\ref{sec:FDT-estimation}): the two methods yield consistent results and we report only the qG-FDT in the main text. Second, we restrict our analysis to the ensemble mean response, as variance responses are dominated by statistical noise at this sample size. Third, as anticipated in Sec.~\ref{sec:FDT-Causal}, the empirical FDT response operator degrades at long lead times due to data scarcity. We show this empirically in Appendix~\ref{app:app_B} and SM Sec.~10 and choose a lag of $\tau_\infty = 12$ months to compute the cumulative sensitivities defined in Eq.~\eqref{eq:sensitivity}. This validation is shown in Figure~\ref{fig:validation_ENSO_FDT}.
\begin{itemize}
    \item \textit{Validation of ENSO sensitivity to perturbations in SST.} The FDT-inferred SST sensitivity map in Figure~\ref{fig:validation_ENSO_FDT}(a) reveals both remote Pacific influence and inter-basin connections. The eastern Pacific shows the expected positive sensitivity: warming perturbations in the eastern Pacific will lead to warming in the same location, reflected as a positive sensitivity of the ENSO mode. A prominent negative sensitivity emerges in the tropical Atlantic sector associated with Atlantic Ni\~no/Atlantic zonal-mode variability \cite{AtlanticNino}. This negative sensitivity is consistent with previous work via observational analyses and model experiments, showing that warm tropical Atlantic SST anomalies can alter the Walker circulation and favor opposite-sign SST anomalies in the eastern Pacific, thereby modulating ENSO variability \cite{Fonseca,Keenlyside,Ham}. A positive sensitivity appears in the subtropical North Pacific, broadly consistent with the Pacific Meridional Mode (PMM) pathway \cite{DiLorenzoPMM,ManuPMM}. The PMM has been shown to be an important driver of the equatorial Pacific variability at decadal time scales \cite{ManuPMM}.\\
    Figure~\ref{fig:validation_ENSO_FDT}(b,c): the data-driven model recovers the signs and broad structures of the eastern Pacific and tropical Atlantic responses but underestimates the subtropical North Pacific response associated with the PMM. The dynamics of the PMM involves surface winds and air–sea flux feedbacks \cite{Vimont,DiLorenzoPMM}. Such fast variability is absent in the SST-SSH state vector and it could appear as autocorrelated residual forcing. 
    \item \textit{Validation of ENSO sensitivity to perturbations in SSH.} The FDT-inferred sensitivity map in Figure~\ref{fig:validation_ENSO_FDT}(d) shows a positive sensitivity of the ENSO mode to changes in SSH across the tropical Pacific. The largest positive sensitivity is found in the eastern equatorial Pacific and it extends into the western part of the basin, highlighting an ``equatorial wave guide'' consistent with the ocean recharge paradigm \cite{rechargeI,rechargeII} for ENSO.\\
    Figure~\ref{fig:validation_ENSO_FDT}(e,f): the data-driven model reproduces the broad Pacific sensitivity only qualitatively: its maximum is displaced toward the central Pacific, the western-Pacific amplitude is too weak, and substantial biases remain in the Indian Ocean.
\end{itemize}

Importantly, we note that computing sensitivities over a shorter horizon of $\tau_\infty=3$ months leads to better agreement between emulators and FDT benchmarks. In this case, ENSO sensitivities to both SST and SSH perturbations remain largely confined to the Pacific. These shorter-horizon results are reported in Section~11 of the SM.\\

The validation above shows that the physics-constrained model largely recovers the stationary statistics of the ENSO mode but captures its responses to perturbations only qualitatively. Two observations suggest a common explanation. First, the discrepancy from the FDT benchmark grows with the integration horizon $\tau_\infty$: agreement is close at $\tau_\infty = 3$ months (see Figure~17 in Section~11 of the SM) and degrades at $\tau_\infty = 12$ months. Second, the largest discrepancies involve pathways, notably the PMM in the subtropical North Pacific, whose dynamics depend on fast atmospheric variability absent from the SST--SSH state vector. Both are consistent with unresolved processes leaving memory in the deterministic residuals. In the next section, we test whether a non-Markovian closure for unresolved fast processes can reduce these biases in the modeled ENSO responses.

\subsection{Improving ENSO response predictions with non-Markovian closures}

In this final section, we examine whether augmenting the physics-constrained model with a non-Markovian stochastic closure can reduce the response biases observed in Figure~\ref{fig:validation_ENSO_FDT}. This extension is motivated by a common issue in reduced-order modeling: even after selecting physically meaningful coarse-grained variables, the deterministic residuals may remain temporally correlated. Such memory can be reduced by enlarging the resolved state vector or averaging further in time to filter out fast variability \cite{FabriCoarseGraining}. Alternatively, when the chosen variables provide a useful but only approximately Markovian representation, the residual memory can be modeled stochastically. Given that the stationary statistics are well reproduced while the responses agree only qualitatively, we pursue this second route.\\

To achieve this, we integrate our proposed physics-constrained models with the multilayer stochastic modeling (MSM) strategy. The MSM framework was first proposed by Kravtsov et al. (2005) \cite{KRAVTSOV} and later generalized by Kondrashov et al. (2015) \cite{KONDRASHOV201533}. These models augment a reduced-order dynamics with additional residual layers that approximate unresolved memory effects. Here, we use the MSM as a stochastic extension of our physics-constrained emulator.

\subsubsection{Merging MSM with the constrained emulator}
Following previous work of MSM on ENSO \cite{KondraSeasonality,EMREnso} and idealized ocean models \cite{Agarwal}, we augment our discrete-time constrained neural models with two stochastic memory layers:
\begin{equation}
\begin{aligned}
\mathbf{x}_{t+1} &=
\mathbf{F} + \mathbf{M}\mathbf{Q}(\mathbf{x}_t)\mathbf{x}_t
+
\mathbf{r}_t, \\
\mathbf{r}_{t+1} &=
\mathbf{A}\mathbf{x}_t
+
\mathbf{B}\mathbf{r}_t
+
\mathbf{e}_t, \\
\mathbf{e}_{t+1} &=
\mathbf{C}\mathbf{x}_t
+
\mathbf{D}\mathbf{r}_t
+
\mathbf{E}\mathbf{e}_t
+
\mathbf{\Sigma}\bm{\xi}_t .
\end{aligned}
\label{eq:memory_model_2}
\end{equation}
Here, the term $\mathbf{F} + \mathbf{M}\mathbf{Q}(\mathbf{x}_t)\mathbf{x}_t$ denotes our proposed deterministic Markovian map. After fitting this map, we define the first residual as $\mathbf{r}_t = \mathbf{x}_{t+1} - \mathbf{f}(\mathbf{x}_t)$. The linear operators $\mathbf{A}$ and $\mathbf{B}$ in the first memory layer are then fitted by ordinary least squares (OLS). The variable $\mathbf{e}_t = \mathbf{r}_{t+1} - \mathbf{A}\mathbf{x}_t - \mathbf{B}\mathbf{r}_t$ represents the residual of this first layer: it can be closed directly with white noise or further modeled by adding another memory layer \footnote{We also evaluated a single-level closure; while effective, adding a second layer yields marginal yet significant improvements in capturing the system's autocorrelations.}. Here, we choose the latter and fit $\mathbf{e}_{t+1} = \mathbf{C}\mathbf{x}_t + \mathbf{D}\mathbf{r}_t + \mathbf{E}\mathbf{e}_t + \bm{\eta}_t$ again with OLS. The final residual $\bm{\eta}_t$ is then used to formulate a white-noise closure term $\bm{\eta}_t=\mathbf{\Sigma}\bm{\xi}_t$, where $\mathbf{\Sigma}$ is obtained as the Cholesky factor of the covariance of $\bm{\eta}_t$.\\ 
This closure allows for a two-way interaction between resolved (i.e. $\mathbf{x}_t$) and hidden (i.e. $\mathbf{r}_t,\mathbf{e}_t$) variables. It is the discrete-time analogue of the MSM strategy and it provides a practical approximation of the Mori-Zwanzig formalism in statistical mechanics \cite{KONDRASHOV201533,LucariniChekroun}.

\subsubsection{Effect on stationary statistics and response validation against the FDT}

The stochastic memory closure produces a modest but significant improvement in stationary statistics, most
clearly in the ENSO autocorrelation function; see Appendix~\ref{app:app_C}. As also shown in Appendix~\ref{app:app_C}, this improvement is systematic across the $n = 20$ degrees of freedom. In Section~10 of the SM we also report corrections to the mean response operator compared to the qG-FDT benchmark.\\

\begin{figure}
\centering
\includegraphics[width=1\linewidth]{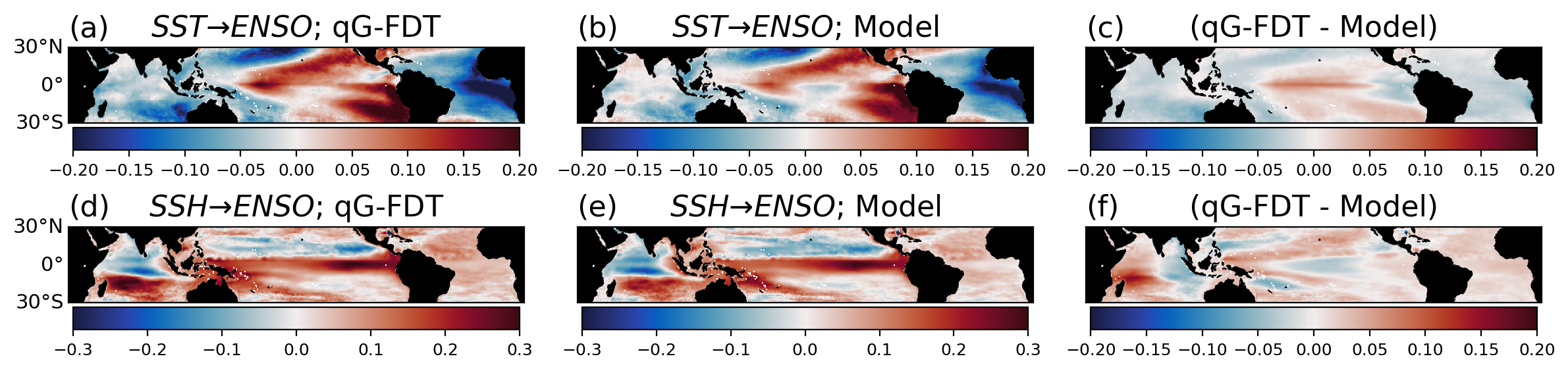}
\caption{Same as Figure~\ref{fig:validation_ENSO_FDT} but for the model with non-Markovian closure in Eq.~\eqref{eq:memory_model_2}.}
\label{fig:validation_ENSO_FDT_nonMarkov}
\end{figure}

Figure~\ref{fig:validation_ENSO_FDT_nonMarkov} shows the cumulative sensitivity maps produced by the non-Markovian model in Eq.~\eqref{eq:memory_model_2}. Relative to the Markovian results in Figure~\ref{fig:validation_ENSO_FDT}, these maps exhibit substantially improved quantitative agreement with the FDT benchmark. For SST sensitivities, shown in Figure~\ref{fig:validation_ENSO_FDT_nonMarkov}(a-c), the non-Markovian closure strengthens the sensitivity over the subtropical North Pacific, bringing it closer to the FDT estimate, while leaving largely unchanged the eastern Pacific and tropical Atlantic sensitivities already captured by the Markovian model. For SSH sensitivities, shown in Figure~\ref{fig:validation_ENSO_FDT_nonMarkov}(d-f), the maximum value shifts eastward and the western-Pacific response strengthens, both in agreement with the FDT benchmark. The Indian-Ocean sensitivity bias is further reduced.\\ 
We note that, although ground-truth ENSO sensitivities are unavailable, this improved agreement provides meaningful validation because the FDT benchmark and the emulator responses are obtained through methodologically independent routes.

\subsubsection{Predicted 10-year cumulative ENSO sensitivities}

\begin{figure}
\centering
\includegraphics[width=1\linewidth]{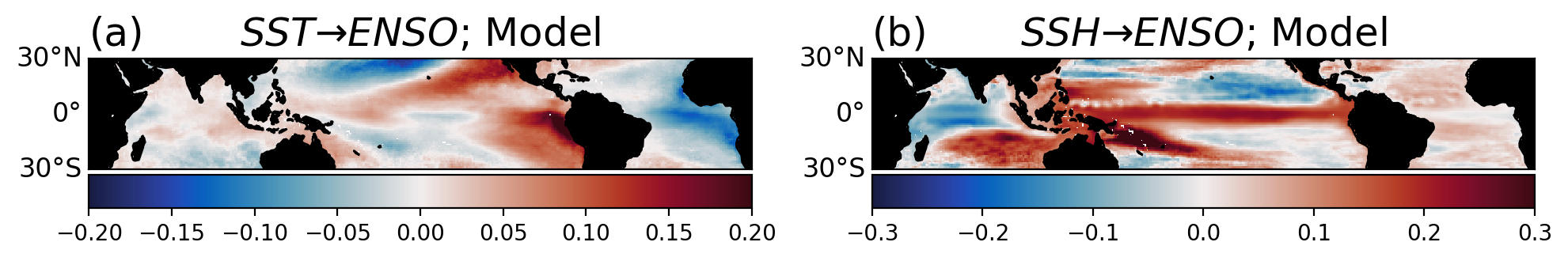}
\caption{Long-time ($\tau_\infty = 120$ months) ENSO sensitivities predicted by the non-Markovian model in Eq.~\eqref{eq:memory_model_2}. Left panel: sensitivity of the ENSO mode to perturbations in the SST field. Right panel: sensitivity to perturbations in the SSH field. Sign conventions and interpretation as in Figure~\ref{fig:validation_ENSO_FDT}.}
\label{fig:long_time_responses_memory}
\end{figure}

The agreement between emulator sensitivities and the methodologically independent FDT benchmark over the 12-month validation horizon motivates extending the cumulative sensitivity calculation to $\tau_\infty = 120$ months (10 years). Beyond approximately one year, finite-sample FDT estimates become increasingly dominated by sampling noise and become unreliable; see Appendix \ref{app:app_B}. At these longer horizons, we therefore estimate responses through perturbation experiments with the response-validated emulator, using large ensembles to obtain statistically significant sensitivity estimates. This analysis is shown in Figure \ref{fig:long_time_responses_memory}. The main signals in ENSO sensitivities to perturbations in the SST field are already established in the one-year window analyzed in the previous section. Extending the cumulative-response analysis to 10 years further underscores the importance of three main SST regions to ENSO variability: the eastern Pacific, the subtropical North Pacific, and the tropical Atlantic. Sensitivities in the SSH show larger changes but a similar structure: the positive eastern-Pacific sensitivities broaden across the full equatorial Pacific and maximize in the western-Pacific region. SSH is a proxy for upper-ocean heat content and thermocline depth and positive SSH anomalies correspond to a recharged ocean state with a deeper thermocline. A maximum SSH sensitivity in the western-Pacific indicates positive, deeper thermocline anomalies driving subsequent ENSO growth through delayed ocean adjustment, consistent with the ocean recharge paradigm \cite{rechargeI,rechargeII}.\\

Indian-Ocean SST does not emerge as a dominant 10-year cumulative sensitivity in this representation,
despite some recent literature documenting Indian-Ocean influence on ENSO \cite{Pantropical}. The relevant pathway may instead be mediated by subsurface variability captured by SSH (for which our model shows significant Indian Ocean sensitivities). Alternatively, sign-changing SST responses may cancel in the time integral, or previously reported associations may reflect a statistical correlation rather than a causal link. Distinguishing among these possibilities requires analysis of the full time-dependent response operator across different fields and provides an important avenue for future work.

\section{Conclusions} \label{sec:conclusions}
We introduced a flexible framework for physics-constrained reduced-order neural modeling of turbulent dynamical systems. We first formulated an abstract discrete-time representation of turbulent dynamics based on finite-time flow maps with strict energy-conserving nonlinearities, tailored to coarse-grained, discrete observational data. We then parameterized this structure with neural networks, yielding physics-constrained reduced-order emulators capable of stable long-time integration. We further leveraged the fluctuation--dissipation theorem (FDT) in two ways: as a response-theory benchmark for validating the emulators' forced responses using only unperturbed data and as a tool for identifying candidate direct causal links. These links can be imposed through regularization to suppress spurious dependencies, but only after strict validation. Thus, while the energy-conserving architecture provides an \textit{a priori} geometric constraint, causal regularization remains optional and data-driven. Across the numerical experiments, the constrained emulators remained stable, reproduced stationary statistics, and predicted mean and variance responses to weak and strong forcings despite being trained entirely on unperturbed trajectories. By contrast, unconstrained baselines reproduced stationary statistics but yielded less reliable results when perturbed, and, in some cases, unstable long-time behavior, consistent with recent evidence that machine-learning emulators can match unperturbed statistics while misrepresenting forced responses \cite{DAemulators,Senne,Bosong}.\\

The application to tropical climate dynamics demonstrates the relevance of the framework for modeling realistic, partially observed turbulent systems. Because the short ocean-reanalysis record was insufficient to infer a reliable binary causal graph, we focused only on the physics-constrained model. The model remained stable and recovered important statistics of the El Ni\~no--Southern Oscillation (ENSO). Its cumulative responses showed qualitative agreement with the independent FDT benchmark inferred from the same unperturbed data. The remaining discrepancies indicated that the selected coarse-grained variables provide only an approximately Markovian representation of tropical dynamics. Predicting forced responses is more demanding than reproducing the stationary statistics of the ENSO mode because perturbations propagate through the full resolved state, whose effective dynamics can retain memory of unresolved processes. We therefore combined the proposed physics-constrained model with the non-Markovian stochastic closure of Kondrashov et al. (2015) \cite{KONDRASHOV201533}, substantially improving quantitative agreement with the FDT benchmark. The resulting response-validated model was then used to predict 10-year cumulative sensitivity maps characterizing the long-term causal drivers of ENSO variability. In doing so, we also provided a practical procedure for projecting sensitivity experiments from reduced coordinates into interpretable physical-space maps.\\

The framework is flexible and modular rather than tied to a single architecture: physics constraints provide the structural foundation of our strategy, causal regularization can be added when supported by validation criteria, stochastic closures can be introduced when memory persists in the deterministic residuals, and the FDT is leveraged to extend evaluation of data-driven models beyond stationary statistics. Future work will explore using FDT response operators directly as \textit{a posteriori} calibration targets for pretrained emulators. This would complement the proposed causal regularization by constraining model sensitivity at the level of responses of observables rather than local Jacobian structure. Overall, the proposed methodology provides a practical strategy for stable, response-validated reduced-order modeling of complex turbulent systems.

\paragraph*{Code availability.}
The code used to generate the results is available at\\ \href{https://github.com/FabriFalasca/Physics-and-causally-constrained-neural-models/}{https://github.com/FabriFalasca/Physics-and-causally-constrained-neural-models/}.
\paragraph*{Acknowledgments.}
This research was supported by Schmidt Sciences, LLC, through the M$^2$LInES project. Computational resources and support were provided in part by the NYU IT High Performance Computing facilities, services, and staff. F.F. is grateful to Rory Basinski, Matthieu Blanke, and Andre Souza for enriching discussions on this subject.

\appendix
\section{Discrete-time, physics-constrained model: Building geometric intuition with the Lorenz-63 system} \label{app:app_A}

To build intuition for the proposed physics-constrained discrete-time framework, we illustrate how a continuous deterministic chaotic system with nonlinearities of the form in Eq.~\eqref{eq:majda_generalized} maps onto Eq.~\eqref{eq:discrete_case} for small $\Delta t$. The extension to stochastic systems is detailed in Section 2 of the SM. Here, we focus on the deterministic Lorenz-63 equations \cite{Lorenz}, which belong to the class of models derived as Galerkin truncations of turbulent flows and therefore respect the abstract formulation in Eq.~\eqref{eq:majdaQuadratic}. The continuous equations governing the state $\mathbf{x} = (x, y, z)^\top$ are given by:
\begin{equation}
\dot{x} = \sigma(y - x), ~ \dot{y} = x(\rho - z) - y, ~  \dot{z} = xy - \beta z.
\label{eq:lorenz63}
\end{equation}
We can decompose these dynamics exactly into a constant linear operator $\mathbf{A}$ and a state-dependent, strictly skew-symmetric matrix $\mathbf{S}(\mathbf{x})$:
\begin{equation}
    \dot{\mathbf{x}} = \underbrace{\begin{bmatrix} -\sigma & \sigma & 0 \\ \rho & -1 & 0 \\ 0 & 0 & -\beta \end{bmatrix}}_{\mathbf{A}} \begin{bmatrix} x \\ y \\ z \end{bmatrix}
    + \underbrace{\begin{bmatrix} 0 & 0 & 0 \\ 0 & 0 & -x \\ 0 & x & 0 \end{bmatrix}}_{\mathbf{S}(\mathbf{x})} \begin{bmatrix} x \\ y \\ z \end{bmatrix}.
    \label{eq:lorenz63_decomposition}
\end{equation}
The nonlinear interactions are entirely isolated within $\mathbf{S}(\mathbf{x})\mathbf{x}$. Because $\mathbf{S}(\mathbf{x})^\top = -\mathbf{S}(\mathbf{x})$, the nonlinear term is strictly energy-preserving, satisfying $\mathbf{x}^\top \mathbf{S}(\mathbf{x}) \mathbf{x} = 0$. Given a small $\Delta t$, a first-order operator splitting procedure \cite{LieTrotter} applied to Eq.~\eqref{eq:lorenz63_decomposition} yields the discrete-time approximation: 
\begin{equation}
    \mathbf{x}_{t+1} = \underbrace{\exp(\Delta t \mathbf{A})}_{\mathbf{M}} \underbrace{\exp(\Delta t \mathbf{S}(\mathbf{x}_t))}_{\mathbf{Q}(\mathbf{x}_t)} \mathbf{x}_t.
    \label{eq:lorenz63_splitting}
\end{equation}
A detailed formal derivation of this splitting is provided in Section 2 of the SM. Here, $\mathbf{x}_t = (x_t, y_t, z_t)^\top$ denotes the discrete state vector at time $t$. Since the infinitesimal generator $\mathbf{S}(\mathbf{x}_t)$ is real and skew-symmetric, its matrix exponential $\mathbf{Q}(\mathbf{x}_t)$ is orthogonal and has unit determinant; hence $\mathbf{Q}(\mathbf{x}_t)\in SO(3)$. The nonlinear update matrix $\mathbf{Q}(\mathbf{x}_t)$ of the Lorenz-63 system can be calculated analytically as:
\begin{equation}
    \mathbf{Q}(\mathbf{x}_t) = \exp(\Delta t \mathbf{S}(\mathbf{x}_t)) = \begin{bmatrix} 1 & 0 & 0 \\ 0 & \cos(x_t \Delta t) & -\sin(x_t \Delta t) \\ 0 & \sin(x_t \Delta t) & \cos(x_t \Delta t) \end{bmatrix}.
    \label{eq:q_lorenz}
\end{equation}
The linear operator $\mathbf{M}$ can be computed numerically once a $\Delta t$ is specified. Here, we focus on its analytical approximation. For small $\Delta t$, we approximate $\mathbf{M}$ via a second-order Taylor expansion:
\begin{equation}
\begin{aligned}
\mathbf{M} &= \exp(\Delta t \mathbf{A}) \approx \mathbf{I} + \Delta t \mathbf{A} + \frac{\Delta t^2}{2} \mathbf{A}^2 \\ 
&= \begin{bmatrix} 
1 - \sigma \Delta t + \frac{\Delta t^2}{2}(\sigma^2 + \sigma\rho) & \sigma \Delta t - \frac{\Delta t^2}{2}(\sigma^2 + \sigma) & 0 \\ 
\rho \Delta t - \frac{\Delta t^2}{2}(\sigma\rho + \rho) & 1 - \Delta t + \frac{\Delta t^2}{2}(\sigma\rho + 1) & 0 \\ 
0 & 0 & 1 - \beta \Delta t + \frac{\Delta t^2}{2}\beta^2 \end{bmatrix}.
\label{eq:M_taylor_lorenz}
\end{aligned}
\end{equation}
Given an initial condition $\mathbf{x}_{t=0}$, the discrete mapping of the Lorenz-63 system in Eq.~\eqref{eq:lorenz63_splitting} explicitly factors the dynamics into two sequential geometric operations: 
\begin{enumerate}
    \item \textit{Energy-preserving rotation.} First, the state-dependent orthogonal update $\mathbf{v}_t = \mathbf{Q}(\mathbf{x}_t)\mathbf{x}_t$ applies an energy-preserving rotation strictly in the $y$-$z$ plane. The rotation angle $\theta = x_t \Delta t$ is state-dependent and dictated by the instantaneous value of the $x$ coordinate at time $t$, so nearby trajectories experience different rotations.
    \item \textit{Linear transformation.} Second, the linear transformation $\mathbf{x}_{t+1} = \mathbf{M}\mathbf{v}_t$ stretches and contracts the rotated vector $\mathbf{v}_t$.
\end{enumerate}
Thus, the Lorenz-63 dynamics consists of repeatedly applying the same two-stage mechanism: a state-dependent, energy-preserving rotation followed by a fixed linear deformation. As discussed in Section \ref{sec:physics_constrained}, this two-stage mechanism is a general feature of a large class of fluid flows when projected onto orthogonal basis functions. The Charney–DeVore \cite{charneyDeVore}, Lorenz-96 \cite{Lorenz96}, and Lorenz-84 \cite{Lorenz84} models, as well as quasi-geostrophic (QG) equations projected on orthogonal bases \cite{FranzkeMajda}, all admit this interpretation.\\

Crucially, this geometric factorization is representative of how our general data-driven framework operates on larger, high-dimensional physical systems: the nonlinear rotation acts as a generalized, energy-conserving nonlinearity analogous to advective terms in fluid dynamics, while the linear operator $\mathbf{M}$ accounts for the system's underlying linear dynamics. The reader can verify that the formulation above allows one to reconstruct the Lorenz attractor for small $\Delta t$.\\

Importantly, unlike a purely dissipative operator, the Lorenz-63 linear operator $\mathbf{M}$ is not contractive in its norm: it expands some directions while contracting others. In this sense, Lorenz-63 is a special case relative to the class of turbulent systems considered in this study, where the linear operator is decomposed into energy-conserving dispersive and strictly dissipative components. Therefore, this example should be interpreted only as a pedagogical illustration of the proposed splitting structure for small $\Delta t$. For large sampling intervals, there is no reason to expect that the system can be represented, or learned from data, by simple effective operators. In Section 2.2 of the SM, we demonstrate that our framework can indeed learn stable, coarse-grained effective dynamics at large sampling scales if an underlying effective dynamics actually exists. We showcase this using the Charney-DeVore model, demonstrating stable simulations from severely subsampled data in a regime where standard numerical integrators and traditional continuous-time data-driven models fail.\\

\section{Numerical estimation of model response operators} \label{app:app_A_1}

To estimate the response operator from a model we follow the steps proposed in \cite{FabriCoarseGraining}. Given a model $\mathbf{x}_{t+1} = \mathbf{f}(\mathbf{x}_t) + \mathbf{\Sigma}\bm{\xi}(t)$, we build the impulse response operator as follows:

\begin{enumerate}
    \item We simulate a very long trajectory from a random initial condition and remove an initial transient. We then sample $N_e$ random points from the simulation to define an ensemble of $N_e$ initial conditions on the model’s attractor. $N_e$ should be very large (i.e. $N_e \gg 1$) in order to sample the whole attractor and approximate averages over the invariant distribution.
    \item For each one of the $N_e$ initial conditions, we impose an impulse perturbation $\Delta_j$ to the degree of freedom $x^{(j)}$ at time $t = 0$. The amplitude of the perturbation  should be theoretically infinitesimally small, ensuring linearity of the response even in nonlinear systems. In practice we do as follows: we consider the long time series of $x^{(j)}_t$ from the long control integration above and define $\Delta_j = 10^{-1} \sigma_j$, where $\sigma_j$ is the standard deviation of time series $x^{(j)}_t$. 
    \item Therefore, for a given initial condition $\mathbf{x}_0$, we simulate two trajectories: a control trajectory without perturbation and a perturbed one, where an impulse perturbation $\Delta_j$ has been imposed on the $j$-th degree of freedom at time $t = 0$. This procedure is repeated in parallel for all initial conditions, resulting in an ensemble of $N_e$ pairs of control and perturbed trajectories. Importantly, for each pair of control and perturbed trajectories we set the same noise process by fixing the random seed.
    \item At each time $t$ we then estimate the time-dependent mean $\langle \mathcal{A}(x^{(k)}_t) \rangle$ of an observable $\mathcal{A}(x^{(k)}_t)$ for both the perturbed and unperturbed runs. We refer to the perturbed and control ensemble averages as $\langle \mathcal{A}(x^{(k)}_t) \rangle_{\text{p}}$ and $\langle \mathcal{A}(x^{(k)}_t) \rangle$, respectively. The observables considered in this study are: (i) $\mathcal{A}(x^{(k)}_t) = x^{(k)}_t$ and (ii) $\mathcal{A}(x^{(k)}_t) = (x^{(k)}_t - \mu^{(k)}_t)^2$, where $\bm{\mu}_t$ represents the time-dependent mean of the distribution. These observables quantify the responses of the ensemble mean and variance, respectively.
    \item We define the impulse response operator for observable $\mathcal{A}(x^{(k)}_t)$ as
    \begin{equation*}
        R^{k,j}_t = \frac{\langle \mathcal{A}(x^{(k)}_t) \rangle_{\text{p}} - \langle \mathcal{A}(x^{(k)}_t) \rangle}{\Delta_j}.
    \end{equation*}
\end{enumerate}

\section{Noise estimation of response operators from data at long time scales} \label{app:app_B}

The accuracy of FDT estimations from observational data is inherently limited by sample size: the statistical error of the FDT estimator grows rapidly with lead time \cite{nonEqStatMech}. This complicates the estimation of the mean response from short datasets and renders variance responses highly unreliable. To illustrate this issue, we examine the practical application from Section \ref{sec:application}, where the state vector has dimension $n = 20$ but the dataset consists of only $T = 792$ discrete points. We compute the time-dependent mean response operator $R^{(2,1)}_t$, representing the response of the second mode $x^{(2)}_t$ to an impulse perturbation applied to the first mode (the SST ENSO mode) $x^{(1)}$ at $t = 0$. This example is representative of the broader issue; we refer the reader to Section~10 of the SM for additional cases.\\

\begin{figure}
\centering
\includegraphics[width=0.7\linewidth]{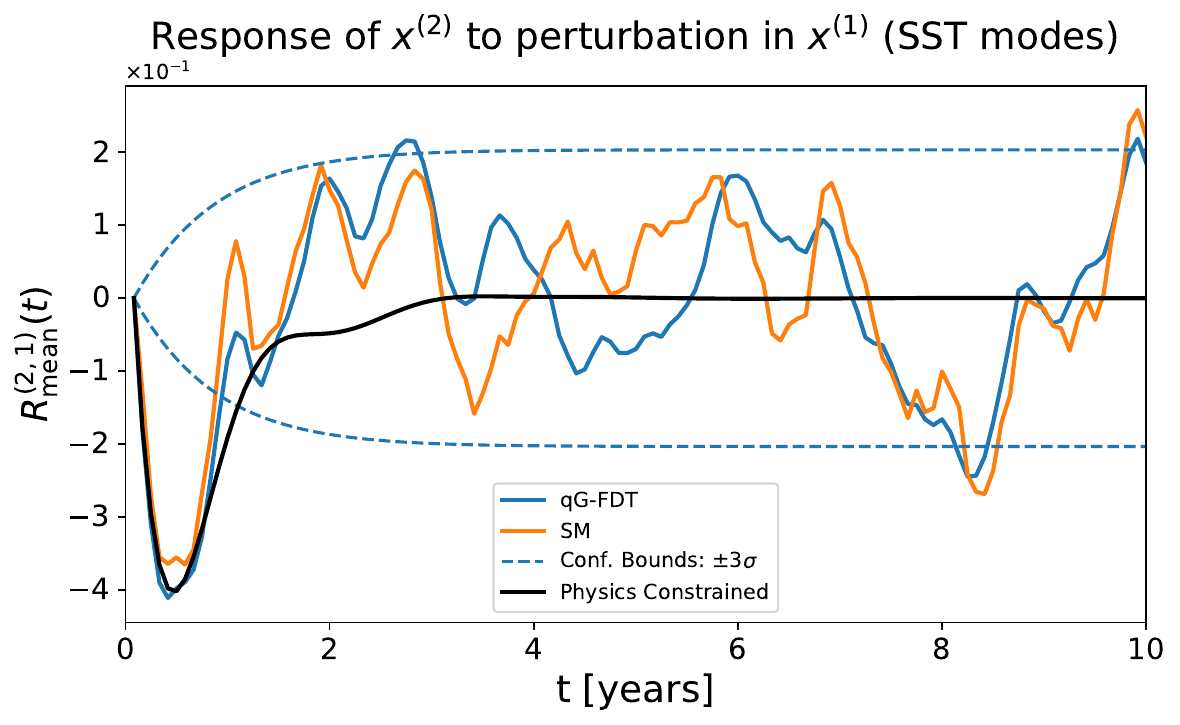}
\caption{Time-dependent mean response of mode $x^{(2)}_t$ to a small impulse perturbation imposed on $x^{(1)}$ at time $t = 0$ (i.e., the mean response operator $R^{(2,1)}_t$). Blue: computed using the qG-FDT approximation in Eq.~\eqref{eq:quasi-Gaussian}. Orange: computed using the full FDT formula in Eq.~\eqref{eq:response_general}, where the score has been approximated via the score-matching (SM) procedure introduced in Section \ref{sec:FDT-estimation}. Black: computed by simulating responses over a 1000-member ensemble using the physics-constrained model proposed in this study. Dashed black curves: theoretical confidence bounds at the $\pm 3 \sigma$ level. FDT estimates (Blue and Orange curves) falling inside these bounds are statistically indistinguishable from spurious finite-sample noise.}
\label{fig:response_data}
\end{figure}

Figure \ref{fig:response_data} compares this response computed via: (a) the quasi-Gaussian FDT (qG-FDT), (b) the score-matching FDT (SM-FDT), and (c) the physics-constrained reduced-order model proposed in this work. While the qG-FDT and SM-FDT yield nearly identical results, and all three methods largely agree at short time scales, the empirically estimated FDT responses are very noisy at longer lead times. This is a direct artifact of estimating temporal averages from short, finite time series. In contrast, the reduced-order emulator computes responses by averaging over a 1000-member ensemble. Consequently, the emulator correctly captures the physical expectation that the response to a small impulse perturbation must eventually decay to zero at long time scales. To further formalize this, we evaluate the empirical FDT estimates against the analytical confidence bounds proposed in Falasca et al. (2024) \cite{FabCausal} (Eq. (8) in that paper). These bounds quantify the expected spurious response generated by a finite-sample multivariate autoregressive (AR(1)) null model with zero true cross-dependencies. As shown in Figure \ref{fig:response_data}, the long-time fluctuations of the empirical FDT fall entirely within the $\pm 3\sigma$ bounds, confirming they are statistically indistinguishable from spurious red-noise estimates.\\

This justifies our strategy in the main text: for short datasets, the FDT should only be leveraged to validate the short-time responses of the emulator. A successful short-time validation provides evidence supporting the emulator’s use for longer-time predictions.

\section{Effect of non-Markovian closures on autocorrelation functions} 
\label{app:app_C}

Here we assess the impact of augmenting the proposed physics-constrained neural model with the stochastic memory closure in Eq.~\eqref{eq:memory_model_2}. This closure follows the multilevel residual-regression strategy of Kondrashov et al. \cite{KONDRASHOV201533}, adapted here to our discrete-time finite-map formulation. The augmented model reproduces the probability distribution of the ENSO mode with skill comparable to the Markovian physics-constrained model discussed in Section~\ref{sec:stationary_statistics}. The main improvement is visible in the autocorrelation function, which more closely follows the observational estimate. The Fourier spectrum also shifts modestly toward lower frequencies, bringing the upper part of the model ensemble envelope closer to the large observed spectral peak.

\begin{figure}
\centering
\includegraphics[width=1\linewidth]{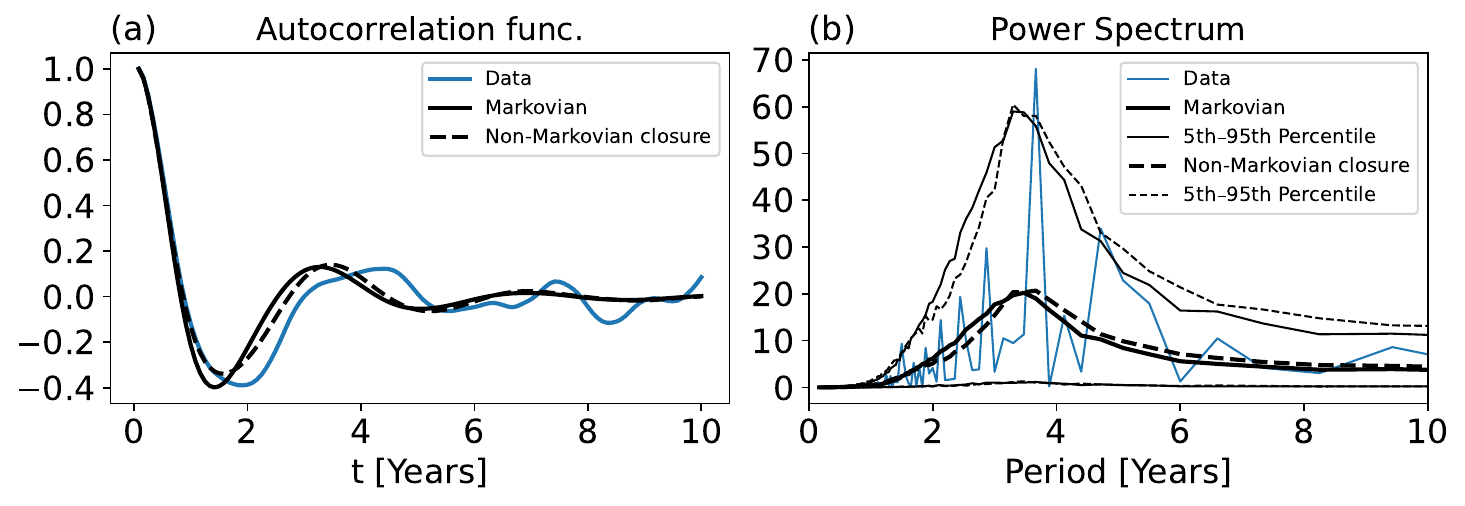}
\caption{Stationary statistics of the ENSO mode for the physics-constrained model with stochastic memory closure. Panel (a): Autocorrelation function. The solid black curve shows the ensemble mean of the Markovian physics-constrained emulator. The dashed black curve shows the ensemble mean of the emulator augmented with the stochastic memory closure in Eq.~\eqref{eq:memory_model_2}. Blue denotes the observational estimate. Panel (b): Same as panel (a), but for the Fourier spectrum.}
\label{fig:ENSO_stationary_memory}
\end{figure}

\begin{figure*}[h!]
\centering
\includegraphics[width=0.9\textwidth]{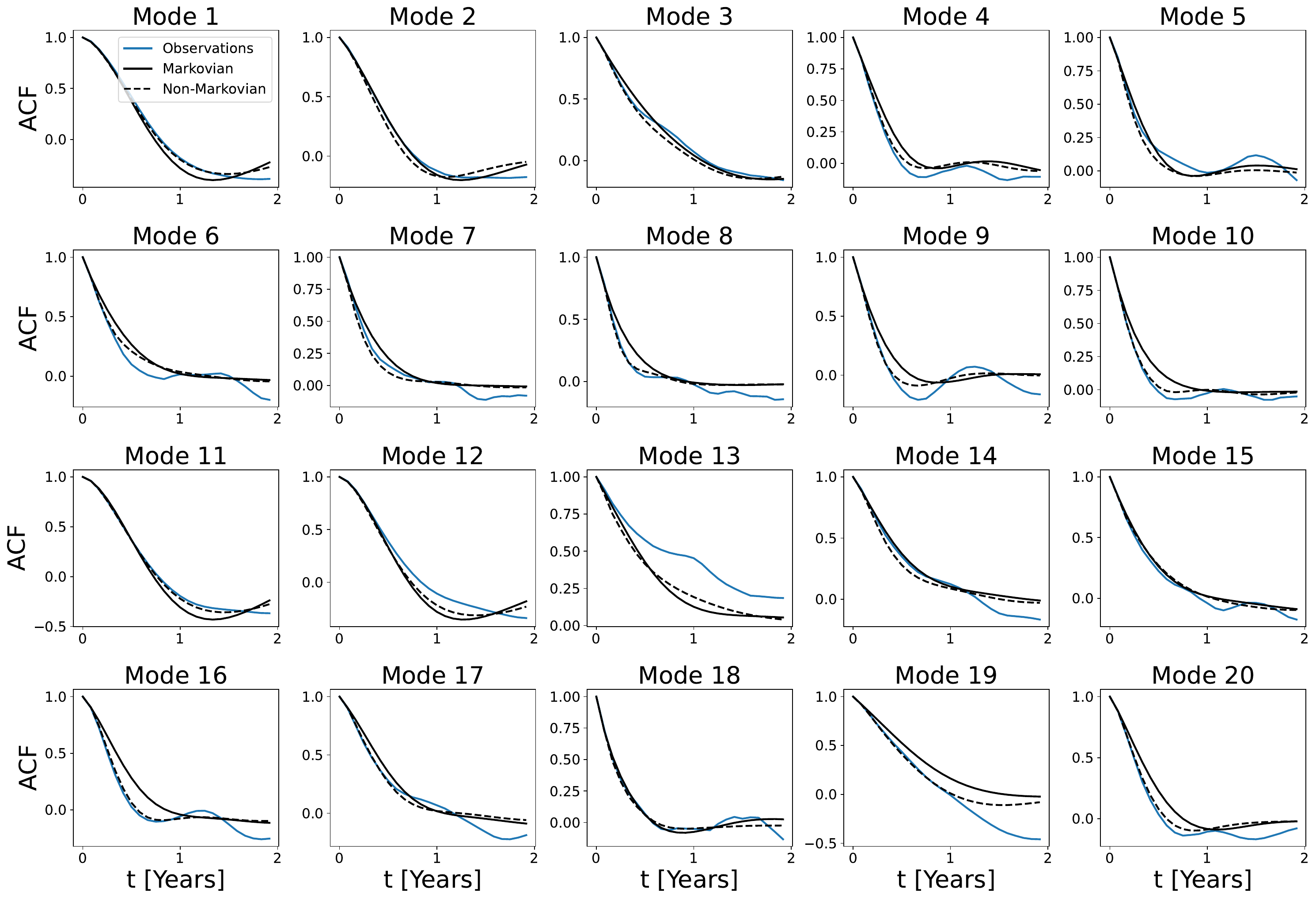}
\caption{Autocorrelation functions (ACFs) for the state variables $x^{(1)}_t, \dots, x^{(20)}_t$ of the reduced-order models. For the models, the ACFs are computed as the ensemble mean over 1000 independent trajectories. Solid black lines denote the model with the Markovian closure, while dashed black lines correspond to the model with the non-Markovian closure.}
\label{fig:all_ACFs}
\end{figure*}

\begin{figure}
\centering
\includegraphics[width=0.4\linewidth]{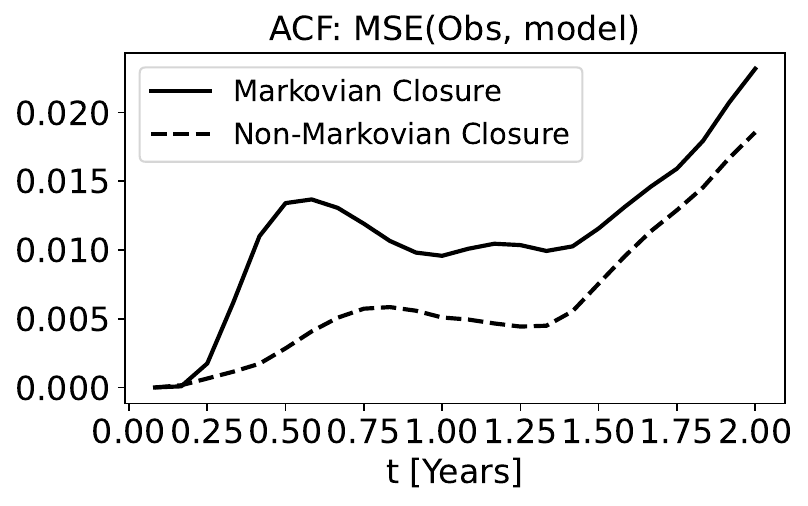}
\caption{Total mean squared error (MSE) computed at each lag, aggregated across all $n = 20$ autocorrelation functions (ACFs), between the ACFs in the observed and emulated data. The non-Markovian memory closure yields a consistently lower error than the Markovian model, demonstrating a systematic correction of the simulated autocorrelations.}
\label{fig:ACFs-MSE}
\end{figure}

To further quantify this improvement across all autocorrelation functions (ACFs), we expand our analysis to all 20 degrees of freedom of the reduced-order models, $x^{(1)}_t, x^{(2)}_t, \dots, x^{(20)}_t$. Figure~\ref{fig:all_ACFs} compares the observational ACFs against those generated by the models utilizing Markovian and non-Markovian closures. We focus on a lag range of the first two years, as the observational statistics are least affected by sampling noise within this window. The model with non-Markovian closure generally enhances the simulated autocorrelations. Finally, to rigorously quantify this improvement, we compute the total mean squared error (MSE) of the emulated ACFs relative to the observations. The MSE is aggregated across all $n=20$ modes and tracked as a function of time lag. This analysis is shown in Figure~\ref{fig:ACFs-MSE}: the model with non-Markovian memory closure yields a consistently lower MSE.

\clearpage

\stopcontents[main]
\clearpage
\startcontents[sm]

\begin{center}
    {\large \bf Supplemental Material for:\\[4pt]
    Physics constraints and response validation in discrete-time reduced-order modeling: from idealized turbulent systems to climate dynamics}\\[8pt]
    Fabrizio Falasca and Laure Zanna\\
    {\small Courant Institute School of Mathematics, Computing and Data Science, New York University}\\[4pt]
    (Dated: \today)
\end{center}

\setcounter{section}{0}
\setcounter{subsection}{0}
\setcounter{subsubsection}{0}
\setcounter{figure}{0}
\setcounter{table}{0}
\setcounter{equation}{0}

\renewcommand{\thesection}{\arabic{section}}
\renewcommand{\thefigure}{\arabic{figure}}
\renewcommand{\thetable}{\arabic{table}}
\renewcommand{\theequation}{\arabic{equation}}

\renewcommand{\theHsection}{SM.\arabic{section}}
\renewcommand{\theHsubsection}{SM.\arabic{section}.\arabic{subsection}}
\renewcommand{\theHsubsubsection}{SM.\arabic{section}.\arabic{subsection}.\arabic{subsubsection}}
\renewcommand{\theHfigure}{SM.\arabic{figure}}
\renewcommand{\theHtable}{SM.\arabic{table}}
\renewcommand{\theHequation}{SM.\arabic{equation}}

\section*{Supplemental Material Contents}
\begingroup
\setcounter{tocdepth}{2}
\printcontents[sm]{}{1}{}
\endgroup

\clearpage

\section{Physics- and causality-constrained models}

In Section 2 of the main text, we introduced a discrete-time formulation based on the abstract class of turbulent dynamical systems considered by Majda and collaborators \cite{MajdaIntroTurbulence}. The formulation takes the following form
\begin{equation}
\mathbf{x}_{t+1} = \mathbf{F} + \mathbf{M}\mathbf{Q}(\mathbf{x}_t)\mathbf{x}_t + \mathbf{\Sigma}\bm{\xi}_t,
\label{eq:discrete_case_SM}
\end{equation}
where the nonlinear operator $\mathbf{Q}(\mathbf{x}_t)$ is constrained to be strictly orthogonal ($\mathbf{Q}(\mathbf{x}_t)^\top \mathbf{Q}(\mathbf{x}_t) = \mathbf{I}$). The isolated nonlinear term $\mathbf{v}_t = \mathbf{Q}(\mathbf{x}_t)\mathbf{x}_t$ contributes no net growth or decay to the energy budget, strictly preserving the $L_2$-norm of the state:
\begin{equation*}
\|\mathbf{v}_t\|^2 = \|\mathbf{Q}(\mathbf{x}_t)\mathbf{x}_t \|^2 = \mathbf{x}_t^\top \mathbf{Q}(\mathbf{x}_t)^\top \mathbf{Q}(\mathbf{x}_t)\mathbf{x}_t = \mathbf{x}_t^\top \mathbf{I} \mathbf{x}_t = \|\mathbf{x}_{t}\|^2.
\end{equation*}
Geometrically, the deterministic update rule acts via two sequential operations: a state-space rotation, $\mathbf{v}_t = \mathbf{Q}(\mathbf{x}_t)\mathbf{x}_t$, followed by a linear transformation $\mathbf{M}\mathbf{v}_t$. The overall energy growth or decay of the system is therefore controlled exclusively by the singular values of the linear operator $\mathbf{M}$.\\

This discrete formulation can be efficiently parameterized by neural networks; in this supplement, we detail its implementation.

\subsection{Details on the emulator implementation}

Given an $n$-dimensional discrete dynamical system represented by a long trajectory $\mathbf{x} \in \mathbb{R}^{n, T}$, with length $T$, we aim to fit: 
\begin{equation}
\begin{split}
\mathbf{x}_{t+1} = \mathbf{f}(\mathbf{x}_t) + \mathbf{\Sigma} \bm{\xi}_t~
\end{split}
\label{eq:functional_form_discrete} 
\end{equation}
where the deterministic drift is
\[
\mathbf{f}(\mathbf{x}_t)=\mathbf{F}+\mathbf{M}\mathbf{Q}(\mathbf{x}_t)\mathbf{x}_t.
\]\\

\paragraph{Fitting the deterministic drift $\mathbf{f}(\mathbf{x}_t)$.} 
We fit the deterministic dynamics such that $\mathbf{x}_{t+1} \approx \mathbf{f}(\mathbf{x}_t)$. We do so by minimizing the mean squared error (MSE) over the trajectory:
\begin{equation}
    \mathcal{L} = \text{MSE}(\mathbf{x}_{t+1}, \mathbf{f}(\mathbf{x}_t)).
\label{eq:vanilla_loss}
\end{equation}

\begin{itemize}
    \item All terms comprising the discrete map, $\mathbf{f}(\mathbf{x}_t) = \mathbf{F} + \mathbf{M}\mathbf{Q}(\mathbf{x}_t)\mathbf{x}_t$, are trained jointly via gradient descent. The primary challenge in this joint optimization is the initialization step. To resolve this, the linear components $\mathbf{F}$ and $\mathbf{M}$ are first pre-computed using Ordinary Least Squares (OLS) to capture the baseline linear dynamics. This deterministic ``first guess'' serves as the initialization for the network parameters. 
    \item The nonlinear operator $\mathbf{Q}(\cdot)$ is parameterized by a Multi-Layer Perceptron (MLP) with a SiLU activation function. To strictly enforce orthogonality ($\mathbf{Q}^\mathrm{T}\mathbf{Q} = \mathbf{I}$) across all forward passes, we employ the following architectural construction:
    \begin{itemize}
    \item The MLP maps a state $\mathbf{x}_t \in \mathbb{R}^{n}$ to $n(n-1)/2$ independent scalar values. 
    \item These values populate the upper-triangular elements of a matrix, which is then used to construct an exact skew-symmetric matrix $\mathbf{S}(\mathbf{x}_t) = -\mathbf{S}^\mathrm{T}(\mathbf{x}_t)$.
    \item We map the skew-symmetric $\mathbf{S}(\mathbf{x}_t)$ to an orthogonal operator via the matrix exponential, $\mathbf{Q}(\mathbf{x}_t) = \exp(\mathbf{S}(\mathbf{x}_t))$, or the computationally efficient Cayley transform, $\mathbf{Q}(\mathbf{x}_t) = (\mathbf{I} - \mathbf{S}(\mathbf{x}_t))^{-1}(\mathbf{I} + \mathbf{S}(\mathbf{x}_t))$. In the main text we use the matrix exponential as default option.
    \end{itemize}
    Crucially, the weights of the final linear layer of the MLP are initialized to exactly zero. Consequently, at the onset of training, the network outputs $\mathbf{S}(\mathbf{x}_t) = \mathbf{0}$, which yields $\mathbf{Q}(\mathbf{x}_t) = \mathbf{I}$. This initialization strategy guarantees that the joint training step begins precisely at the stable OLS baseline before gradually introducing the parameterized nonlinear dynamics.
    \item Finally, once the deterministic discrete map $\mathbf{f}$ is estimated, we close the model with a stochastic forcing. We described this in Section 2 of the main paper.
\end{itemize}

\paragraph{Data standardization.} Data standardization requires careful treatment, as the strict physical constraint of energy conservation (the $L_2$ norm) depends on the geometry of the original variables and is not generally preserved in a shifted and scaled coordinate space. To ensure numerical stability during optimization, the neural network emulator is trained on standardized variables, obtained by transforming each time series in $\mathbf{x}_t$ to zero mean and unit variance:
$$ \tilde{\mathbf{x}}_t = \mathbf{\Sigma}_{\text{std}}^{-1}(\mathbf{x}_t - \bm{\mu}), $$
where $\bm{\mu}$ contains the empirical means and $\mathbf{\Sigma}_{\text{std}}$ is a diagonal matrix of standard deviations. 

The orthogonal constraint on the nonlinear operator $\mathbf{Q}$ is formulated to conserve energy exclusively in the physical space. Accordingly, while the neural network generating $\mathbf{Q}(\tilde{\mathbf{x}}_t)$ takes standardized inputs, the resulting rotation acts on the unstandardized physical state to yield an intermediate rotated state $\mathbf{v}_t$:
$$ \mathbf{v}_t = \mathbf{Q}(\tilde{\mathbf{x}}_t)\big(\mathbf{\Sigma}_{\text{std}}\tilde{\mathbf{x}}_t + \bm{\mu}\big). $$
The standardized variables are used only to parametrize the state-dependent rotation: the neural network takes $\tilde{\mathbf{x}}_t$ as input and outputs $\mathbf{Q}(\tilde{\mathbf{x}}_t)$, but this orthogonal matrix acts on the corresponding unstandardized physical state. To apply the linear dynamics, this physically rotated state must be mapped back into the standardized coordinate space, effectively removing the physical offsets:
$$ \tilde{\mathbf{v}}_t = \mathbf{\Sigma}_{\text{std}}^{-1}(\mathbf{v}_t - \bm{\mu}). $$
Finally, the linear operator $\mathbf{M}$ and the effective deterministic forcing $\mathbf{F}$ (both parameterized in the standardized space) act on this standardized rotated state to yield the full discrete update:
$$ \tilde{\mathbf{x}}_{t+1} = \mathbf{F} + \mathbf{M}\tilde{\mathbf{v}}_t. $$
This careful separation guarantees that the nonlinear interactions remain strictly energy-conserving with respect to the original physical variables, while the linear dissipation (and the neural network optimization) operate in a well-conditioned standardized space. Finally, we note that by virtue of standardizing the state variables to zero mean, the constant forcing vector $\mathbf{F}$ is essentially zero at initialization and remains negligibly small during training.\\

In realistic applications, the state vector may contain several physical fields with different units. In this case, the concatenated Euclidean norm is not meaningful unless the fields are first put into comparable nondimensional units.
We therefore recommend an initial field-wise normalization before applying the standardization procedure described above. For example, suppose that the state contains two fields, $\mathbf{y}_t\in\mathbb{R}^{n_y}$ and $\mathbf{z}_t\in\mathbb{R}^{n_z}$, with different physical units. We first compute one scalar standard deviation for each field, denoted by $\sigma_y$ and $\sigma_z$, using all spatial degrees of freedom and all training snapshots. The fields are then nondimensionalized as $\widehat{\mathbf{y}}_t = \frac{\mathbf{y}_t}{\sigma_y}$ and $\widehat{\mathbf{z}}_t = \frac{\mathbf{z}_t}{\sigma_z}$.
The model state is then formed as
$$
\mathbf{x}_t =
\begin{pmatrix}
\widehat{\mathbf{y}}_t \\
\widehat{\mathbf{z}}_t
\end{pmatrix}.
$$
The component-wise centering and standardization described above are then applied to this nondimensionalized state vector. With this convention, the orthogonal operator $\mathbf{Q}$ preserves the Euclidean norm in the nondimensional physical state space.

\subsection{Causal constraints}

The physics-constrained networks detailed above can be augmented via a causal regularization term by adding a quadratic penalty to the MSE loss:
\begin{equation}
\begin{aligned}
\mathcal{L} &= \mathcal{L}_{\text{MSE}} + \mathcal{L}_{\text{Causal}}\\ 
&= \text{MSE}(\mathbf{x}_{t+1}, \mathbf{f}(\mathbf{x}_t)) + \lambda \sum_{(k,j) \in \mathcal{S}} \left(\frac{\partial f_k}{\partial x^{(j)}}\right)^2,
\end{aligned}
\label{eq:causal_loss_sm}
\end{equation}
where the gradients $\frac{\partial f_k}{\partial x^{(j)}}$ are efficiently computed during training via automatic differentiation \cite{PyTorch}.\\

In the ideal case of recovering the exact causal graph, the purely quadratic penalty in Eq.~\eqref{eq:causal_loss_sm} suffices to enforce the correct sparsity pattern in the discrete map's Jacobian. However, in high-dimensional systems with finite data, causal estimates are inherently uncertain, and False Negatives (i.e. true causal couplings $x^{(j)} \rightarrow x^{(k)}$ mistakenly classified as non-causal) may occur. Under a strict quadratic penalty, such errors would severely bias the learned dynamics by artificially suppressing the required gradients. 

To mitigate this, we propose a robust, capped penalty in the main text:
\begin{equation}
\begin{aligned}
 \mathcal{L}_{\text{Causal}}^{\text{Robust}} = \lambda \sum_{(k,j) \in \mathcal{S}} 
\min\!\left(\left(\frac{\partial f_k}{\partial x^{(j)}}\right)^2, \gamma\right).
\end{aligned}
\label{eq:causal_loss_robust_SM}
\end{equation}
This capped formulation strictly bounds the penalty. Once the squared gradient reaches the threshold $\gamma$, the derivative of the regularizer with respect to the network weights vanishes. Consequently, if the MSE loss strongly dictates that a specific coupling is necessary to predict $\mathbf{x}_{t+1}$ (i.e., a mistakenly excluded interaction), the MSE term will easily overcome the bounded penalty. False Positives, by contrast, do not induce structural bias, as they are simply excluded from the set $\mathcal{S}$ and thus unpenalized.

\subsubsection{Heuristic for the ``cap'' parameter $\gamma$} \label{sec:heuristic_gamma}

We propose a heuristic to estimate the cap parameter $\gamma$ directly from the discrete impulse response operator $R^{k,j}_1$ obtained via the FDT. The impulse response $R^{k,j}_t$ measures the change in variable $x^{(k)}$ at time step $t$ following a small perturbation applied to $x^{(j)}$ at step $0$. As described in the main text, direct causal links $x^{(j)} \rightarrow x^{(k)}$ are identified from the unit-step responses $R^{k,j}_1$ by: (i) applying a logarithmic transformation $\ln |R^{k,j}_1|$ for $k \neq j$, (ii) partitioning these values using $k$-means clustering ($k=2$), and (iii) assigning a causal link to the cluster associated with the larger centroid. Let $\tilde{R}^{k,j}_1$ denote the subset of responses belonging to this cluster, corresponding to statistically significant (non-spurious) interactions. We define the cap parameter $\gamma$ as a low quantile of these squared significant responses; in our experiments, we use the $0.1$ quantile. This choice establishes a strict lower bound on the magnitude of reliably detectable causal interactions. If during training the MSE term drives a penalized gradient beyond this threshold, i.e. $\left(\partial f_k / \partial x^{(j)}\right)^2 \ge \gamma$), this indicates that the interaction is likely inconsistent with the assumed absence of a causal link (i.e. we are dealing with a False Negative).  In this regime, the capped loss effectively deactivates the penalty, allowing the dynamics to be governed by the MSE term rather than by a potentially incorrect causal constraint.\\

\subsubsection{Implementation details for the causal constraint}

Enforcing the causal penalty defined in Eq.~\eqref{eq:causal_loss_robust_SM} requires computing the Jacobian of the deterministic forward map in Eq. \eqref{eq:discrete_case_SM}. However, backpropagation through $\mathbf{Q}(\mathbf{x}_t) = \exp(\mathbf{S}(\mathbf{x}_t))$ is computationally expensive.\\

Therefore, to enforce the causal constraints efficiently, we construct a first-order proxy for the forward map through the skew-symmetric infinitesimal generator $\mathbf{S}(\mathbf{x}_t)$. By expanding the orthogonal map to first order, $\mathbf{Q}(\mathbf{x}_t) \approx \mathbf{I} + \mathbf{S}(\mathbf{x}_t)$, the deterministic update $\mathbf{M}\mathbf{Q}(\mathbf{x}_t)\mathbf{x}_t$ can be evaluated for the purpose of the Jacobian penalty as:
\begin{equation}
\tilde{\mathbf{f}}(\mathbf{x}_t) = \mathbf{M}\mathbf{x}_t + \mathbf{M}\mathbf{S}(\mathbf{x}_t)\mathbf{x}_t.
\label{eq:fast_path_generator}
\end{equation}
Constant forcing terms are ignored in this proxy, as their derivative with respect to the state is zero. Crucially, this proxy is evaluated strictly to compute the gradients $\partial \tilde{f}_k / \partial x^{(j)}$ for the regularization term $\mathcal{L}_{\text{Causal}}^{\text{Robust}}$. The primary state prediction $\mathbf{x}_{t+1}$, and its associated MSE loss, are always computed using the exact, energy-conserving orthogonal map $\mathbf{M}\mathbf{Q}(\mathbf{x}_t)\mathbf{x}_t$. This dual-path approach allows us to bypass the differentiation of the matrix exponential entirely.

\section{Mathematical origin of the discrete model and a few examples}

The data-driven model proposed in this work is formulated and solved as a discrete-time map and it is fundamentally designed to simulate coarse-grained, \textit{effective} dynamics directly from data. To establish the theoretical grounding of this discrete formulation, we first demonstrate its mathematical connection to continuous-time systems. We further reformulate a stochastic triad model via the proposed formulation. Finally, we test the framework by learning the effective dynamics of severely sub-sampled trajectories of the Charney-DeVore dynamical system considered in the main text, targeting a regime where data-driven continuous models can fail due to numerical instability. In summary, this Section targets the following objectives:

\begin{itemize}
    \item \textit{Formulation in the continuous limit.} To build mathematical intuition, we first examine the limit of small time steps ($\Delta t \to 0$). We show that the general discrete architecture proposed in the main text arises naturally from the continuous equations via a splitting procedure \cite{LieTrotter}. We then further highlight this connection by reformulating a known triad model with energy-conserving nonlinearities into the proposed framework.
    \item \textit{Effective coarse-grained dynamics (Charney--DeVore).} Finally, we demonstrate the primary use case of our method: stable learning from severely coarse-grained data. Using the Charney--DeVore model presented in the main text, we train the framework on trajectories subsampled every 100 and 500 integration steps. In this limit, both standard numerical integrators and physics-constrained, continuous data-driven models inevitably go unstable. By contrast, our learned discrete map acts as a stable emulator of the effective finite-time dynamics.
\end{itemize}

\subsection{Connection to the continuous case: general stochastic processes} \label{sec:Lie-Trotter-stuff}

We consider the general abstract formulation arising when many turbulent flows are projected onto orthogonal bases, as discussed in the main text and at the start of this SM. We write this formulation as:
\begin{equation}
d\mathbf{x} = [\mathbf{F} + \mathbf{A}\mathbf{x} + \mathbf{S}(\mathbf{x})\mathbf{x}]dt + \mathbf{\Sigma}d\mathbf{W},
\label{eq:functional_form_general_nonlinear-wiener}
\end{equation}
where $\mathbf{W}$ is a standard Wiener process and $\mathbf{S}(\mathbf{x})$ is strictly skew-symmetric, leading to energy conservation of the nonlinear term. For details on the various terms, see Section 2.1 in the main text. To map this continuous system to our proposed discrete-time architecture over a time step $\Delta t$, we apply a first-order operator splitting \cite{LieTrotter}. This splitting procedure consists of two steps: we first decompose the dynamics into a purely nonlinear, energy-conserving step, followed by a linear, forced, and stochastic step.\\

\paragraph*{Step (a): Nonlinear rotation.} We first isolate the nonlinear advection term:
\begin{equation}
    d\mathbf{x}^{(1)} = \mathbf{S}(\mathbf{x}^{(1)})\mathbf{x}^{(1)} dt.
\end{equation}
We approximate the solution over the interval $[t, t+\Delta t]$ by freezing the state dependence of $\mathbf{S}$ at the beginning of the interval. Using the initial condition $\mathbf{x}^{(1)}(t) = \mathbf{x}_t$, this yields the approximate nonlinear update:
\begin{equation}
    \mathbf{v}_t = \exp\big(\Delta t \mathbf{S}(\mathbf{x}_t)\big) \mathbf{x}_t \equiv \mathbf{Q}(\mathbf{x}_t)\mathbf{x}_t.
\end{equation}
Because $\mathbf{S}(\mathbf{x}_t)$ is skew-symmetric, its matrix exponential $\mathbf{Q}(\mathbf{x}_t)$ is strictly orthogonal ($\mathbf{Q}^\mathrm{T}\mathbf{Q} = \mathbf{I}$). Thus, this step represents a pure rotation in state space that strictly preserves the $L_2$-norm (energy). Analytically, this step incurs a $\mathcal{O}(\Delta t^2)$ truncation error.\\

\paragraph*{Step (b): Linear, forced, and stochastic flow.} Next, we use the rotated intermediate state $\mathbf{v}_t$ as the initial condition for the remaining terms over the same time interval, leading to the linear SDE:
\begin{equation}
    d\mathbf{x}^{(2)} = \big[ \mathbf{A}\mathbf{x}^{(2)} + \mathbf{F} \big] dt + \mathbf{\Sigma} d\mathbf{W}_t, \quad \text{with } \mathbf{x}^{(2)}(t) = \mathbf{v}_t.
\label{eq:linear_sde_splitting}
\end{equation}
Its exact solution over one time step is:
\begin{equation}
    \mathbf{x}_{t+1} = e^{\mathbf{A}\Delta t} \mathbf{v}_t + \int_{0}^{\Delta t} e^{\mathbf{A}(\Delta t - s)} \mathbf{F} \, ds + \int_{0}^{\Delta t} e^{\mathbf{A}(\Delta t - s)} \mathbf{\Sigma} \, d\mathbf{W}_s.
\end{equation}
We identify the following terms:
\begin{itemize}
    \item \textit{Linear Operator:} $\mathbf{M} = e^{\mathbf{A}\Delta t}$. The term $e^{\mathbf{A}\Delta t} \mathbf{v}_t$ in Eq. \eqref{eq:functional_form_discrete} then represents the sequential update $\mathbf{M}\mathbf{Q}(\mathbf{x}_t)\mathbf{x}_t$ considered in the previous Sections.
    \item \textit{Effective Deterministic Forcing:} $\mathbf{F}_{\text{discrete}} = \int_{0}^{\Delta t} e^{\mathbf{A}(\Delta t - s)} \mathbf{F} \, ds$. Assuming $\mathbf{A}$ is invertible and that $\mathbf{F}$ is a constant forcing, the solution of this integral is:
    \begin{equation}
    \mathbf{F}_{\text{discrete}} = \mathbf{A}^{-1} \big(e^{\mathbf{A}\Delta t} - \mathbf{I}\big) \mathbf{F} = \mathbf{A}^{-1} (\mathbf{M} - \mathbf{I}) \mathbf{F}.
    \end{equation}
    \item \textit{Effective Stochastic Forcing:} $\mathbf{\Sigma}_{\text{discrete}}\bm{\xi}_t = \int_{0}^{\Delta t} e^{\mathbf{A}(\Delta t - s)} \mathbf{\Sigma} \, d\mathbf{W}_s$. The covariance matrix of this process can be written as:
    \[
    \mathbf{C} = \mathbf{\Sigma}_{\text{discrete}}\mathbf{\Sigma}_{\text{discrete}}^\mathrm{T} = \int_{0}^{\Delta t} e^{\mathbf{A}(\Delta t - s)} \mathbf{\Sigma} \mathbf{\Sigma}^\mathrm{T} e^{\mathbf{A}^\mathrm{T}(\Delta t - s)} \, ds,
    \]
    where $\mathbf{\Sigma}_{\text{discrete}}$ is the lower-triangular Cholesky factor of $\mathbf{C}$, and $\bm{\xi}_t \sim \mathcal{N}(\mathbf{0}, \mathbf{I})$. Importantly, $\mathbf{C}$ satisfies the Lyapunov matrix equation \cite{TIAN200844}:
    \begin{equation}
    \mathbf{A}\mathbf{C} + \mathbf{C}\mathbf{A}^\mathrm{T} = \mathbf{M}\mathbf{\Sigma}\mathbf{\Sigma}^\mathrm{T}\mathbf{M}^\mathrm{T} - \mathbf{\Sigma}\mathbf{\Sigma}^\mathrm{T}, \qquad \mathbf{M}=e^{\mathbf{A}\Delta t}.
    \label{eq:lyapunov_eq}
    \end{equation}
    Thus, $\mathbf{C}$ can be identified by numerically solving the Lyapunov equation above. The term $\mathbf{\Sigma}_{\text{discrete}}$ is then extracted via Cholesky decomposition of $\mathbf{C}$.
    \end{itemize}

Combining these steps yields a split numerical approximation of the forward dynamics:
\begin{equation}
\mathbf{x}_{t+1} = \mathbf{F}_{\text{discrete}} + \mathbf{M}\mathbf{Q}(\mathbf{x}_t)\mathbf{x}_t + \mathbf{\Sigma}_{\text{discrete}}\bm{\xi}_t.
\label{eq:approximation-splitting}
\end{equation}
This derivation is formally first-order accurate, with a local $\mathcal{O}(\Delta t^2)$ splitting error and it precisely mirrors the discrete model employed in the main text. In contrast, the data-driven formulation does not impose any explicit requirement on the temporal resolution and is tailored to fit the \textit{effective} dynamics of the system at coarse-grained time scales, which can differ substantially from the underlying continuous dynamics. At the same time, it enforces the resolution-independent orthogonality constraint of the nonlinear update $\mathbf{Q}(\mathbf{x}_t)\mathbf{x}_t$, guaranteeing that the nonlinear term cannot induce spurious energy growth or decay.

\subsubsection{Stochastic regime. Pedagogical example with triad models}

We now consider the stochastic triad model proposed in \cite{majdaSIAM} and formulated as: 
\begin{equation}
\begin{aligned}
dx_1 &= L_2 x_3 - L_3 x_2 - d_1 x_1 + B_1 x_2 x_3 + F_1 +\sigma_1 dW_1, \\
dx_2 &= L_3 x_1 - L_1 x_3 - d_2 x_2 + B_2 x_1 x_3 + F_2 +\sigma_2 dW_2\\
dx_3 &= L_1 x_2 - L_2 x_1 - d_3 x_3 + B_3 x_1 x_2 + F_3 +\sigma_3 dW_3.
\end{aligned}
\label{eq:triad}
\end{equation}
This model serves as a fundamental building block for complex turbulent dynamical systems, as three-dimensional Galerkin truncations of various fluid equations naturally yield this structural form (see Chapter 2 of Ref.~\cite{MajdaIntroTurbulence}). The linear operator $\mathbf{A}$ can be decomposed as $\mathbf{A} = \mathbf{L} + \mathbf{D}$, where $\mathbf{L}$ is a skew-symmetric matrix representing dispersion, and $\mathbf{D}$ is a symmetric negative-definite matrix representing dissipation. In the specific case in Eq. \eqref{eq:triad}, $\mathbf{D}$ is diagonal and the components $d_i$ are strictly positive. $\mathbf{W}$ is the standard Wiener process. The quadratic nonlinearities are energy conserving, satisfying the condition $\mathbf{x} \cdot \mathbf{B}(\mathbf{x}, \mathbf{x}) = 0$. This property explicitly imposes the constraint $B_1 + B_2 + B_3 = 0$ on the nonlinear coefficients.\\

The model can be naturally cast into the general stochastic differential equation form introduced in Eq. \eqref{eq:functional_form_general_nonlinear-wiener}:

\begin{equation}
\begin{aligned}
\begin{bmatrix} dx_1 \\ dx_2 \\ dx_3 \end{bmatrix} &= \Big(\underbrace{\begin{bmatrix} F_1 \\ F_2 \\ F_3 \end{bmatrix}}_{\mathbf{F}} + \underbrace{\begin{bmatrix} -d_1 & -L_3 & L_2 \\ L_3 & -d_2 & -L_1 \\ -L_2 & L_1 & -d_3 \end{bmatrix}}_{\mathbf{A}} \begin{bmatrix} x_1 \\ x_2 \\ x_3 \end{bmatrix} + \underbrace{\begin{bmatrix} 0 & B_1 x_3 & 0 \\ -B_1x_3 & 0 & -B_3x_1 \\ 0 & B_3x_1 & 0 \end{bmatrix}}_{\mathbf{S}(\mathbf{x})} \begin{bmatrix} x_1 \\ x_2 \\ x_3 \end{bmatrix} \Big)dt  + \underbrace{\begin{bmatrix} \sigma_1 & 0 & 0 \\ 0 & \sigma_2 & 0 \\ 0 & 0 & \sigma_3 \end{bmatrix}}_{\mathbf{\Sigma}} \begin{bmatrix} dW_1 \\ dW_2 \\ dW_3 \end{bmatrix},
\end{aligned}
\label{eq:triad_functional_form_general_nonlinear}
\end{equation}
with $B_2 = -(B_1 + B_3)$. $\mathbf{S}(\mathbf{x})$ is skew-symmetric, and therefore energy conserving.\\

\paragraph*{Performance of the approximation in a regime with dual energy cascade.} We now test the relevance of the splitting procedure in Eq. \eqref{eq:approximation-splitting} in the stochastic context. We follow Majda and Qi \cite{majdaSIAM} and consider a set of parameters leading to a dual energy cascade across modes $x_1$, $x_2$, and $x_3$. Specifically, we set $d_1 = 1$, $d_2 = d_3 = 2$; $\sigma_1^2 = 10$, $\sigma_2^2 = \sigma_3^2 = 0.01$. The nonlinear coefficients are $B_1 = 2$, $B_2 = B_3 = -1$. The linear interactions are set to $L_1 = 0.09$, $L_2 = 0.06$, and $L_3 = -0.03$. The deterministic forcing is applied only to the $x_2$ and $x_3$ modes, such that $F_1 = 0$, $F_2 = -1$, and $F_3 = 1$. Given the chosen parameters, the $x_1$ mode is subject to strong stochastic forcing, while $x_2$ and $x_3$ are less energetic. The quadratic nonlinear coupling $\mathbf{B}(\mathbf{x},\mathbf{x})$ redistributes this energy, creating a cascade from the highly energetic $x_1$ mode to the less energetic $x_2$ and $x_3$ modes. Conversely, the deterministic forcing applied to $x_2$ and $x_3$ drives a backward energy cascade towards $x_1$. Therefore, these parameters place the system in a complex, dual energy cascade regime.\\

\begin{figure*}[h!]
\centering
\includegraphics[width=1\textwidth]{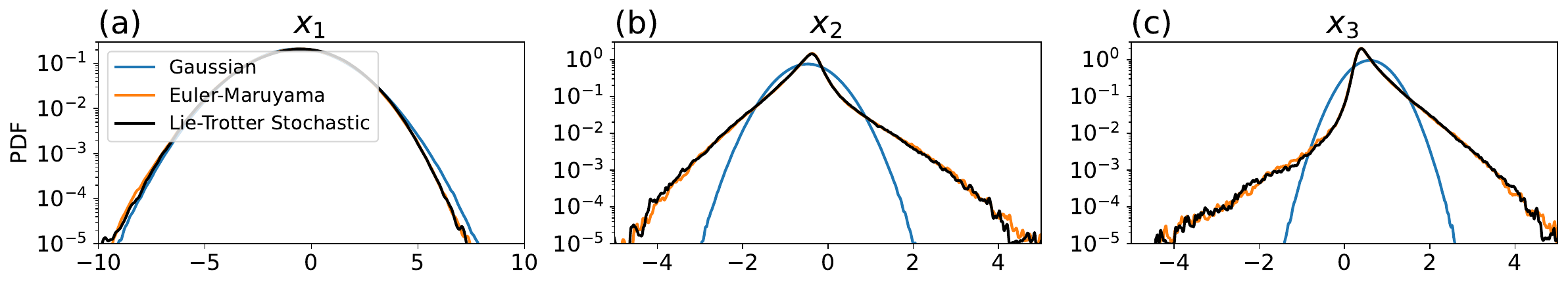}
\caption{Stationary distributions (PDFs) of $x_1$, $x_2$, and $x_3$ in the triad model, simulated by a standard Euler-Maruyama scheme and our Lie-Trotter splitting procedure $\mathbf{x}_{t+1} = \mathbf{F}_{\text{discrete}} + \mathbf{M}\mathbf{Q}(\mathbf{x}_t)\mathbf{x}_t + \mathbf{\Sigma}_{\text{discrete}}\bm{\xi}_t$. The deterministic and stochastic forcings, $\mathbf{F}_{\text{discrete}}$ and $\mathbf{\Sigma}_{\text{discrete}}\bm{\xi}_t$, are derived analytically via the steps proposed in Section \ref{sec:Lie-Trotter-stuff}. PDFs are plotted on a logarithmic scale. A Gaussian distribution with the same mean and variance as the Euler-Maruyama results is provided for reference.}
\label{fig:Triad_dual_cascade}
\end{figure*}

We simulate the dynamics using the proposed splitting procedure for stochastic systems as in Eq. \eqref{eq:approximation-splitting} and compare it to a standard Euler-Maruyama scheme. The simulation length is $5 \times 10^7$ time steps with a step size of $\Delta t = 0.001$. Importantly, the effective deterministic forcing $\mathbf{F}_{\text{discrete}}$ and the effective stochastic forcing $\mathbf{\Sigma}_{\text{discrete}}\bm{\xi}_t$ in Eq. \eqref{eq:approximation-splitting} are derived following the steps proposed in Section \ref{sec:Lie-Trotter-stuff}: $\mathbf{F}_{\text{discrete}} = \mathbf{A}^{-1} (\mathbf{M} - \mathbf{I}) \mathbf{F}$, and the covariance for $\mathbf{\Sigma}_{\text{discrete}}$ is obtained by solving the associated continuous Lyapunov equation. In Figure \ref{fig:Triad_dual_cascade}, we show the stationary distributions of the $x_1$, $x_2$, and $x_3$ modes, plotted against a Gaussian distribution with the same mean and variance for reference. The two simulations produce identical stationary statistics and are both in good agreement with the results of Majda and Qi (see Figure 4.1C in Ref.~\cite{majdaSIAM}).\\

This result further motivates Eq. \eqref{eq:approximation-splitting} as a robust scheme to reproduce trustworthy statistics in the continuous limit $\Delta t \rightarrow 0$ beyond deterministic systems, successfully extending to forced stochastic systems. In the next section, we return to the data-driven setting and demonstrate that learning neural models constrained by the formulation in Eq. \eqref{eq:approximation-splitting} yields skillful emulators even from severely sub-sampled data, a regime where both standard numerical integrators and (physics-constrained) continuous-time data-driven models fail.

\subsection{Data-driven emulation of the Charney-DeVore model from subsampled data}

\begin{figure*}[h!]
\centering
\includegraphics[width=1\textwidth]{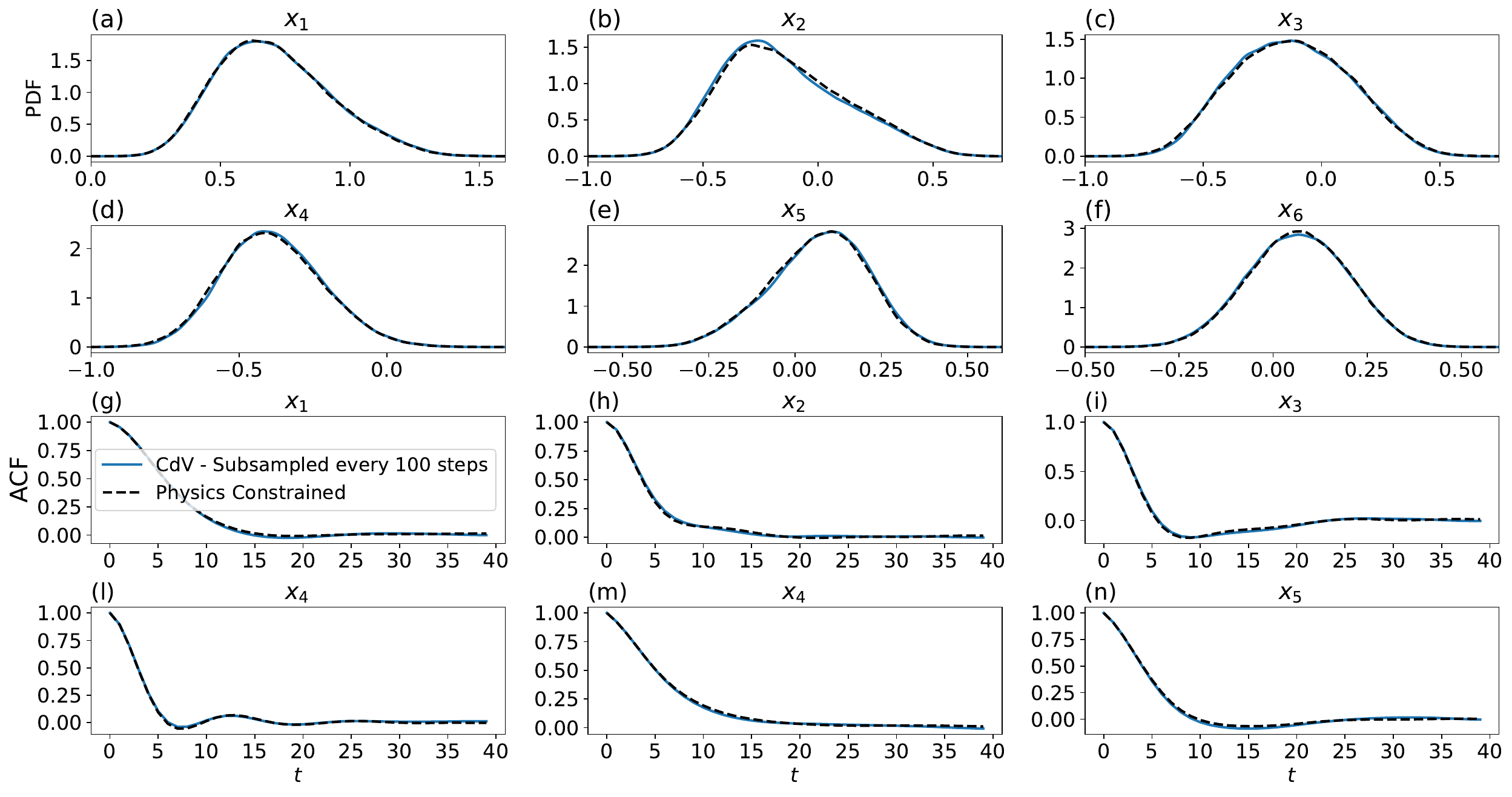}
\caption{Experiment with subsampling size 100. In the first and second rows, we show the stationary distributions (PDFs) of the variables $\tilde{x}^{(1)}_t, \tilde{x}^{(2)}_t, \dots, \tilde{x}^{(n)}_t$ of the physics-constrained emulator and the subsampled time series $\tilde{\mathbf{x}}_t$. The third and fourth rows show the same comparison for the autocorrelation functions (ACFs).}
\label{fig:CdV-Discrete-100}
\end{figure*}

\begin{figure*}[h!]
\centering
\includegraphics[width=1\textwidth]{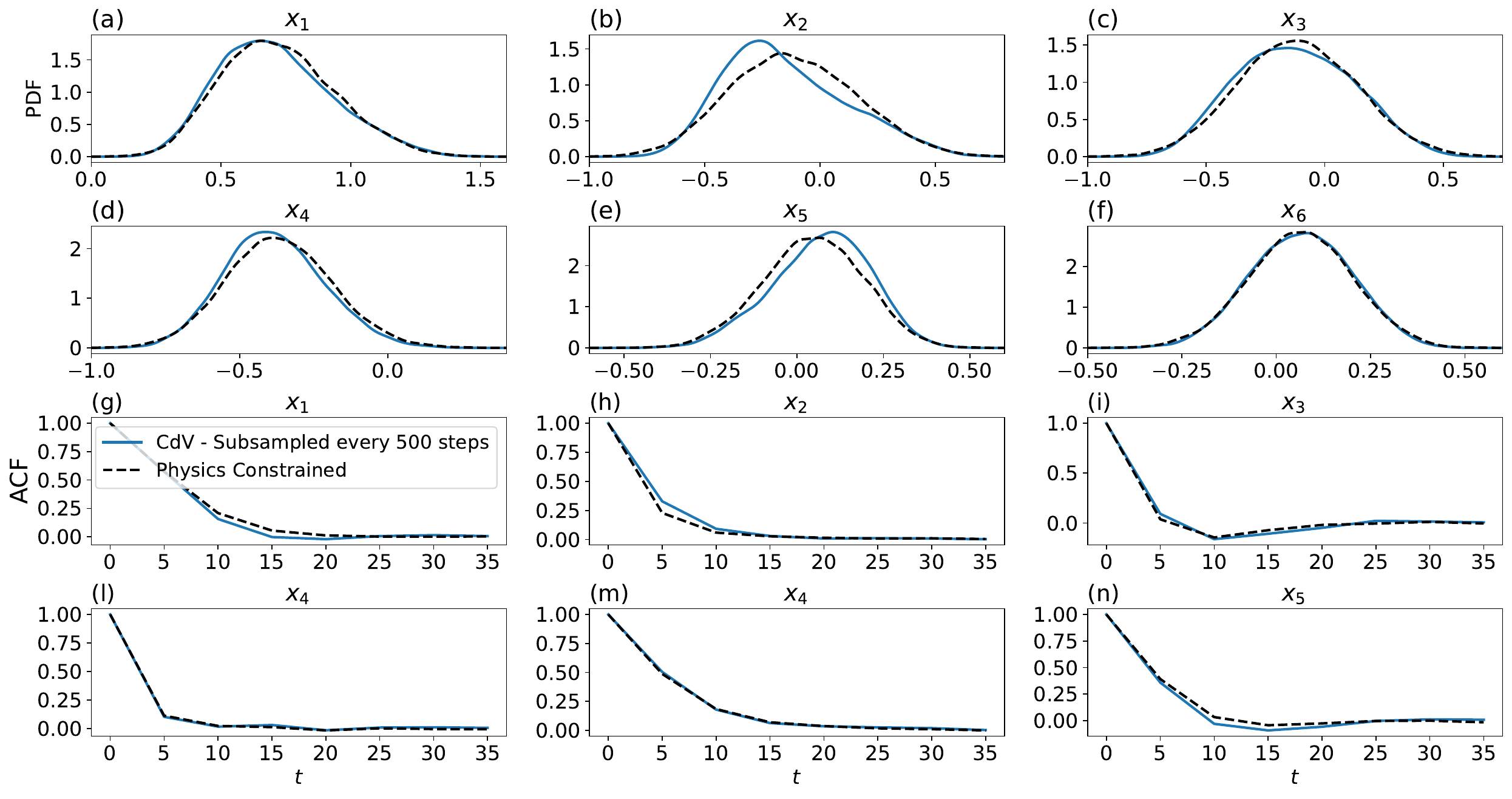}
\caption{Same as Fig. \ref{fig:CdV-Discrete-100} but after subsampling the original Charney-DeVore integration every 500 time steps (corresponding to $dt = 5$).}
\label{fig:CdV-Discrete-500}
\end{figure*}

We return to the Charney-DeVore model \cite{charneyDeVore} introduced in Eq. (8) of the main text. In the main manuscript, the emulator was learned from a well-sampled trajectory to allow for a direct statistical comparison with the continuous model. However, our proposed neural model can operate in sparse observational regimes where standard numerical integrators fail. To demonstrate this, we evaluate our architecture in two coarse-grained regimes:

\begin{itemize}
    \item \textit{Subsampling every 100 time steps.} We consider a long integration ($T = 10^7$ time steps) $\mathbf{x}_t \in \mathbb{R}^{n,T}$ of the Charney-DeVore model as in the main text. This integration was generated with an Euler-Maruyama scheme and $\Delta t = 0.01$. We then subsample the data by considering every 100th time step. This corresponds to observing the data every $\Delta t = 1$. We refer to this new subsampled trajectory as $\tilde{\mathbf{x}}_t$. We train the proposed discrete neural emulator in Eq. \eqref{eq:discrete_case_SM}, with energy-conserving nonlinearities, from $\tilde{\mathbf{x}}_t$. We then assess the stationary distributions (PDFs) and autocorrelation functions (ACFs) of variables $\tilde{x}^{(1)}_t, \tilde{x}^{(2)}_t, \dots, \tilde{x}^{(6)}_t$ and show the results in Figure \ref{fig:CdV-Discrete-100}. Our discrete emulator remains strictly stable and successfully learns the effective discrete mapping. On the other hand, the numerical integrator blows up with $\Delta t = 1$. Furthermore, data-driven models with physics constrained defined in the continuous limit also become unstable. 
    
    \item \textit{Subsampling every 500 time steps.} We further challenge the framework by subsampling the original time series every 500 steps. This reduces the amount of available data for training to 20,000 time points. Despite this, the PDFs and ACFs of the learned discrete mapping remain generally well-approximated (Figure \ref{fig:CdV-Discrete-500}). The largest error is found for the PDFs of the $x_2$ degree of freedom. Most importantly, the neural emulator guarantees stability, providing a robust finite-time flow map in a regime where traditional integration schemes fail.
\end{itemize}

\section{Fluctuation-Dissipation Theorem (FDT) from data: quasi-Gaussian and score matching} \label{sec:FDT}

\paragraph*{The FDT.} Consider an $n$-dimensional system $\mathbf{x}_t = (x^{(1)}_t, x^{(2)}_t, ..., x^{(n)}_t)$. The Fluctuation-Dissipation Theorem (FDT) states that the time-dependent response of an observable $\mathcal{A}(x^{(k)}_t)$ to a small impulse perturbation $\delta x^{(j)}_0$ imposed on $x^{(j)}_{t=0}$ at time $t=0$, can be retrieved solely from stationary statistics of the system as:
\begin{equation}
R^{k,j}_t = \lim_{\delta x^{(j)}_0\to0} \frac{\delta \langle \mathcal{A}(x^{(k)}_t) \rangle}{\delta x^{(j)}_0} = - \Big\langle \mathcal{A}(x^{(k)}_t) \frac{\partial \ln \rho(\mathbf{x})}{\partial x^{(j)}} \Big|_{\mathbf{x}_0} \Big\rangle .
\label{eq:response_general-SM} 
\end{equation}
where $\rho(\mathbf{x})$ represents the invariant probability distribution of the system \cite{FALCIONI,Marconi}. The FDT establishes a link between perturbed quantities and stationary statistics, i.e. the invariant measure of the system \cite{MajdaBook,Marconi}. The brackets $\langle \cdot \rangle$ represent ensemble averages, in practice computed using temporal averages under the assumption of ergodicity. In this case, Eq. \ref{eq:response_general-SM} represents the impulse response in the ensemble mean.\\ 

\subsection{The FDT from data}

\paragraph*{Practical considerations and correction procedure.} The score-matching procedure introduced by Hyvärinen (2005) \cite{Aapo} (see Section 3.1 of the main text) provides an elegant way to estimate the score $s(\mathbf{x})$ directly on the system’s inertial manifold. In practice, however, this estimation can be prone to several sources of error. First, ensemble averages must be replaced by empirical averages computed from finite datasets. Second, the inferred score inevitably depends on the chosen neural-network architecture and its associated hyperparameters. Nevertheless, the theoretical formalism of the FDT provides analytical constraints on the instantaneous response $\mathbf{R}_{t=0}$. These constraints can be used for an \textit{a posteriori} calibration of the estimated score evaluations, removing violations of exact instantaneous-response identities and improving the numerical consistency of the resulting FDT estimates. Because these analytical constraints determine only selected empirical moments of the score, the resulting corrections are non-unique. Here, we construct separate response-specific corrections of the same score estimate for the mean and variance calculations. We denote the ideal score by $s(\mathbf{x})$, the score inferred through score matching by $\tilde{s}(\mathbf{x})$, and the corrected score by $s_c(\mathbf{x})$. The system’s dynamics is encoded in a data matrix $\mathbf{x} \in \mathbb{R}^{T, n}$, where $T$ is the number of temporal samples and $n$ the system’s dimensionality. The correction procedure is dependent on the type of response analyzed. 

\begin{itemize}
    \item \textit{Correction for the ensemble mean response.} If the interest is in the response in ensemble mean, i.e., for the observable $\mathcal{A}(x^{(k)}_t) = x^{(k)}_t$ in Eq.~\eqref{eq:response_general-SM}, then it is possible to use the correction already outlined in \cite{ResponseScore}. The response operator in the ensemble mean at time $t = 0$ of any system is:
    \begin{equation}
    \mathbf{R}_0 = \mathbf{I}
    \label{eq:R-mean-t0}
    \end{equation}
    $\mathbf{I}$ being the identity matrix. The empirical estimation of the instantaneous response operator with the inferred score $\tilde{s}(\mathbf{x})$ will lead to small errors, so that
    \begin{equation}
    \tilde{\mathbf{R}}_0 = -\frac{1}{T}\mathbf{x}^\text{T} \tilde{s}(\mathbf{x}) = \mathbf{I} - \boldsymbol{\epsilon}
    \label{eq:R-mean-t0-epsilon}
    \end{equation}
    where $\tilde{\mathbf{R}}_0$ denotes the estimated response. It is then possible to correct the inferred score as:
    \begin{equation}
    s_c(\mathbf{x}) = \tilde{s}(\mathbf{x})\cdot(\mathbf{I} - \boldsymbol{\epsilon})^{-1}.
    \label{eq:score-mean-correction}
    \end{equation}
    The corrected score $s_c(\mathbf{x})$ defined in Eq. \eqref{eq:score-mean-correction} guarantees the identity in Eq. \eqref{eq:R-mean-t0} and it is the one used in the computation of responses in ensemble mean in the main text. Note that while this correction is non-unique, different versions lead to small differences as argued in the SM of \cite{GiorginiScore}.
    \item \textit{Correction for the ensemble variance response.} If the interest is in the response of the ensemble variance, the observable is defined as $\mathcal{A}(x^{(k)}_t) = (x^{(k)}_t - \mu^{(k)})^2$, where $\boldsymbol{\mu}$ is the mean of the data. In this case the analytical constraint is that the response to an impulse perturbation at time $t=0$ must be exactly zero:
    \begin{equation}
    \mathbf{R}_0 = \mathbf{0},
    \label{eq:R-var-t0}
    \end{equation}
    where $\mathbf{0}$ is the null matrix. In contrast to the mean responses, this constraint is not automatically satisfied by the quasi-Gaussian approximation. Therefore, in principle a Gaussian score could be corrected using the procedure detailed below: this is not pursued in the current study. The empirical estimation of the instantaneous response operator with the inferred score $\tilde{s}(\mathbf{x})$ will lead to small errors. Therefore, given the centered and squared data matrix $\mathbf{Y} \in \mathbb{R}^{T, n}$ with elements $Y_{i,k} = (x_i^{(k)} - \mu^{(k)})^2$, we have
    \begin{equation}
    \tilde{\mathbf{R}}_0 = -\frac{1}{T}\mathbf{Y}^\text{T} \tilde{s}(\mathbf{x}) \neq \mathbf{0}.
    \label{eq:R-var-t0-empirical}
    \end{equation}
    To enforce the constraint in Eq. \eqref{eq:R-var-t0}, we compute a least-squares correction term $\Delta s(\mathbf{x})$ that minimizes the norm of the adjustment. The correction term is:  
    \begin{equation}
    \Delta s(\mathbf{x}) = \mathbf{Y} (\mathbf{Y}^\text{T} \mathbf{Y})^{-1} \mathbf{Y}^\text{T} \tilde{s}(\mathbf{x}).
    \label{eq:var-delta}
    \end{equation}
    The corrected score for the variance response is then simply obtained by subtracting this projection from the originally inferred score:
    \begin{equation}
    s_c(\mathbf{x}) = \tilde{s}(\mathbf{x}) - \Delta s(\mathbf{x}).
    \label{eq:score-var-correction}
    \end{equation}
    By construction, multiplying $\mathbf{Y}^\text{T}$ by $s_c(\mathbf{x})$ exactly cancels out, guaranteeing that the initial response of the variance $\mathbf{R}_0$ is strictly zero. As with the mean, this correction provides a practical regularization of the data-driven score without requiring modifications to the underlying neural network architecture.
\end{itemize}

Crucially, depending on the data, it may be useful to compute the response $R^{k,j}_t$ on standardized data. This is especially useful when variables span disparate scales, as in principal component representations.\\

\begin{figure*}[h!]
\centering
\includegraphics[width=1\textwidth]{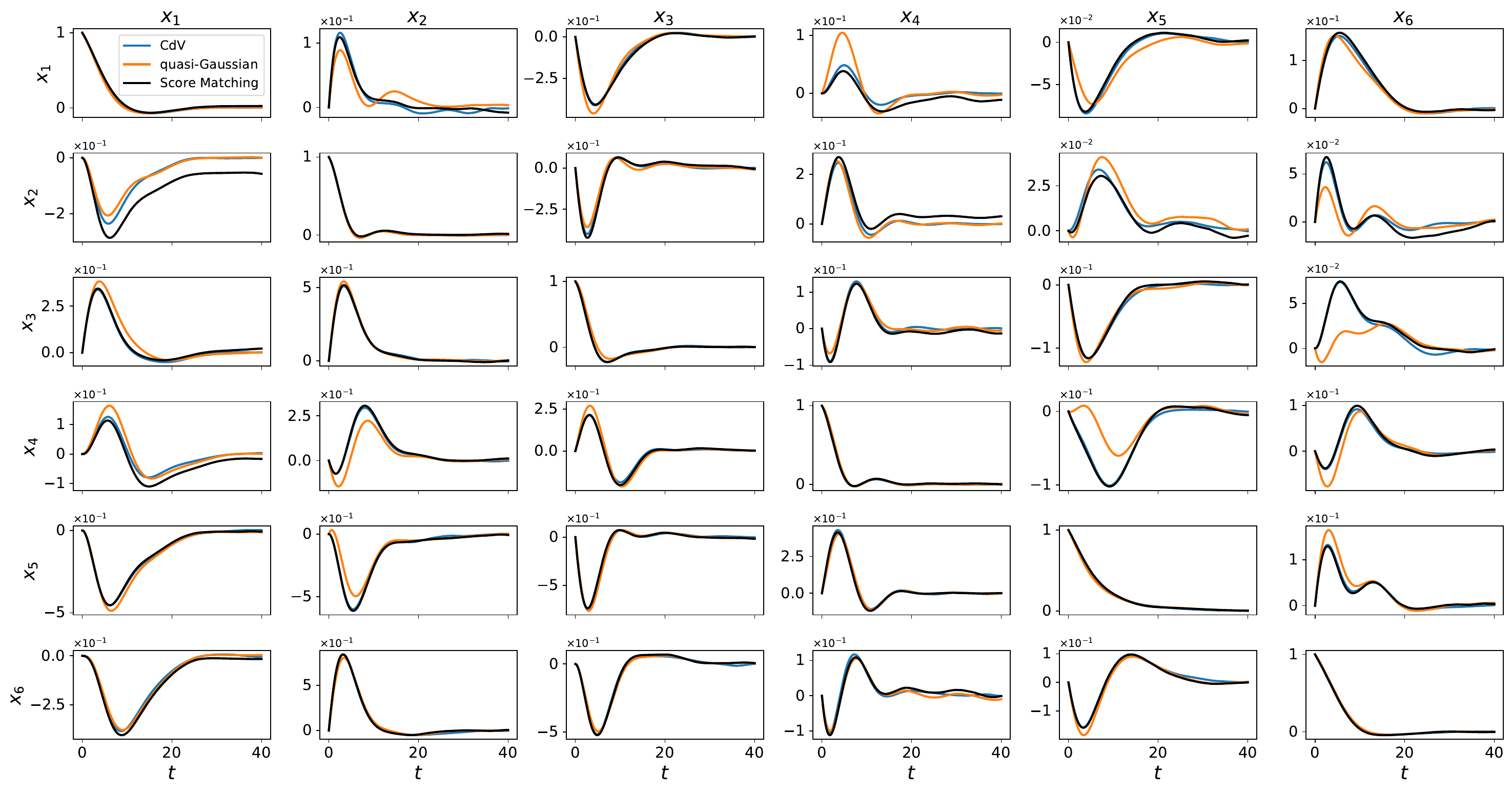}
\caption{Time-dependent responses in ensemble mean $x^{(k)}_t$ to an impulse perturbation in the CdV model (Eq. (8) in the main text) as predicted by the FDT using the quasi-Gaussian and score-matching approximations. Example: column (1), row (2) quantifies the time-dependent mean response of $x^{k = 2}_t$ given a small impulse perturbation imposed on $x^{j = 1}_0$ at time $t = 0$.}
\label{fig:FDT_mean}
\end{figure*}

\paragraph*{Response operator via score matching: response in the ensemble mean.} Given the proposed correction in Eq. \eqref{eq:score-mean-correction}, we now infer the mean response operator for the Charney-DeVore system in Eq. (8) in the main text. We fit the qG-FDT approximation and the score-matching based FDT estimation on the long trajectory of the Charney-DeVore system used in the main text. We compare the FDT predictions against the numerical ground truth and show the results in Figure \ref{fig:FDT_mean}. First, we note that the quasi-Gaussian approximation leads to good, first-order results even in a strongly nonlinear model such as the Charney–DeVore model. The FDT inferred through score matching is generally more accurate than the quasi-Gaussian approximation.

\begin{figure*}[h!]
\centering
\includegraphics[width=1\textwidth]{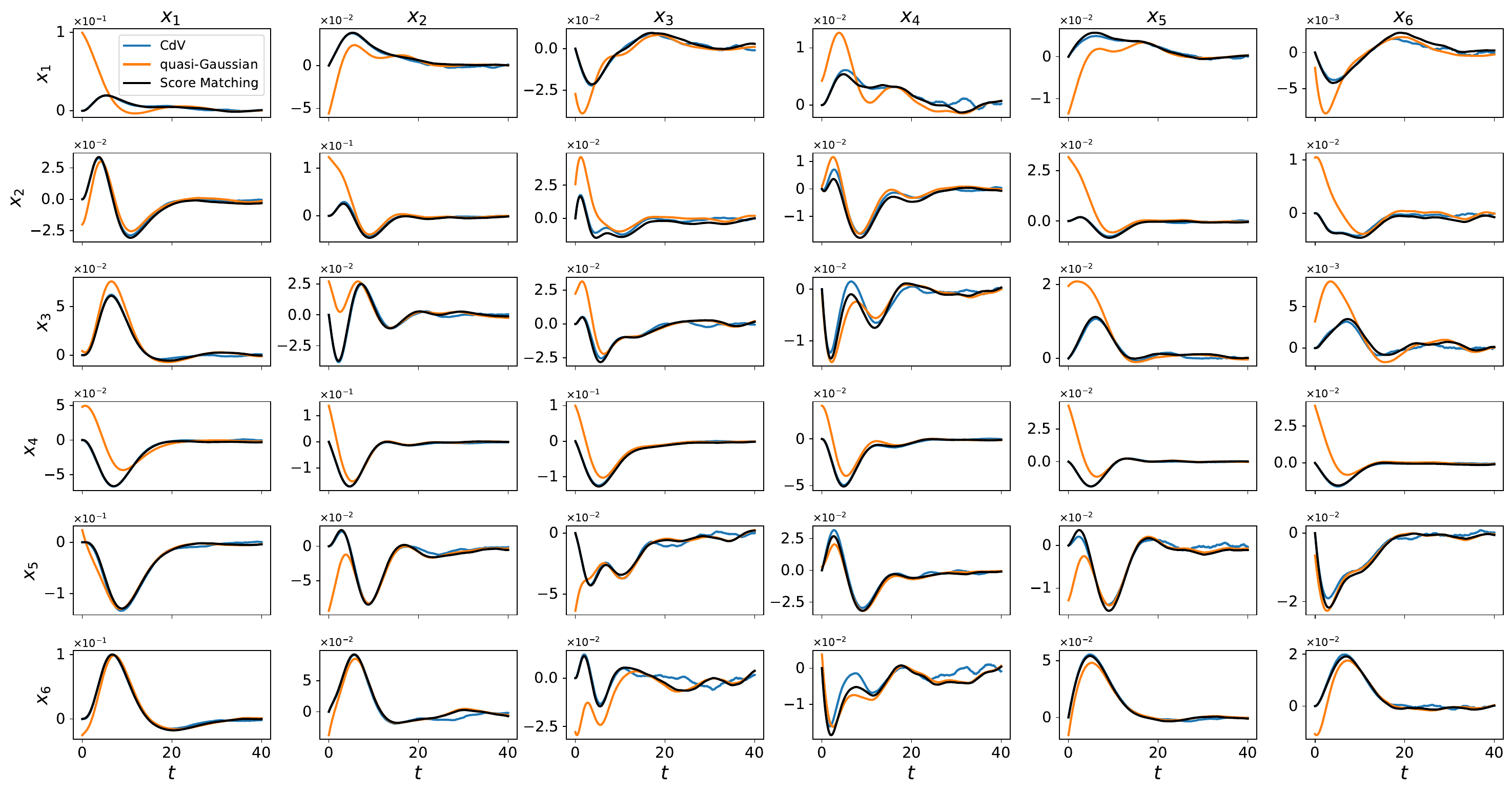}
\caption{Time-dependent responses in ensemble variance $(x^{(k)}_t - \mu^{(k)})^2$ to an impulse perturbation in the CdV model (Eq. (8) in the main text) as predicted by the FDT using the quasi-Gaussian and score-matching approximations. Example: column (1), row (2) quantifies the time-dependent variance response of $x^{k = 2}_t$ given a small impulse perturbation imposed on $x^{j = 1}_0$ at time $t = 0$.}
\label{fig:FDT_var}
\end{figure*}

\paragraph*{Response operator via score matching: response in the ensemble variance.} Given the proposed correction in Eq. \eqref{eq:score-var-correction}, we now infer the variance response operator for the Charney-DeVore system in Eq. (8) in the main text. We fit the qG-FDT approximation and the score-matching based FDT estimation on the long trajectory of the Charney-DeVore system used in the main text. We compare the FDT predictions against the numerical ground truth and show the results in Figure \ref{fig:FDT_var}. The FDT inferred through score matching provides a more accurate response estimate than the quasi-Gaussian approximation.

\section{Possible limitations of the causal constraints and practical considerations} 

Causal relations in high-dimensional stochastic dynamical systems are most naturally, and practically, identified at the level of ensemble-averaged quantities \cite{FabriCoarseGraining}. Accordingly, the causal constraints introduced in the main text exploit vanishing entries of the ensemble-averaged Jacobian, $\langle \partial f_k(\mathbf{x})/\partial x^{(j)} \rangle = 0$, to suppress the corresponding state-dependent couplings $\partial f_k(\mathbf{x})/\partial x^{(j)}$ in the neural network loss. A complication can in principle arise in idealized systems with exact symmetries, such as the Lorenz ’63 and Lorenz ’96 systems~\cite{Lorenz,Lorenz96}. In such cases $R^{k,j}_1 = 0$ can arise from exact cancellations in the ensemble average, even though $\partial f_k(\mathbf{x}) / \partial x^{(j)} \neq 0$ pointwise. One possible strategy in such idealized settings is to explicitly reduce the symmetries of the underlying dynamical system, following symmetry-reduction approaches advocated in~\cite{Predrag}. For example, Ref.~\cite{MIRANDA1993105} proposed a nonlinear coordinate transformation to quotient the $Z_2$ symmetry of the Lorenz ’63 system~\cite{Lorenz}, leading to the so-called ``proto-Lorenz'' system. The limitation discussed above does not constitute a practical obstacle for reduced-order modeling of realistic, complex turbulent dynamical systems. In these realistic flows, symmetries are always broken.

\section{Charney De-Vore model: parameters} \label{app:CdV}

The values of the parameters used for the Charney-DeVore model (Eq. (8) in the main text) are reported in Table~\ref{tab:params} below. All parameters are chosen with standard values as discussed in \cite{de1988low,crommelin2004mechanism,dorrington2023interaction,LudovicoFabriAndre}. The system is integrated using a Euler-Maruyama scheme with $dt = 0.01$.

\begin{table}[h]
\centering
\caption{Model coefficients for the six-dimensional stochastic Charney-DeVore model.}
\label{tab:params}
\begin{tabular}{lll}
\toprule
\textbf{Parameter} & \textbf{Value} & \textbf{Description} \\
\midrule
$C$                & 0.1        & Newtonian relaxation rate \\
$x_1^*$            & 0.95       & Zonal background forcing (mode 1) \\
$x_4^*$            & $-0.76095$ & Zonal background forcing (mode 4) \\
$\gamma$           & $0.2$ & Topographic height \\
$b$           & $1.6$ & Channel aspect ratio \\
$\alpha_m$         & $\frac{8\sqrt{2}}{\pi} \frac{m^2}{4m^2-1} \frac{b^2 + m^2 -1}{b^2 + m^2}$ & Nonlinear advection (mode $m$) \\
$\beta_m$          & $\frac{\beta b^2}{b^2 + m^2}$ & Coriolis effects (mode $m$) \\
$\delta_m$         & $\frac{64 \sqrt{2}}{15 \pi} \frac{b^2 + m^2 +1}{b^2 + m^2}$ & Triad interaction (mode $m$) \\
$\gamma_m$         & $\gamma \frac{4m^3}{4m^2-1} \frac{\sqrt{2}b}{\pi(b^2+m^2)}$ & Orographic damping (mode $m$) \\
$\tilde{\gamma}^{*}_m$ & $\gamma \frac{4m}{4m^2-1} \frac{\sqrt{2}b}{\pi}$ & Orographic forcing (mode $m$) \\
$\varepsilon$      & $\frac{16\sqrt{2}}{5\pi}$ & Wave-wave interaction \\
$\sigma$           & 0.05       & Noise amplitude \\
$dt$      & $0.01$ & Integration time step \\
\bottomrule
\end{tabular}
\end{table}

\section{Neural emulators of the Charney-DeVore model: stationary statistics} \label{sec:stationary_density}

Stationary density and autocorrelation functions of the Charney-deVore system as given by the numerical model, the vanilla emulator and the causal emulator.

\begin{figure*}[h!]
\centering
\includegraphics[width=1\textwidth]{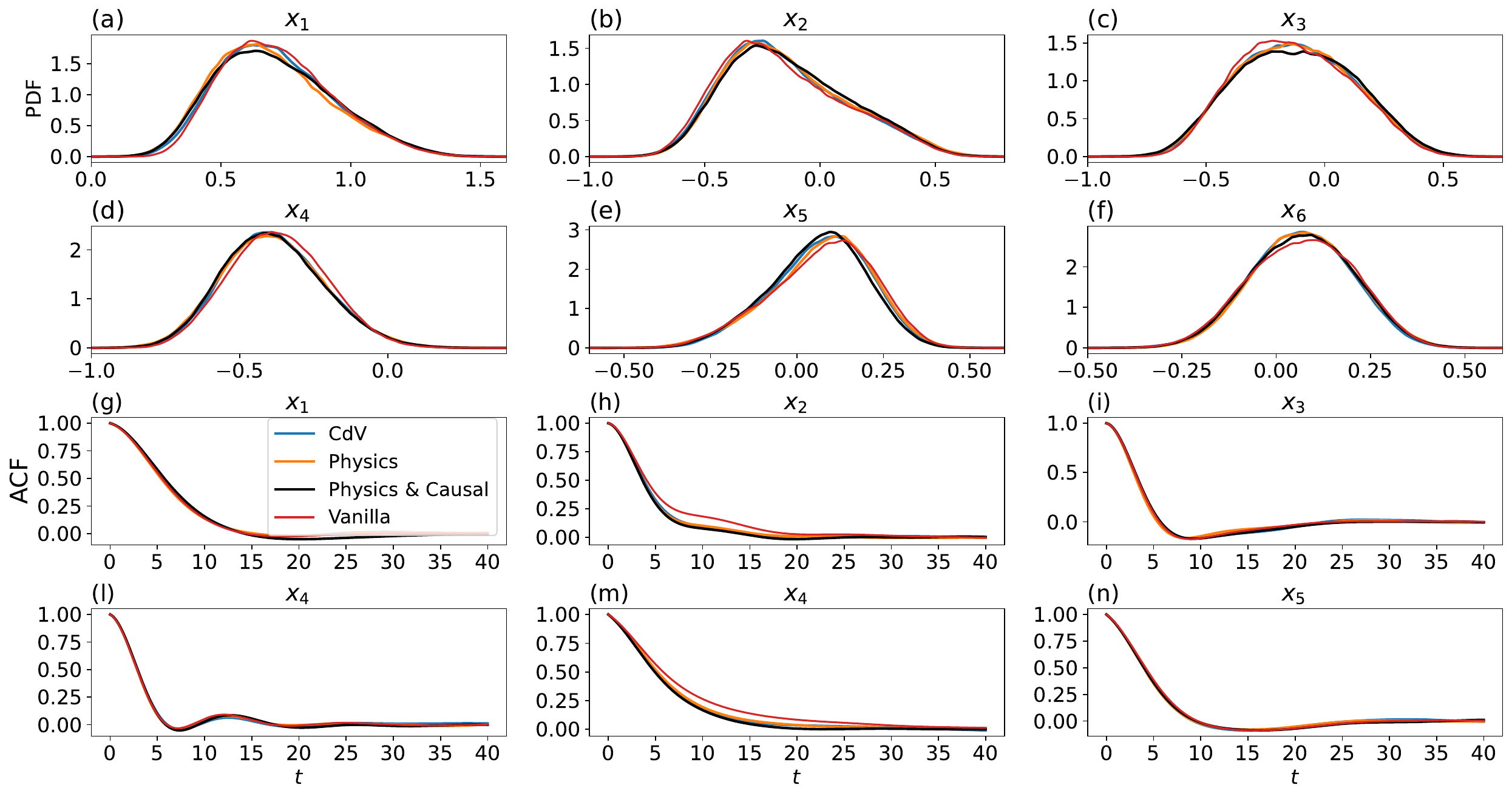}
\caption{Stationary statistics. First and second rows: stationary distributions (PDFs) of the variables $x^{(1)},...,x^{(6)}$ of the CdV model. Labels: ``CdV'' refers to the numerical model; ``Physics'' denotes the model with only physics constrained; ``Physics $\&$ Causal'' denotes the model constrained by both physics and causal information; ``Vanilla'' denotes the unconstrained model. Third and fourth rows: same as the first and second rows but for the autocorrelation functions (ACFs).}
\label{fig:Stationary_Statistics}
\end{figure*}

\section{Response operator for the Charney-DeVore model} \label{sec:response_operator_CdV}

\begin{figure*}[h!]
\centering
\includegraphics[width=1\textwidth]{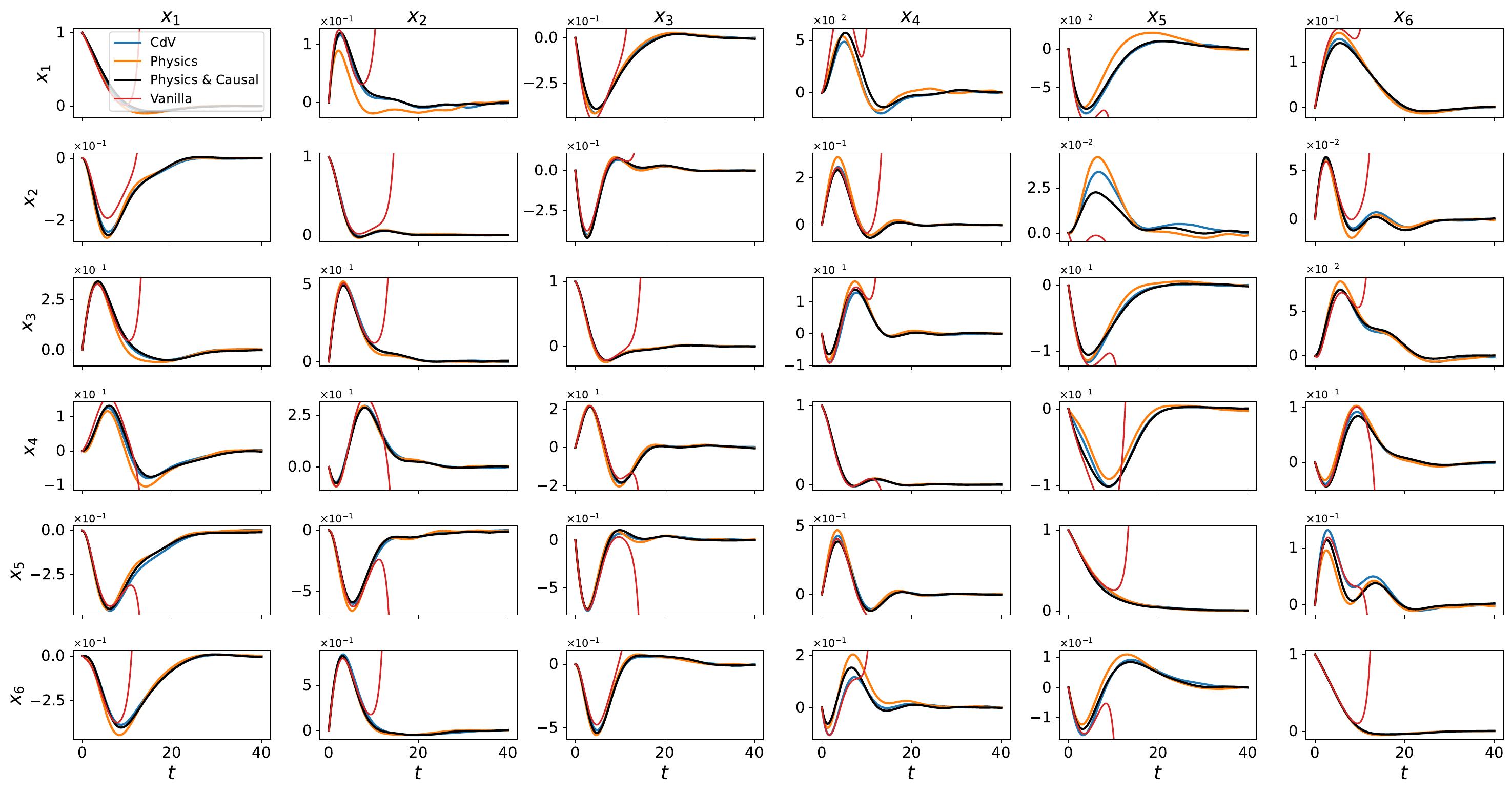}
\caption{Time-dependent responses in ensemble mean of variable $x^{(k)}_t$ to an impulse perturbation in the CdV model in Eq. (8) in the main text as predicted by the unconstrained and causality constrained emulators. Example: column (1), row (2) shows the time-dependent response of $x^{(k=2)}_t$ given a small impulse perturbation imposed on $x^{(j=1)}_0$ at time $t = 0$. Labels: ``CdV'' refers to the numerical model; ``Physics'' is the model with only physics constrained; ``Physics $\&$ Causal'' is the model constrained by both physics and causal information; ``Vanilla'' is the unconstrained model.
}
\label{fig:emulator_FDT_mean}
\end{figure*}

\begin{figure*}[h!]
\centering
\includegraphics[width=1\textwidth]{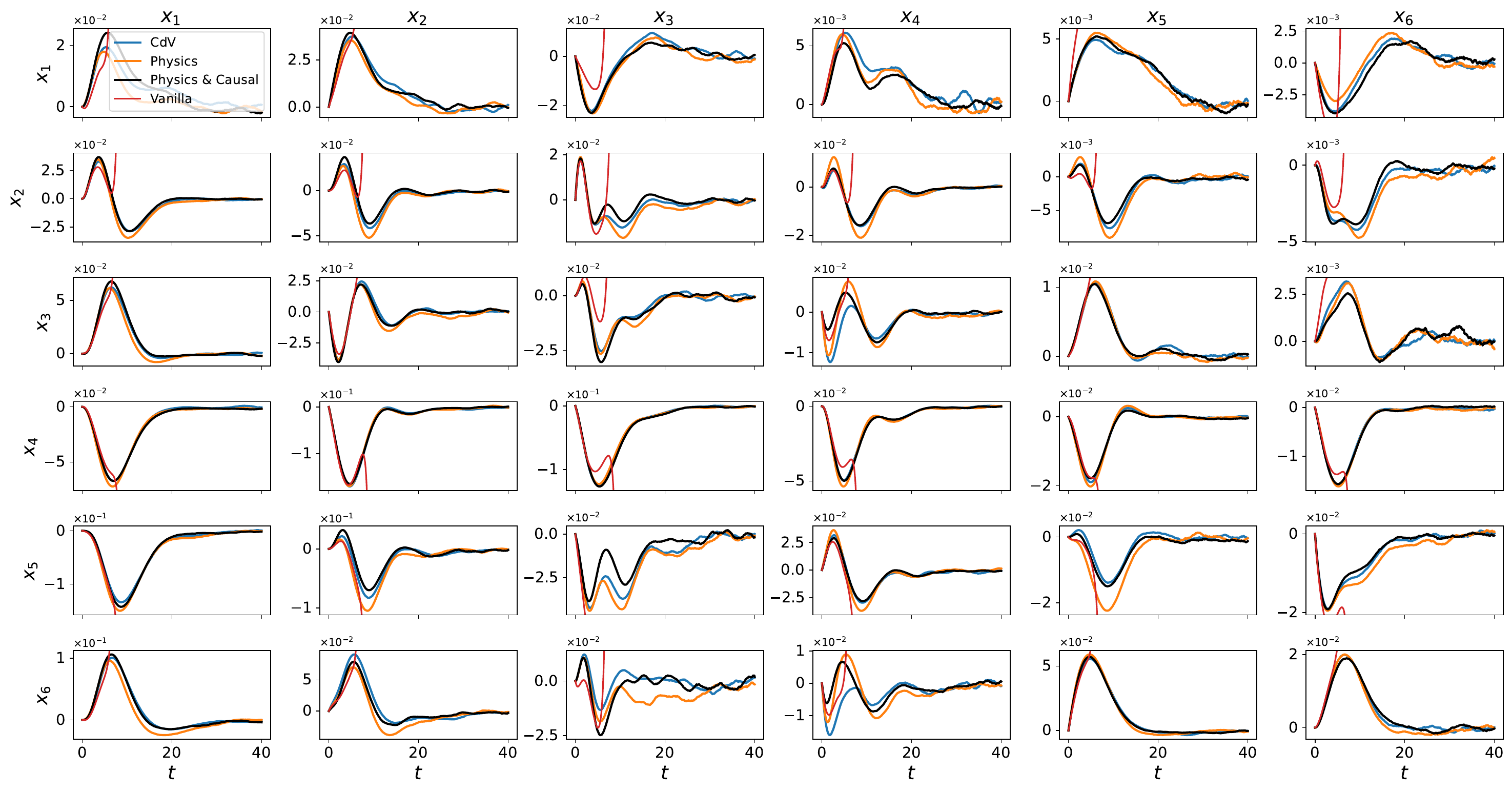}
\caption{Time-dependent responses in ensemble variance of variable $x^{(k)}_t$ to an impulse perturbation in the CdV model in Eq. (8) in the main text as predicted by the unconstrained and causality constrained emulators. Example: column (1), row (2) shows the time-dependent response of $x^{(k = 2)}_t$ given a small impulse perturbation imposed on $x^{(j = 1)}_0$ at time $t = 0$. Labels: ``CdV'' refers to the numerical model; ``Physics'' is the model with only physics constrained; ``Physics $\&$ Causal'' is the model constrained by both physics and causal information; ``Vanilla'' is the unconstrained model.
}
\label{fig:emulator_FDT_var}
\end{figure*}

\begin{figure*}
\centering
\includegraphics[width=1\textwidth]{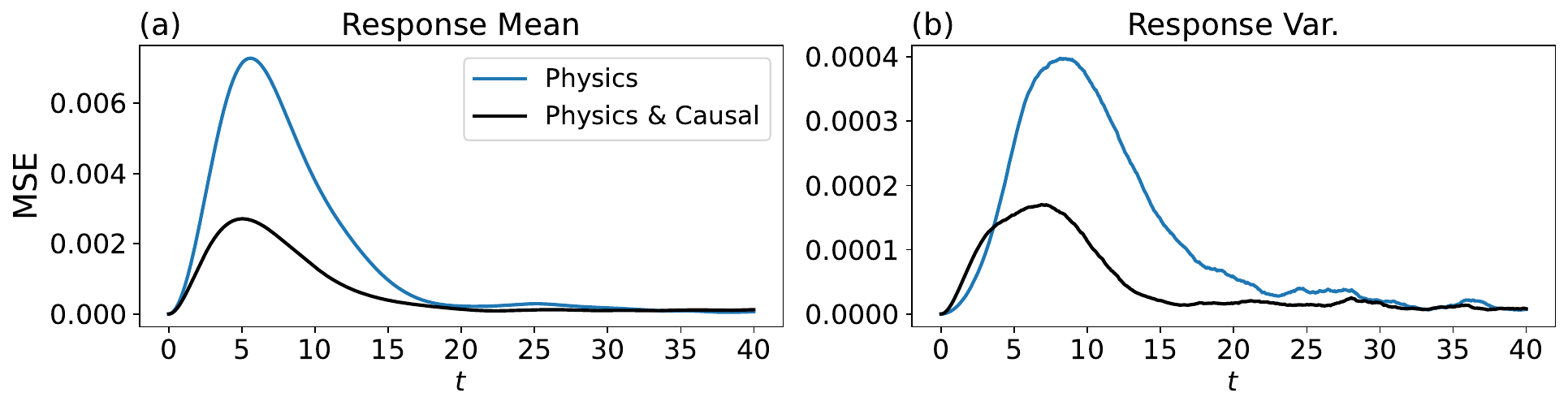}
\caption{Perturbed statistics in the linear regime. Panel (a): Total MSE over time for the mean response, computed from the full response operator $\mathbf{R}_t\in\mathbb{R}^{6, 6}$ for each model. Panel (b): Same as Panel (a) but the response in ensemble variance.}
\label{fig:MSE_impulse_response}
\end{figure*}

In Figure \ref{fig:emulator_FDT_mean} and Figure \ref{fig:emulator_FDT_var}, we report the estimation of the full response operator in ensemble mean and variance as obtained by the numerical model, the constrained emulators and the unconstrained ``vanilla'' emulator. The response operators have been computed with the method presented in Appendix B of the main paper and using $N_e = 10^6$ ensemble members. The physics constrained emulator provides a very good representation of responses to impulse perturbations, as expected given the large training dataset and Markovianity of the system \cite{FabriCoarseGraining}. The causality constrained emulator yields a systematically improved representation of both mean and variance responses. This is further quantified by the total time-dependent MSE, aggregated over all $j$ and $k$, between the emulators and the ground truth. This analysis is shown in Figure \ref{fig:MSE_impulse_response}. The vanilla emulator becomes unstable under impulse perturbations, despite achieving a training MSE comparable to that of the physics-constrained model.

\section{Response to large step function forcings in the Charney-DeVore model} \label{sec:response_step_CdV}

\begin{figure*}[h!]
\centering
\includegraphics[width=1\textwidth]{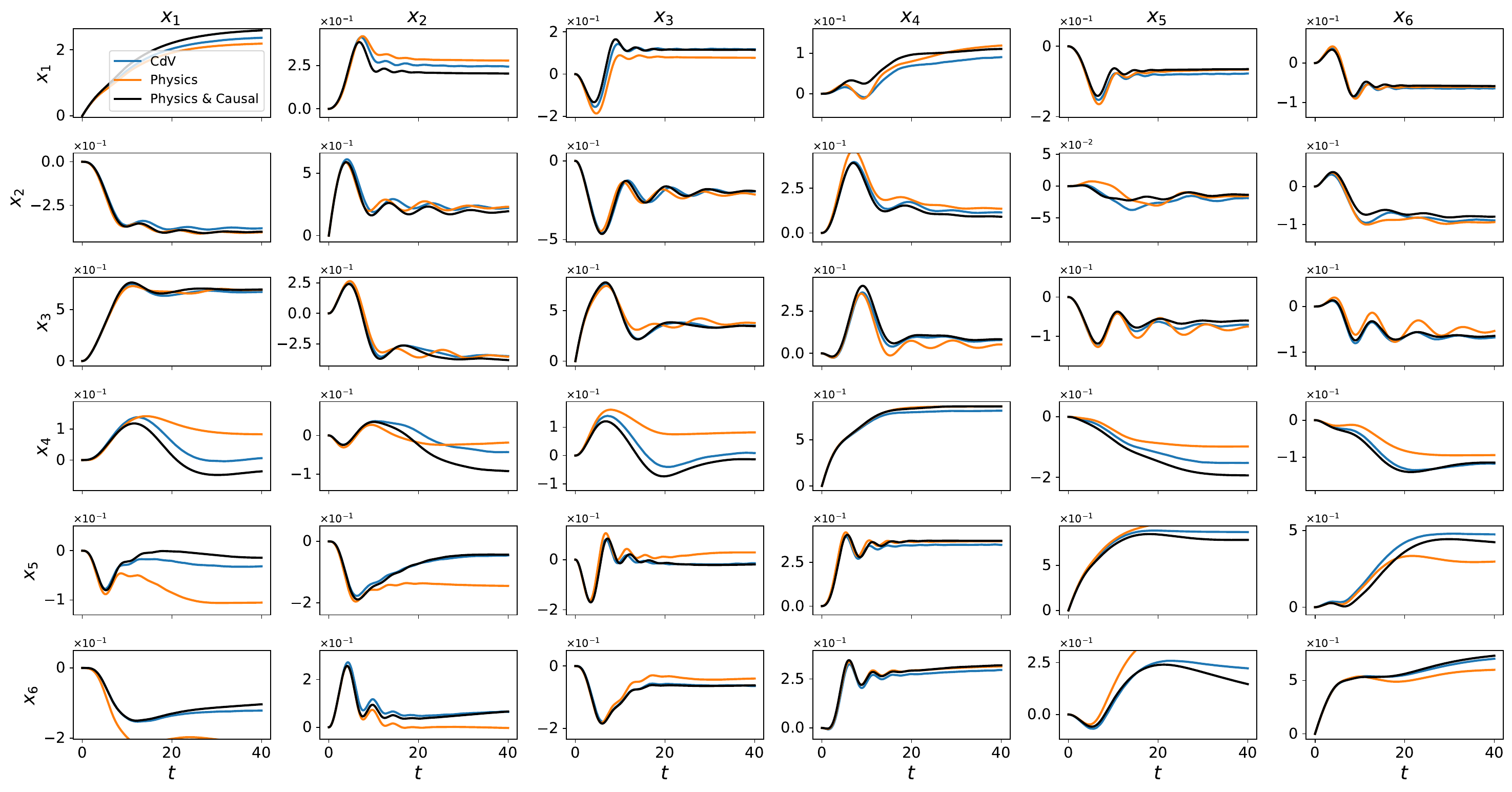}
\caption{Perturbed statistics in the nonlinear regime. Row (i) shows the response in ensemble mean to a step function forcing applied to $x_i$; e.g., row (1): response in ensemble mean to a step function forcing $\mathbf{F} = (\sigma_1,0,0,0,0,0)$ applied for $t \ge 0$, where $\sigma_1$ is the standard deviation of $x_1$. Row (2): same as Row (1) but using $\mathbf{F} = (0,\sigma_2,0,0,0,0)$. Labels: ``CdV'' refers to the numerical model; ``Physics'' is the model with only physics constrained; ``Physics $\&$ Causal'' is the model constrained by both physics and causal information.
}
\label{fig:emulator_step_mean}
\end{figure*}

\begin{figure*}[h!]
\centering
\includegraphics[width=1\textwidth]{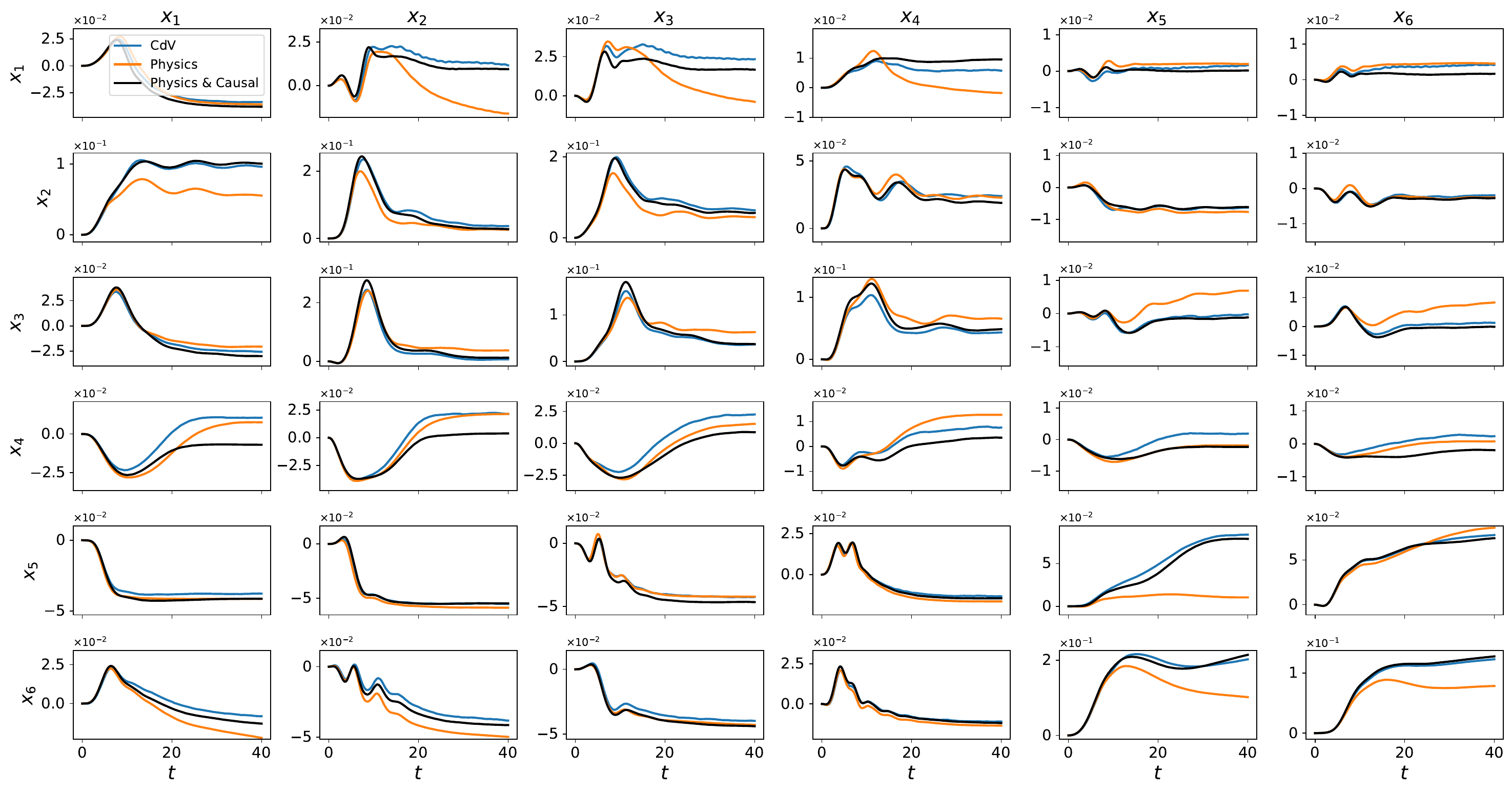}
\caption{Perturbed statistics in the nonlinear regime. Row (i) shows the response in ensemble variance to a step function forcing applied on $x_i$; e.g., row (1): response in ensemble variance to a step function forcing $\mathbf{F} = (\sigma_1,0,0,0,0,0)$ applied for $t \ge 0$, where $\sigma_1$ is the standard deviation of $x_1$. Row (2): same as Row (1) but using $\mathbf{F} = (0,\sigma_2,0,0,0,0)$. Responses are computed using an ensemble of $N_e = 10^5$ members. Labels: ``CdV'' refers to the numerical model; ``Physics'' is the model with only physics constrained; ``Physics $\&$ Causal'' is the model constrained by both physics and causal information.
}
\label{fig:emulator_step_var}
\end{figure*}

\begin{figure*}
\centering
\includegraphics[width=1\textwidth]{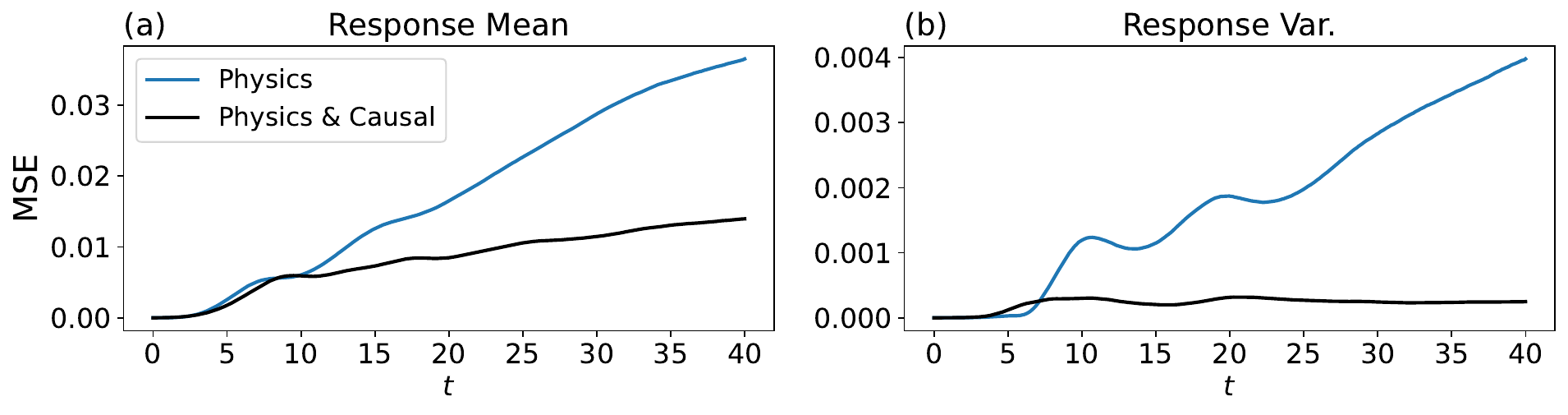}
\caption{Perturbed statistics in the nonlinear regime. Panel (a): Total MSE over time for the ensemble mean response. Panel (b): Same as Panel (a) but the response in ensemble variance.}
\label{fig:MSE_step_response}
\end{figure*}

We now consider the nonlinear response regime by adding a step function forcing $\mathbf{F}$ to both the numerical (CdV) and neural models. In the main text we focused on the case $\dot{\mathbf{x}} = \mathbf{F} + \mathbf{f}(\mathbf{x}) + \mathbf{\Sigma} \bm{\xi}(t)$, where $\mathbf{F} = (0,0,0,0,0,\sigma_6)$ for $t \geq 0$, $\sigma_6$ being the std. dev. of the $x_6$ variable. We stress that this is a very large forcing, imposed on a model that had only access to unperturbed variability. Here, we perform six forcing experiments, applying a step forcing to each degree of freedom in turn with amplitude equal to its standard deviation: $\mathbf{F}=(\sigma_1,0,\ldots,0)$, $(0,\sigma_2,0,\ldots,0)$, $\ldots$, $(0,\ldots,0,\sigma_6)$. For each case, we analyze the time-dependent response of the ensemble mean and variance of the full system. Results are shown in Fig.~\ref{fig:emulator_step_mean} (mean) and Fig.~\ref{fig:emulator_step_var} (variance). The emulator constrained by both physics and causality generally leads to a better representation of the responses. This is clearly shown in the total (aggregated over all responses and perturbations) MSE in Figure \ref{fig:MSE_step_response}, where the ``physics and causal'' emulator shows lower error in the long-time response.

\section{Symmetry-Broken Lorenz-96 (L96) system} \label{sec:L96}

\paragraph{Random forcing $F_j$.} In this work, we introduce a time-independent forcing in the L96 system defined as $F_j = F + \eta_j$, where $\eta_j \sim \mathcal{N}(0,\sigma_F^2)$ with $\sigma_F = 10$. Due to the random component, we report the specific realization used in our experiments for reproducibility: $F_j = $ (6.1087865 , 12.32213349, 28.87925261, 17.93974419, 25.202309  , 21.77103791,  9.63536354, 21.4195222 , 12.83404549, 12.77610884, 16.97167319,  4.        , 27.92166104,  9.28910325, 26.0026942 , 17.36321124, 31.3203308 ,  9.40030586, 12.88205144, 19.37769127).

\subsection{Stationary statistics} \label{sec:L96-stationary}

\begin{figure*}[h!]
\centering
\includegraphics[width=1\textwidth]{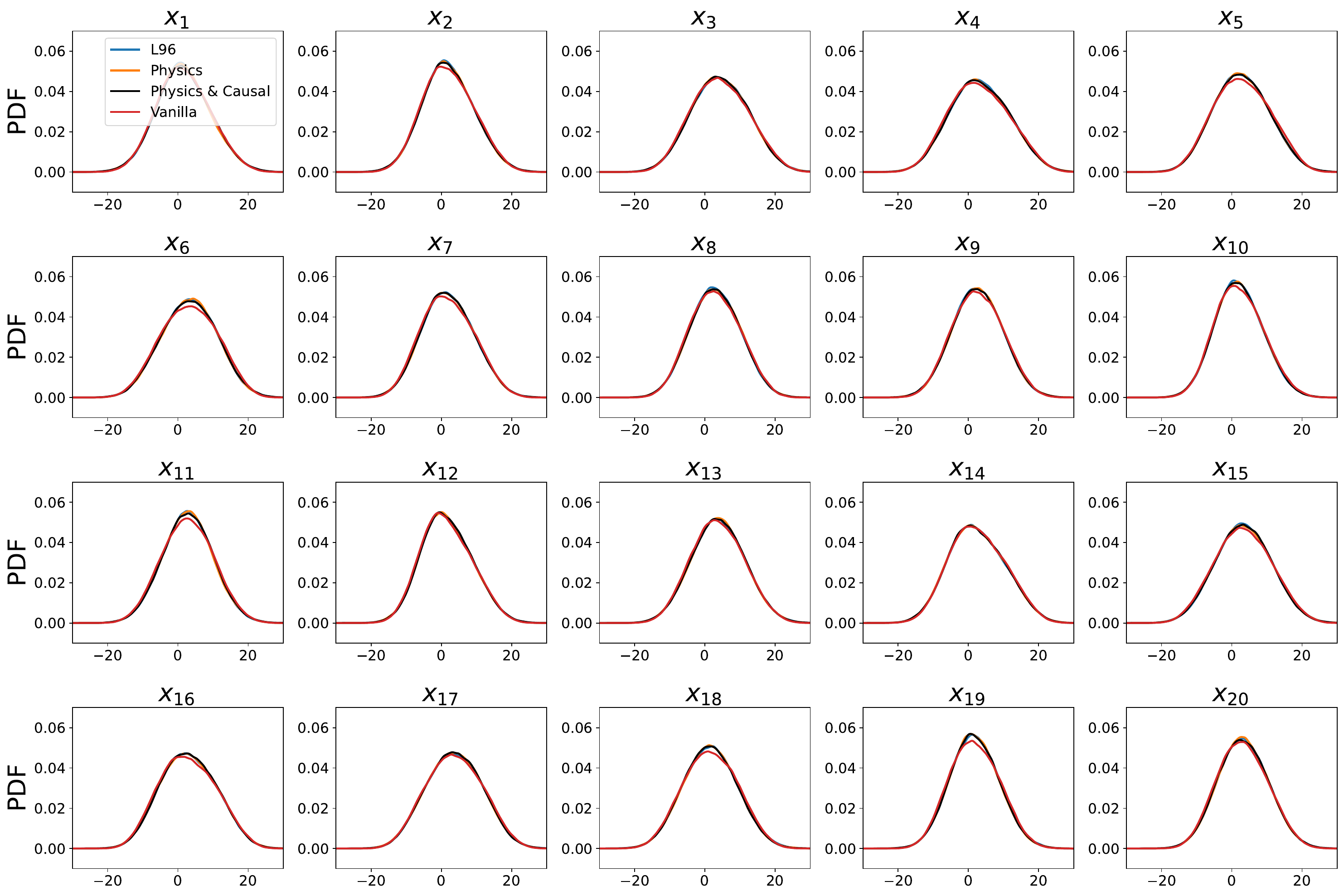}
\caption{Stationary statistics. Stationary distributions (PDFs) of the variables $x_1,...,x_{20}$ of the L96 system considered in the main text. Labels: ``L96'' refers to the numerical model; ``Physics'' denotes the model with only physics constrained; ``Physics $\&$ Causal'' is the model constrained by both physics and causal information; ``Vanilla'' denotes the unconstrained emulator.}
\label{fig:Stationary_Statistics_L96}
\end{figure*}

\begin{figure*}[h!]
\centering
\includegraphics[width=1\textwidth]{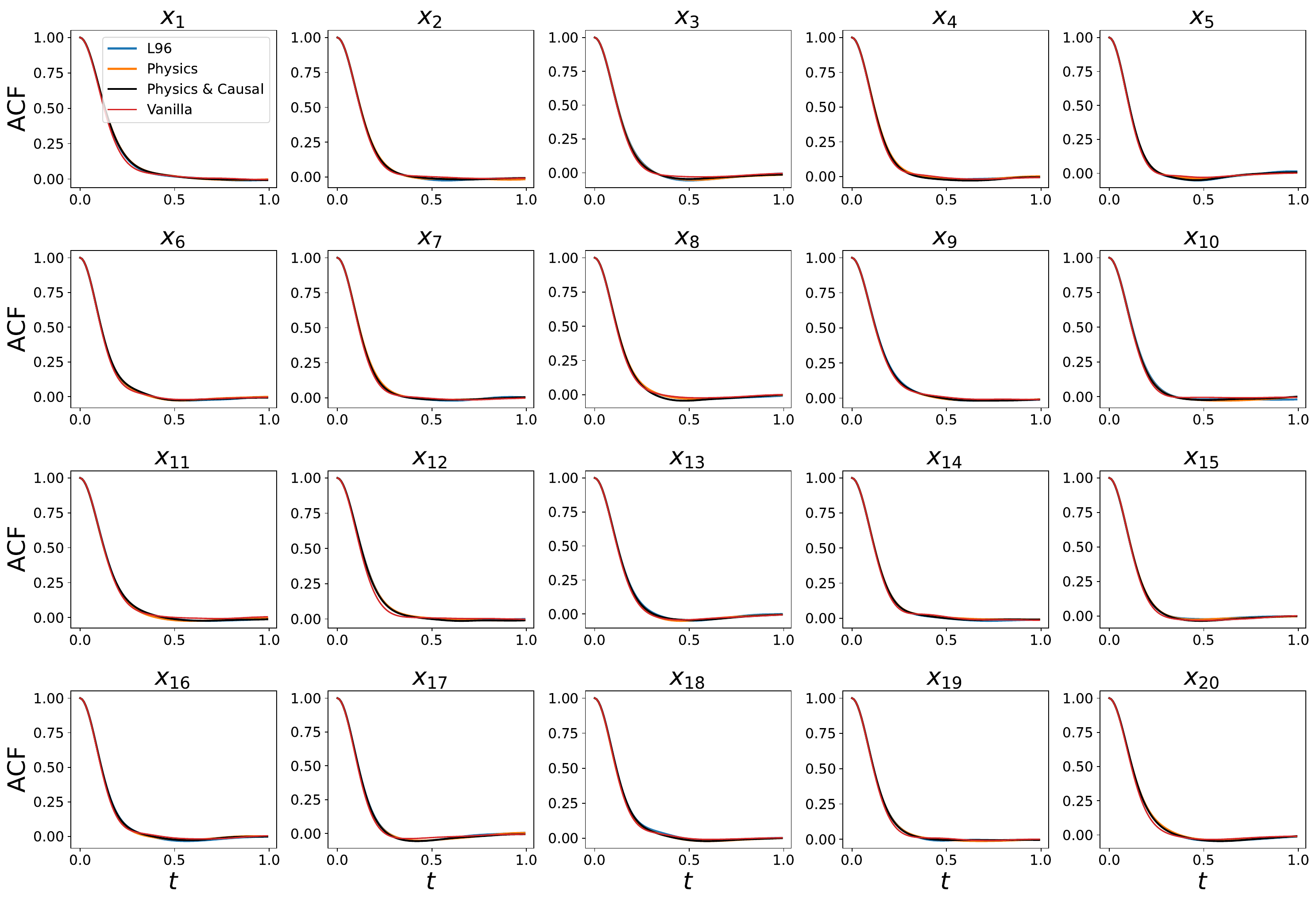}
\caption{Autocorrelation functions (ACFs) of the variables $x_1,...,x_{20}$ of the L96 system considered in the main text. Labels: ``L96'' refers to the numerical model; ``Physics'' is the model with only physics constrained; ``Physics $\&$ Causal'' is the model constrained by both physics and causal information.}
\label{fig:ACFs_L96}
\end{figure*}

The marginals of the invariant distribution of the L96 system predicted by the two constrained emulators and by the unconstrained emulator are shown in Fig.~\ref{fig:Stationary_Statistics_L96}. The corresponding autocorrelation functions are reported in Fig.~\ref{fig:ACFs_L96}.

\subsection{Dependence on the penalty parameter $\lambda$} \label{sec:lambda-dependence}

\begin{figure*}[h!]
\centering
\includegraphics[width=1\textwidth]{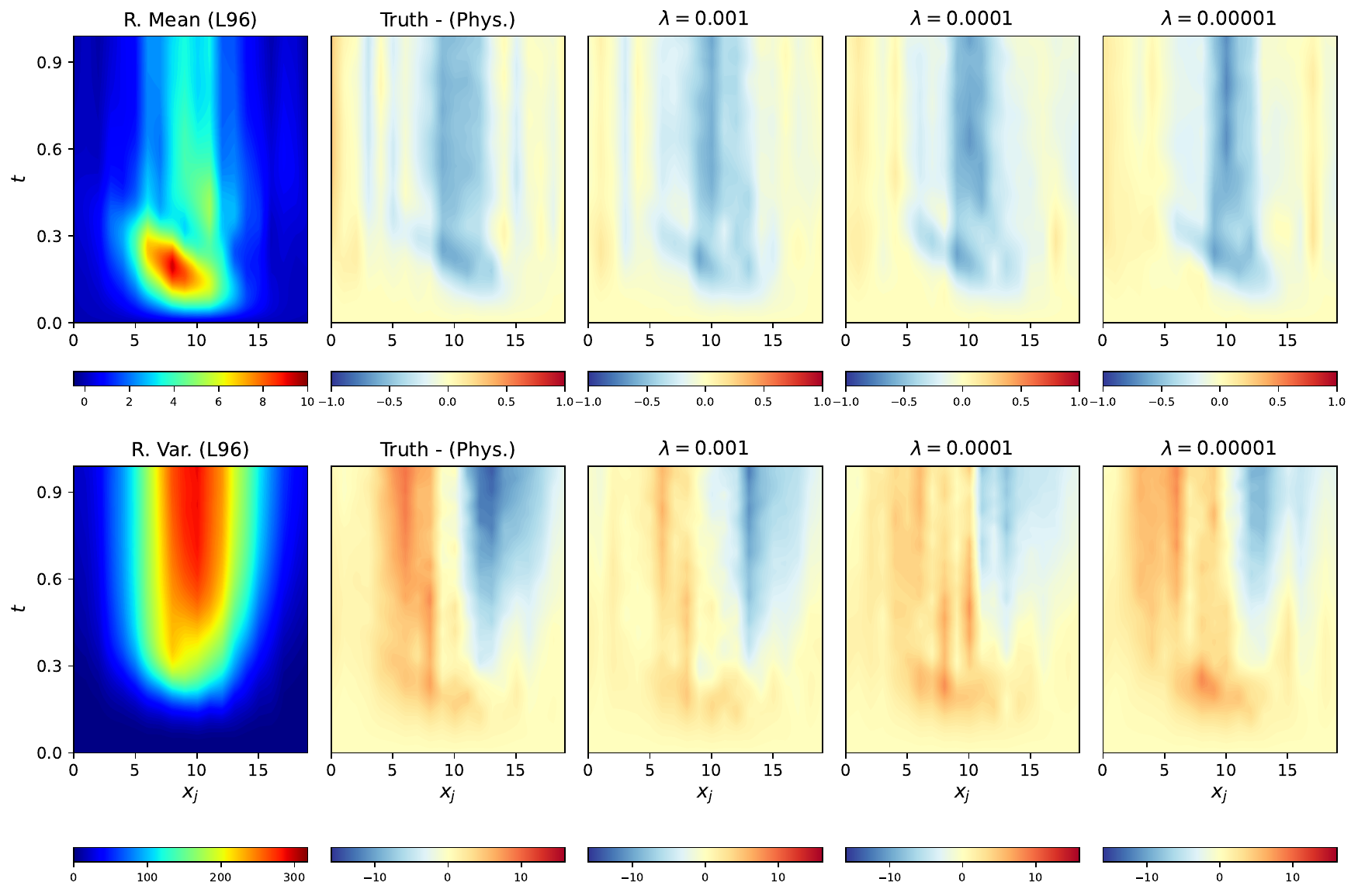}
\caption{Sensitivity of the Lorenz--96 response prediction to the causal-penalty strength $\lambda$. The system is forced by the large Gaussian forcing in Eq.~(10) of the main text, applied and held constant for $t \geq 0$. Top row: ensemble-mean response. Bottom row: ensemble-variance response. Left column: ground-truth response. Second column: difference between the physics-constrained emulator and the truth. Third, fourth, and fifth columns: difference between the physics- and causality-constrained emulator and the truth for $\lambda = 10^{-3}, 10^{-4}, 10^{-5}$, respectively.}
\label{fig:lambda_sensitivity}
\end{figure*}

The strength of the causal penalty in Eqs.~(6) and (7) of the main text is controlled by the parameter $\lambda$. This parameter determines the relative weight assigned to suppressing dependencies that are excluded by the inferred causal graph. When the inferred graph is accurate, the causal penalty does not significantly interfere with model fitting by the MSE loss. This is the case for the Charney--DeVore example in Section~4.1, where the identified causal links are exact. Consequently, the results for this system are only weakly dependent on the precise value of $\lambda$. The role of $\lambda$ becomes more important when the inferred causal graph contains false positives or false negatives, as in the Lorenz-96 example in Section~4.2. In this case, the causal penalty in Eqs.~(6) and (7) can partially compete against the data-fitting objective given by the MSE loss. If $\lambda$ is chosen too large, the penalty may suppress physically relevant dependencies, thereby degrading the fitted model. If $\lambda$ is chosen too small, the causal constraint becomes ineffective and the model approaches the purely physics-constrained emulator. In the Lorenz-96 experiments reported in the main text, we used the causal penalty in Eq.~(7) with $\lambda = 10^{-4}$. The parameter $\gamma$ was first estimated using the heuristic described in Section~\ref{sec:heuristic_gamma} and then adjusted empirically. We find that the stationary statistics are largely insensitive to the values of $\lambda$ considered here. The main dependence appears in the response to external perturbations, which is the more stringent diagnostic for this work. To assess this sensitivity, we repeat the Lorenz-96 analysis of Section~4.2 for three values of the causal-penalty strength,
\[
\lambda = 10^{-3}, \qquad 10^{-4}, \qquad 10^{-5}.
\]
For each value, we train a physics- and causality-constrained emulator and evaluate its response to the large Gaussian forcing in Eq.~(10) of the main text, applied and held constant for $t \geq 0$. Figure~\ref{fig:lambda_sensitivity} shows the resulting responses of the ensemble mean and variance. Across this range of $\lambda$, the response predictions are largely unchanged. In particular, the variance response is consistently improved by the causality-constrained model relative to the purely physics-constrained emulator. These results indicate that the conclusions of Section~4.2 are robust to variations of the causal-penalty strength across two orders of magnitude.

\clearpage

\section{Visualization of response operator associated with SST modes alone} \label{sec:noisy_r_operator_SM}

In Appendix C of the main text, we showed that empirical FDT estimations from the observational data considered in Section 5 are inherently limited by sample size, in agreement with theoretical expectations \cite{nonEqStatMech}. This limitation motivates our strategy of validating the emulator's performance via the FDT on yearly time scales. As a representative example, Appendix C examined the response operator $R^{(2,1)}_t$, representing the time-dependent mean response of $x^{(2)}_t$ to an impulse perturbation applied to the first mode (the SST ENSO mode) $x^{(1)}_{t=0}$ at $t = 0$. 

\begin{figure}
\centering
\includegraphics[width=1\linewidth]{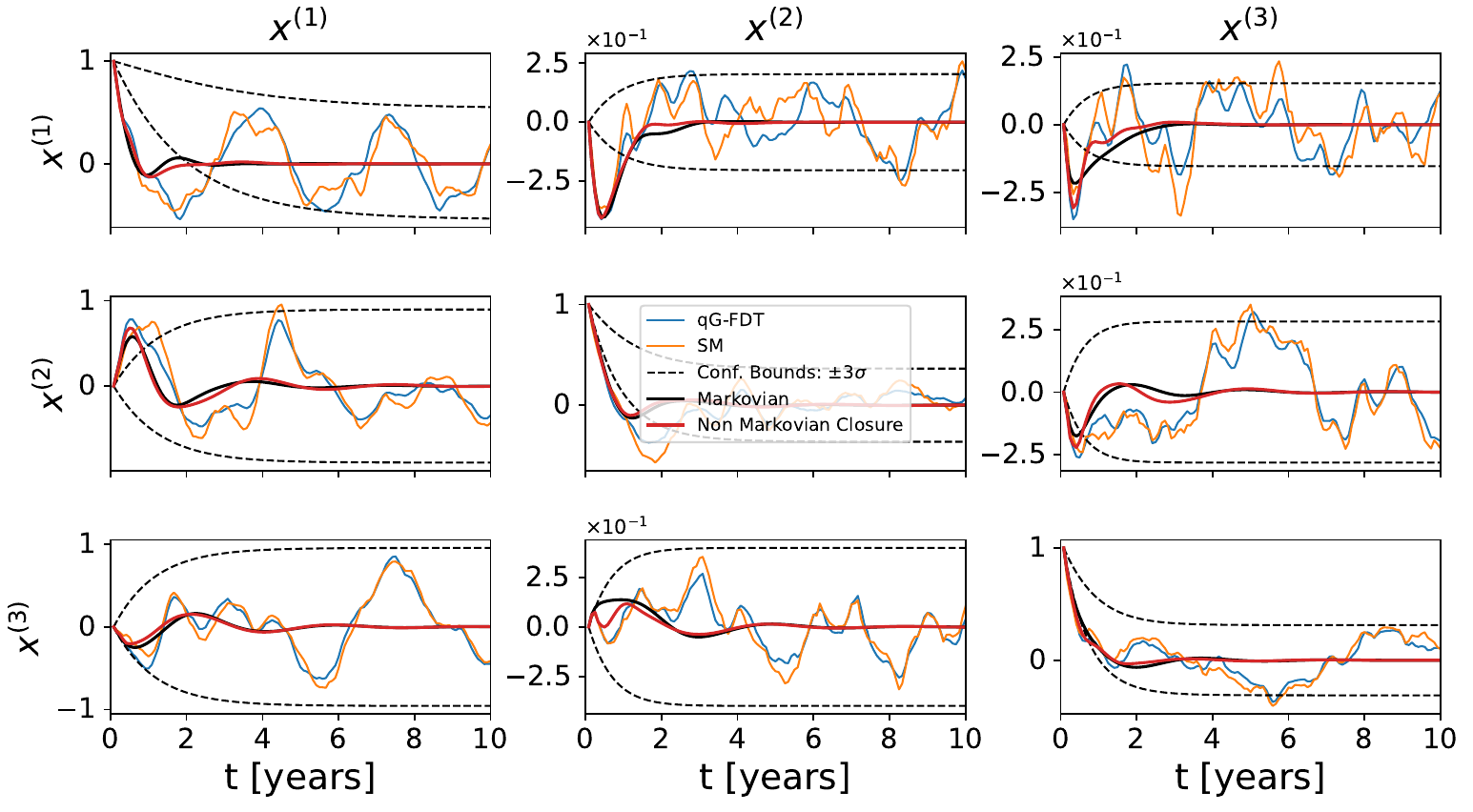}
\caption{Same as Figure 10 in the main text, but expanded to show the time-dependent mean response operators $R^{(k,j)}_t$ for the leading modes $k,j \in \{1,2,3\}$. In blue: qG-FDT prediction; In black: prediction of the Markovian model; In red: prediction of the non-Markovian model. x-axis: years.}
\label{fig:response_data_SST}
\end{figure}

Here, we expand this analysis to the $3 \times 3$ response matrix $R^{(k,j)}_t$ by also including the model with the non-Markovian closure. While the full state vector analyzed in the main text is 20-dimensional (which would yield a full response tensor of $20 \times 20 \times 120$ months), visualizing this $3 \times 3$ subset provides a sufficient and representative illustration of the finite-sample noise problem. Figure \ref{fig:response_data_SST} displays the response operators $R^{(k,j)}_t$ for $k,j \in \{1,2,3\}$. Figure~\ref{fig:response_data_SST} also highlights how the non-Markovian closure consistently corrects the responses of the modeled degrees of freedom. This correction is more visible in the full
$20 \times 20$ response matrix, which includes the faster modes explaining less variance; we do not
show it here for clarity, as it would comprise $20 \times 20 \times 120$ values.

\section{Response validation against the FDT integrated over $\tau_\infty = 3$ months} \label{sec:sensitivity_3_months}

\begin{figure}
\centering
\includegraphics[width=1\linewidth]{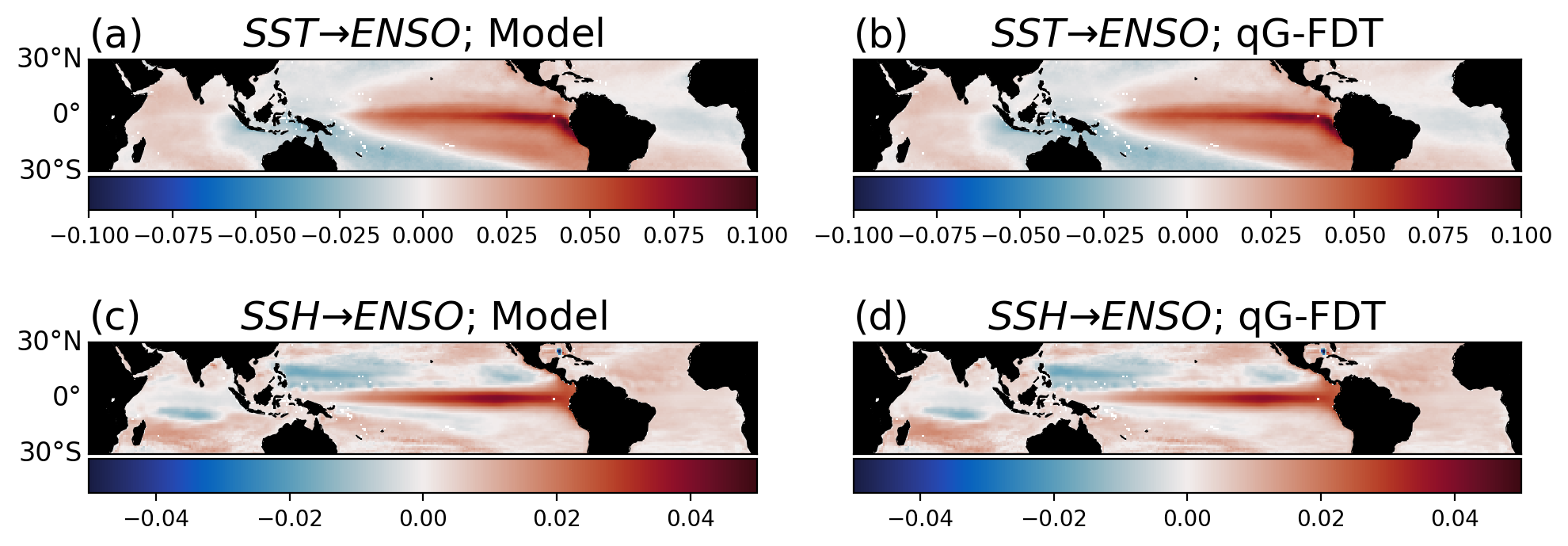}
\caption{Same as Figures 7 and 8 in the main text, but the sensitivities have been computed by integrating responses over $\tau_\infty = 3$ months.}
\label{fig:validation_ENSO_FDT_3months}
\end{figure}

Comparison between the FDT-based sensitivity benchmark and the Markovian emulator. Sensitivities have been computed by integrating response patterns over a period of $\tau_\infty = 3$ months. The analysis is shown in Figure \ref{fig:validation_ENSO_FDT_3months}. The model’s short-time sensitivities agree well with the FDT predictions. In both cases, the analysis reveals that on short time scales, the ENSO mode is primarily sensitive to local processes, as expected. The largest sensitivities are found in the SST field and confined to the tropical Pacific, followed closely by the SSH field, which is also confined to the Pacific


\clearpage

\stopcontents[sm]

\bibliographystyle{unsrtnat}
\bibliography{references}

\end{document}